\documentclass[nobibstyle]{twcccreport}
\reportnumber{2024-01}
\usepackage{color}
\usepackage{amsmath,amssymb,amsfonts}
\usepackage{algorithm,algorithmic}
\usepackage{graphicx}
\usepackage{hyperref}
\hypersetup{hidelinks=true}
\usepackage{textcomp}
\usepackage[shortlabels]{enumitem}
\usepackage{comment}

\usepackage{amsthm}
\usepackage{cleveref}

\usepackage{braket}
\usepackage{mathtools}

\usepackage[numbers,sort&compress]{natbib}
\usepackage{bibentry}

\usepackage{ifthen}
\newboolean{LongVersion}
\setboolean{LongVersion}{true} % true or false
\newboolean{OneColumn}
\setboolean{OneColumn}{true} % true or false
\usepackage{multirow}

\newcommand{\pubidadjcol}[1]{#1}

\newtheorem{theorem}{Theorem}
\newtheorem{conjecture}{Conjecture}
\newtheorem{proposition}{Proposition}

\newtheorem{lemma}{Lemma}

\theoremstyle{definition}
\newtheorem{definition}{Definition}
\newtheorem{assumption}{Assumption}
\newtheorem{remark}{Remark}

\Crefformat{assumption}{#2Assumption~\textup{#1}#3}
\Crefrangeformat{assumption}{Assumptions~\textup{#3#1#4--#5#2#6}}
\Crefmultiformat{assumption}{Assumptions~\textup{#2#1#3}}{ and \textup{#2#1#3}}
{, \textup{#2#1#3}}{, and \textup{#2#1#3}}
\Crefrangemultiformat{assumption}{Assumptions~\textup{#3#1#4--#5#2#6}}%
{ and \textup{#3#1#4--#5#2#6}}{, \textup{#3#1#4--#5#2#6}}{, and
  \textup{#3#1#4--#5#2#6}}

\DeclareMathOperator*{\argmin}{argmin}

\newcommand{\allint}{\mathbb{I}}

\newcommand{\posint}{\mathbb{I}_{>0}}
\newcommand{\intinterval}[2]{\mathbb{I}_{#1:#2}}
\newcommand{\real}{\mathbb{R}}
\newcommand{\nnegreal}{\mathbb{R}_{\geq 0}}

\newcommand{\realm}[2]{\real^{#1\times #2}}
\newcommand{\symm}[1]{\mathbb{S}^{#1}}
\newcommand{\nnegdefm}[1]{\mathbb{S}^{#1}_+}
\newcommand{\posdefm}[1]{\mathbb{S}^{#1}_{++}}

\newcommand{\cplx}{\mathbb{C}}
\newcommand{\cplxm}[2]{\cplx^{#1\times #2}}
\newcommand{\herm}[1]{\mathbb{H}^{#1}}
\newcommand{\nnegdefhm}[1]{\mathbb{H}^{#1}_+}
\newcommand{\posdefhm}[1]{\mathbb{H}^{#1}_{++}}

\newcommand{\trilm}[1]{\mathbb{L}^{#1}}

\newcommand{\postrilm}[1]{\mathbb{L}^{#1}_{++}}

\newcommand{\diag}{\textnormal{diag}}
\newcommand{\vect}{\textnormal{vec}}
\newcommand{\vecs}{\textnormal{vecs}}
\newcommand{\tr}{\textnormal{tr}}

\newcommand{\norm}{\mathcal{N}}
\newcommand{\iid}{\overset{\textnormal{iid}}{\sim}}

\newcommand*{\email}[1]{\href{mailto:#1}{\nolinkurl{#1}} }

\makeatletter
\newcommand{\pleaseciteas}[1]{%
  \begin{twmcc@abskw}{Please cite this work as:}{-1ex}
    #1
  \end{twmcc@abskw}}
\makeatother

\begin{document}
\title{Maximum likelihood identification of linear models with integrating
  disturbances for offset-free control%
  \thanks{This report is an extended version of a published work. Please cite
    the published version. The code is made available
    at~\href{https://github.com/rawlings-group/mlid_2024}{\url{https://github.com/rawlings-group/mlid_2024}}.
    This work was supported by the National Science Foundation (NSF) under Grant
    2138985. (e-mail: \email{skuntz@ucsb.edu}; \email{jbraw@ucsb.edu})} %
  \thanks{Version 2 (November 4, 2024): The text and case studies are updated.
    The main technical results remain unchanged.} %
  \thanks{Version 3 (\today): A reference to the published version of this work
    was added. Corrected typos, unclear language, and readability issues
    throughout.}}
\author{Steven J.~Kuntz and James B.~Rawlings \\
  Department of Chemical Engineering \\
  University of California, Santa Barbara}

\maketitle

\vfill

\pleaseciteas{%
  S. J. Kuntz and J. B. Rawlings, “Maximum Likelihood Identification of Linear
  Models with Integrating Disturbances for Offset-Free Control,” \emph{IEEE
    Trans. Auto. Cont.}, vol. 70, no. 9, pp. 5675–5689, 2025, doi:
  \href{https://doi.org/10.1109/TAC.2025.3547607}{10.1109/TAC.2025.3547607}.
  %% NOTE bibentry seems to be broken for the IEEEtran style so the reference is
  %% manually written above.
  % \bibentry{kuntz:rawlings:2025c}%
}

\vfill

\begin{abstract}%
  This \ifthenelse{\boolean{LongVersion}}{report}{article} addresses
  the maximum likelihood identification of models for offset-free model
  predictive control, where linear time-invariant models are augmented
  with (fictitious) uncontrollable integrating modes, called integrating
  disturbances. The states and disturbances are typically estimated with a
  Kalman filter. The disturbance estimates effectively provide integral control,
  so the quality of the disturbance model (and resulting filter) directly
  influences the control performance. We implement eigenvalue constraints to
  protect against undesirable filter behavior (unstable or marginally stable
  modes, high-frequency oscillations). Specifically, we consider the class of
  linear matrix inequality (LMI) regions for eigenvalue constraints. These LMI
  regions are open sets by default, so we introduce a barrier function method to
  create tightened, but closed, eigenvalue constraints. To solve the resulting
  nonlinear semidefinite program, we approximate it as a nonlinear program using
  a Cholesky factorization method that exploits known sparsity structures of
  semidefinite optimization variables and matrix inequalities. The algorithm is
  applied to real-world data taken from two physical systems: a low-cost
  benchmark temperature microcontroller suitable for classroom laboratories, and
  an industrial-scale chemical reactor at Eastman Chemical's plant in Kingsport,
  TN.\@
\end{abstract}
\section{Introduction}
%% Outline:
% - MPC, LADMs, and estimator-performance-as-control-performance concept
% - Tuning vs identifying LADMs.
% - Challenges in identifying LADMs.

%% MPC is widely used
\ifthenelse{\boolean{OneColumn}}{%
  Offset-free %
}{%
  \IEEEPARstart{O}{ffset-free} %
}%
model predictive control (MPC) is a widely used advanced control method that
combines regulation, estimation, and steady-state optimization problems to track
prescribed setpoints~\cite{qin:badgwell:2003,rawlings:mayne:diehl:2020}. In
linear offset-free MPC, a stochastic linear time-invariant (LTI) model is
augmented with \emph{uncontrollable} integrating modes, called \emph{integrating
  disturbances}, providing integral action through the estimator (typically a
Kalman filter) and allowing offset-free tracking even in the presence of
plant-model mismatch and persistent
disturbances~\cite{muske:badgwell:2002,pannocchia:rawlings:2003}. We call such a
model a \emph{linear augmented disturbance model} (LADM).

The LADM or its corresponding Kalman filter can either ``tuned'' by hand or
identified automatically from data. Common tuning methods include pole
placement~\cite{%
  wallace:das:mhaskar:house:salsbury:2012,% hvac (sim)
  wallace:mhaskar:house:salsbury:2015,% heat pump (sim)
  % ntouskas:sarimveis:sopasakis:2018, % pharmacokinetic model (sim)
  % wang:zheng:li:sun:deng:2019, % induction motor (real)
  schmid:eberhard:2021,% maglev (sim)
  xu:shi:zhang:xu:2023% maglev (real)
}, covariance matrix selection~\cite{%
  caveness:downs:2005,% CSTR (real)
  huang:patwardhan:biegler:2010,% air separation unit (sim)
  % lara:molina:yanes:borroto:2016, % hvac (sim)
  petersen:poulsen:niemann:utzen:jorgensen:2017% industrial spray dryer (sim)
  % deng:yang:wang:yang:2019, % supply ship positioning (real)
  % molina:bonfitto:galluzzi:2021, % magnetic bearings (real)
  % ge:zhao:ma:guo:2022, % autonomous vehicles (real)
  % ge:zhao:zhong:shan:ma:guo:han:2022, % autonomous vehicles (real)
  % wen:pan:liu:zhang:sun:2024% aeroengine (real)
}, and filter gain selection~\cite{%
  kim:choi:lee:lee:2015,% AC/DC converter (real)
  bender:goltz:braunl:sawodny:2017,% Autonomous hydraulic excavator (real)
  deenen:maljaars:dejager:heijman:grull:heemels:2018% Hyperthermic cancer
  % treatment (sim)
  % xie:su:lu:xie:2018% CSTR (sim)
}. Disturbance models can be identified with autocovariance least squares
estimation~\cite{%
  odelson:lutz:rawlings:2003b%, % lab-scale cstr (real)
} or maximum likelihood (ML) identification~\cite{%
  kuntz:rawlings:2022,% TCLab
  kuntz:downs:miller:rawlings:2023a,% Eastman CSTR
  zagrobelny:rawlings:2015a,% no application
  simpson:ghezzi:asprion:diehl:2023% temperature control
}.

Tuning of integrating disturbance models can be a time-consuming and ad hoc
procedure, requiring simplified parameterizations (e.g., diagonal covariance
matrices). In prior work, we have suggested identification as the preferred
strategy for acquiring LADMs~\cite{%
  kuntz:rawlings:2022,% TCLab
  kuntz:downs:miller:rawlings:2023a% Eastman CSTR
}. In this work, we further develop ML identification because of its desirable
statistical properties (consistency, asymptotic efficiency) and ability to
handle general model structures and constraints~\cite{%
  astrom:1979,% survey; includes state-space model
  ljung:1999% good book?
}. %

\pubidadjcol

Design constraints can be included in tuning procedures to avoid undesirable
filter behaviors (slow response time, fictitious high frequencies) that are
passed to the control performance through the integrating disturbance estimates.
%
% Standard identification methods rarely enforce such constraints.
%
Control-relevant design constraints and prior knowledge have sometimes been
incorporated into identification
problems~\cite{piga:forgione:formentin:bemporad:2019,%
  formentin:chiuso:2021,%
  berberich:scherer:allgower:2023}. However, there are no general approaches to
shaping the closed-loop filter behavior in ML identification.
To address this gap, we consider ML identification with eigenvalue constraints
implemented via the linear matrix inequality (LMI) regions commonly used in
robust control~\cite{chilali:gahinet:1996,chilali:gahinet:apkarian:1999}.

%% TODO \cite{bageshwar:borrelli:2009}: ``the closed loop estimator always has a
%% non-oscillating mode slower than the slowest, stable, positive nominal system
%% real mode and an oscillating mode faster than the slowest, stable, negative
%% nominal system real mode.''

%%
LMI region constraints have been used in subspace
identification~\cite{miller:decallafon:2013}. However, subspace identification
cannot be used for LADM identification as it is not possible to impose the
required disturbance model structure. Open-loop stability constraints have been
included in the expectation-maximization (EM)
algorithm~\cite{umenberger:wagberg:manchester:schon:2018}, but this formulation
is not obviously generalized to filter stability or general LMI region
constraints.

\ifthenelse{\boolean{LongVersion}}{%
  While EM is an algorithm for ML, it does not have strong convergence
  guarantees. While it can be shown that the EM iterates produce, almost surely,
  an increasing sequence of likelihood
  values~\cite{shumway:stoffer:1982,gibson:ninness:2005}, slow convergence at
  low noise levels has been reported on a range of problems~\cite{%
    umenberger:wagberg:manchester:schon:2018,% Linear Gaussian state-space models
    redner:walker:1984,% Gaussian mixture models
    bermond:cardoso:1999,% Independent component analysis
    petersen:winther:hansen:2005,% Errors-in-variables state-space models
    petersen:winther:2005,% Gaussian mixture models
    olsson:petersen:lehn-schioler:2007% Linear Gaussian state-space models
  }. Interior point, and even gradient
  methods~\cite{olsson:petersen:lehn-schioler:2007}, are therefore preferable to
  the EM approach. %
}{}

As originally posed by Chilali and
colleagues~\cite{chilali:gahinet:1996,chilali:gahinet:apkarian:1999}, LMI
regions are strict semidefinite matrix inequalities. While Miller and de
Callafon~\cite{miller:decallafon:2013} used relaxed LMI regions with nonstrict
inequalities, as we show in \Cref{sec:stable}, the constraint sets are not
closed, and thus problematic as optimization constraints. To address this issue,
we formulate tightened LMI region constraints that define a closed constraint
set. This formulation introduces nonlinear matrix inequalities and semidefinite
matrix arguments, making the ML problem a nonlinear semidefinite program (NSDP).

To efficiently convert the NSDP to a nonlinear program (NLP), we generalize the
Burer-Monteiro-Zhang (BMZ)
method~\cite{burer:monteiro:zhang:2002,burer:monteiro:zhang:2002a}, which was
originally used to convert sparse semidefinite matrix arguments into vector
arguments with minimal dimension. An additional advantage of the BMZ method over
standard Cholesky factor substitution is that structural knowledge (e.g.,
flowsheet or network structure) can be efficiently imposed in the model
parameterization. Finally, while this work is primarily motivated by
identification of LADMs and offset-free MPC implementations, we remark that any
linear Gaussian state-space model can be identified, with eigenvalue
constraints, using this approach.

The rest of this \ifthenelse{\boolean{LongVersion}}{report}{article} is
organized as follows. In \Cref{sec:ml}, the ML identification problem is stated.
In \Cref{sec:algo} the algorithm is outlined (\Cref{algo:main}). In
\Cref{sec:stable}, we introduce tightened LMI region constraints show they
define closed sets of system matrices (\Cref{prop:lmi:cont}). In
\Cref{sec:chol}, we present our substitution and elimination scheme for
approximating NSDPs as NLPs %
\ifthenelse{\boolean{LongVersion}}{%
  (\Cref{prop:gbmz:value,prop:gbmz:cont}). %
}{%
  (\Cref{prop:gbmz:cont}). %
}%
In \Cref{sec:casestudies}, we apply the algorithm to two real-world applications
of offset-free MPC:\@ first, a benchmark temperature microcontroller used for
classroom laboratories and prototyping~\cite{park:martin:kelly:hedengren:2020},
and second, an industrial-scale chemical reactor at Eastman Chemical's plant in
Kingsport, TN~\cite{kuntz:downs:miller:rawlings:2023a}. Finally, in
\Cref{sec:conclusion} we discuss broader implications and potential future
research.

%% TODO LADM (or even SLTI) identification as ID for control (LQG/OFMPC)?:
% ``Given a control performance objective, design the identification in such a
% way that the performance achieved by the model-based controller on the true
% system is as high as possible.''

\ifthenelse{\boolean{LongVersion}}{%
  % TODO cite published paper
  This report is an extended version of the published
  work~\cite{kuntz:rawlings:2025c}, and contains additional review, discussion,
  and proofs of minor results that were omitted from the journal version due to
  page limitations. Compared to the journal version, this report contains the
  following additions:
  \begin{itemize}
    %% TODO make sure I listed everything
  \item a longer discussion of problem formulations in \Cref{sec:ml};
  \item a characterization of models that can be converted to innovation form
    (\Cref{prop:dare}, see \Cref{app:dare} for proof);
  \item additional basic LMI regions in \Cref{sec:algo} (see \Cref{lem:lmi:basic});
  \item an explicit counterexample of \cite[Thm.~1]{miller:decallafon:2013},
    which (incorrectly) characterized the eigenvalues of relaxed LMI regions, in
    \Cref{conj:lmi:wrong};
  \item a proof of our (correct) characterization of the eigenvalues of relaxed
    LMI regions (\Cref{prop:lmi:closed}) in \Cref{app:lmi};
  \item a proof of the fact that relaxed LMI regions define neither open nor
    closed sets of system matrices (\Cref{prop:lmi:topo}(b,c)) in
    \Cref{app:lmi:topo};
  \item an additional results on solution uniqueness and minimum/infimum
    equivalences for the BMZ and generalized BMZ method properties
    (\Cref{lem:bmz:J}, \Cref{prop:gbmz}, and \Cref{thm:bmz,prop:gbmz:value}) in
    \Cref{sec:chol};
  \item and additional remarks throughout.
  \end{itemize}
}{%
  % For brevity, proofs of minor results and details on the industrial-scale
  % reactor case study are omitted, but can be found in an extended technical
  % report~\cite{kuntz:rawlings:2024a}.%
}

\paragraph*{Notation}
%% Basic scalar sets
% Denote the integers, nonnegative integers, and intervals of integers by
% \(\allint\), \(\nnegint\), and \(\intinterval{a}{b}=\set{a,a+1,\ldots,b-1,b}\).
% Denote the set of reals, nonnegative reals, positive reals, and \(n\times m\)
% real matrices by \(\real\), \(\nnegreal\), \(\posreal\), and \(\realm{n}{m}\),
% respectively.
For any \(z\in\cplx\), let \(\overline{z}\) denote its complex conjugate.
%% Matrix sets
Denote the set of \(n\times n\) symmetric, positive definite, and positive
semidefinite matrices by \(\symm{n}\), \(\posdefm{n}\), and \(\nnegdefm{n}\).
Denote the set of lower triangular matrices and lower triangular matrices with
positive diagonal entries by \(\trilm{n}\) and \(\postrilm{n}\). Recall
\(M\in\realm{n}{n}\) is positive definite if and only if there exists a unique
\(L\in\postrilm{n}\), called the \emph{Cholesky factor}, such that
\(M=LL^\top\).
\ifthenelse{\boolean{LongVersion}}{%
  A \(2\times 2\) Hermitian matrix \(M =
  \begin{bsmallmatrix} a & b \\ \overline{b} & c \end{bsmallmatrix}\in\herm{2}\)
  is positive (semi)definite if and only if \(a,c>0\) (\(a,c\geq 0\)) and
  \(ac>|b|^2\) (\(ac\geq|b|^2\)). %
}{}%
%% Matrix operations
Denote the matrix direct sum and the Kronecker product by \(\oplus\) and
\(\otimes\), respectively, defined as in~\cite{golub:vanloan:2013}.
%% Eigenvalues, spectral radius/abscissa, stability
Define the set of eigenvalues of a matrix \(A\in\realm{n}{n}\) by
\(\lambda(A)\subset\cplx\). The spectral radius and spectral abscissa are
defined as \(\rho(A)\coloneqq\max_{\lambda\in\lambda(A)}|\lambda|\) and
\(\alpha(A)\coloneqq\max_{\lambda\in\lambda(A)}\mathrm{Re}(\lambda)\),
respectively. We say a matrix \(A\) is Schur (Hurwitz) stable if \(\rho(A)<1\)
(\(\alpha(A)<0\)).
%% Probability, random variables
We use \(\sim\) as a shorthand for ``distributed as'' and \(\iid\) as a
shorthand for ``independent and identically distributed as.''
%% General set stuff
The complement, interior, closure, and boundary of a set \(S\) are denoted
\(S^c\), \(\textnormal{int}(S)\), \(\textnormal{cl}(S)\), and \(\partial S\),
respectively.

\section{Problem statement}\label{sec:ml}
We consider stochastic LTI models in innovation form:
\begin{subequations}\label{eq:sskf}
  \begin{align}
    \hat{x}_{k+1} &= A(\theta)\hat{x}_k + B(\theta)u_k + K(\theta)e_k \\
    y_k &= C(\theta)\hat{x}_k + D(\theta)u_k + e_k \\
    e_k &\iid \norm(0, R_e(\theta))
  \end{align}
\end{subequations}
where \(\hat{x}\in\real^n\) are the model states, \(u\in\real^m\) are the
inputs, \(y\in\real^p\) are the outputs, \(e\in\real^p\) are the innovation
errors, and \(\theta\in\Theta\) are the model parameters. The model functions
\(\mathcal{M}(\cdot) \coloneqq (A(\cdot), B(\cdot), C(\cdot), D(\cdot),
\hat{x}_0(\cdot), K(\cdot), R_e(\cdot))\) are assumed to be known. While the
model \(\mathcal{M}\) is kept fairly general throughout, it is advantageous to
assume
% the functions are twice differentiable (for optimization with interior
% point methods) and
the model is identifiable
%% TODO in the sense of...?
in \(\Theta\).
Last, for brevity, we often drop the dependence on the parameters
\(\theta\in\Theta\) and write the model functions as \(\mathcal{M} = (A, B, C,
D, \hat{x}_0, K, R_e)\).

While the subsequent developments apply to any model of the form \cref{eq:sskf},
our main motivation is to identify the LADM,
\begin{subequations}\label{eq:ladm}
  \begin{align}
    \hat{s}_{k+1} &= A_s(\theta)\hat{s}_k + B_d(\theta)\hat{d}_k +
                    B_s(\theta)u_k + K_s(\theta)e_k \\
    \hat{d}_{k+1} &= \hat{d}_k + K_d(\theta)e_k \\
    y_k
    &= C_s(\theta)\hat{s}_k + C_d(\theta)\hat{d}_k + D(\theta)u_k + e_k \\
    e_k &\iid \norm(0, R_e(\theta))
  \end{align}
\end{subequations}
where \(\hat{s}\in\real^{n_s}\) denote plant states and
\(\hat{d}\in\real^{n_d}\) denote integrating disturbances.
% \((A_s,B_s,C_s)\) denote plant coefficient matrices, \((B_d,C_d)\) denote
% disturbance shaping matrices, and \((K_s,K_d)\) are plant and disturbance
% state filter gains. %
\ifthenelse{\boolean{LongVersion}}{%
  The LADM \cref{eq:ladm} is clearly a special case of \cref{eq:sskf} and can be
  put back into the standard form \cref{eq:sskf} by consolidating the plant and
  disturbance states \(\hat{x}_k \coloneqq \begin{bmatrix} \hat{s}_k^\top &
    \hat{d}_k^\top \end{bmatrix}^\top\) and defining
  \begin{align*}
    A &\coloneqq \begin{bmatrix} A_s & B_d \\ 0 & I \end{bmatrix},
    & B &\coloneqq \begin{bmatrix} B_s \\ 0 \end{bmatrix}, \\
    C &\coloneqq \begin{bmatrix} C_s & C_d \end{bmatrix},
    & K &\coloneqq \begin{bmatrix} K_s \\ K_d \end{bmatrix}.
  \end{align*}
}{%
  The LADM \cref{eq:ladm} is clearly a special case of \cref{eq:sskf} with the
  partition \(\hat{x}_k \coloneqq \begin{bmatrix} \hat{s}_k^\top &
    \hat{d}_k^\top \end{bmatrix}^\top\). %
} %
Typically the LADM \cref{eq:ladm} is parameterized with \((A_s,C_s)\) in
observability canonical form~\cite{denham:1974}, \((B_d,C_d)\)
fixed,\footnote{With \((A_s,B_s,C_s,D)\) fixed, all \((B_d,C_d)\) such that
  \cref{eq:ladm} is observable are equivalent up to a similarity
  transform~\cite{rajamani:rawlings:qin:2009}. Thus, \((B_d,C_d)\) are chosen by
  the practitioner to maximize interpretability of the disturbance estimates.}
\((B_s,K_s,K_d,R_e)\) fully parameterized, and
\((D,\hat{s}_0,\hat{d}_0)=(0,0,0)\). Alternatively, a physics-based model may be
used for \((A_s,B_s,C_s,D)\).

\subsection{Constrained maximum likelihood identification}
The ML identification problem is defined as follows:
\begin{equation}\label{eq:ml}
  \min_{\theta\in\Theta} L_N(\theta) \coloneqq
  \frac{N}{2}\ln\det R_e(\theta) +
  \frac{1}{2}\sum_{k=0}^{N-1} |e_k(\theta)|_{[R_e(\theta)]^{-1}}^2
\end{equation}
where \(N\in\posint\) is the sample size, \(L_N\) is the log-likelihood,
\(e_k(\theta)\) are given
by~\cref{eq:sskf}~\cite[p.~557]{astrom:1979},~\cite[p.~219]{ljung:1999}. To
regularize the estimates with respect to a previous parameter estimate
\(\overline\theta\) or incorporate an available prior distribution
\(p_0(\theta)\), we consider the maximum a posteriori (MAP) identification
problem,
\begin{equation}\label{eq:map}
  \min_{\theta\in\Theta} L_N(\theta) + R_0(\theta)
\end{equation}
where \(R_0(\theta) \propto -\ln p_0(\theta)\) is the regularization term,
typically chosen as a distance from
\(\overline\theta\)~\cite{sjoberg:mckelvey:ljung:1993,johansen:1997}.

\ifthenelse{\boolean{LongVersion}}{%
  For a Gaussian prior or generalized \(\ell_2\) penalty, we use
  \begin{equation}\label{eq:map:l2}
    R_0(\theta) \coloneqq \frac{1}{2}|\vect(\theta)-\vect(\overline\theta)|_{V^{-1}}^2
  \end{equation}
  where \(\overline\theta\in\Theta\) is the prior estimate, \(\vect\) is a
  vectorization operator, \footnote{The vectorization operator may depend on the
    parameterization, as \(\theta\) may contain both a vector portion and a
    sparse (semidefinite) matrix portion. The vectorization should only preserve
    the uniquely definednonzero elements of the sparse matrix.} and \(V\succ 0\)
  is the prior estimate variance. Such penalties are useful for model updating
  and re-identification. We typically use the penalty \cref{eq:map:l2} with
  \(V=\rho^{-1}I\). Later on, we transform the parameters \(\theta\) into a more
  convenient space for optimization and find it more convenient to define the
  prior directly in the transformed space. }{}

%% NOTE \cite{douc:moulines:2012} consider only \emph{stationary} models.
% For plants of the form \cref{eq:sskf}, the ML estimates are consistent and
% asymptotically efficient~\cite{astrom:1979}.
In a standard identification setting, the plant takes the form~\cref{eq:sskf}
with \(A-KC\) stable, the estimates found by~\cref{eq:ml} are
consistent~\cite{astrom:1979}, and, with sufficient data, the identified filter
is stable. However, the plant is intentionally misspecified by the
LADM~\cref{eq:ladm}, so this property breaks down in our problem.

Under certain regularity assumptions, we can show consistency with respect to
the estimates nearest in \emph{relative entropy rate}, denoted \(\theta^*\),
taken between the plant and model measurement
distributions\ifthenelse{\boolean{LongVersion}}{, %
  \[
    \theta^* \coloneqq \min_{\theta\in\Theta} N^{-1}\mathbb{E}[L_N(\theta)]
  \]
  where the expectation is taken over the true distribution of measurements
  \((y_k)_{k=0}^{N-1}\)%
}{}%
~\cite{white:1984,douc:moulines:2012}. The optimal estimates \(\theta^*\) may
not represent a stable LADM filter, and the identified LADM filter may be
unstable even in the large-sample limit. Moreover, there are fundamental
limitations to the performance of the Kalman filter for LADMs, as shown
in~\cite{bageshwar:borrelli:2009}. Specifically, the largest real filter
eigenvalue is often bounded from below by the largest real open-loop eigenvalue
(as we also find in \Cref{sec:casestudies}). Thus, it is necessary to design
\(\Theta\) to guarantee high-performance offset-free control.

\subsection{Constraints}
The constraint set \(\Theta\) should capture both estimator design
specifications and system knowledge. At a bare minimum, we require nondegeneracy
of the innovation errors,
\begin{equation}\label{cons:cov}%\tag{C1}
  R_e(\theta) \succ 0
\end{equation}
and stability of the estimator,
\begin{equation}\label{cons:stable:filter}%\tag{C2}
  \rho(A(\theta) - K(\theta)C(\theta)) < 1.
\end{equation}
Other useful constraints include spectral abscissa bounds,
\begin{equation}\label{cons:pos}%\tag{C3}
  \alpha(-\tilde A(\theta)) < 0,
\end{equation}
and bounds on the argument of the eigenvalues,
\begin{align}\label{cons:conic}%\tag{C4}
  0< |\textrm{Im}(\mu)|/\textrm{Re}(\mu) &< s,%\tan(\omega),
  & \forall \mu &\in \lambda(\tilde{A}(\theta))
\end{align}
where \(s>0\), for either the open-loop or estimator stability matrices
(\(\tilde{A}=A\) or \(\tilde{A}=A-KC\)), to eliminate artificial high-frequency
dynamics that may affect the control performance.

Chemical processes exhibit sparse interactions between units (mass and energy
flows), especially for large-scale plants~\cite{daoutidis:tang:allman:2019,%
  tang:carrette:cai:williamson:daoutidis:2023}. Sparse parameterization of
\((A,B,C,D,K)\) is easily encoded, but that of \(R_e\) is less obviously
accomplished. While, in general, \(R_e\) is dense even for sparse plants,
correlations between distant units are small~\cite{motee:jadbabaie:2009}. Thus,
it suffices to consider only nearest-neighbor correlations, e.g.,
\begin{equation}\label{eq:cov:sparse}
  R_e =
  \begin{bmatrix} R_{1,1}
    & R_{1,2} \\ R_{1,2}^\top & R_{2,2} & \ddots \\
    & \ddots & \ddots & R_{N_u-1,N_u} \\
    && R_{N_u-1,N_u}^\top & R_{N_u,N_u}
  \end{bmatrix}
\end{equation}
where \(R_{i,j}\in\realm{p_u}{p_u}\) is the covariance between the innovations
of the \(i\)-th and \(j\)-th process unit innovations. In \cref{eq:cov:sparse},
the sparse formulation introduces just \(O(N_up_u^2)\) variables compared to
\(O(N_u^2p_u^2)\) variables for the dense formulation.
% Another algorithm goal is
% to simultaneously and efficiently enforce both \cref{cons:cov,eq:cov:sparse}.
% Finally, we remark that such
Such constraints can be applied to the ML identification of any networked system
with a time-invariant topology, as in~\cite{zamani:ninness:aguero:2015}.

% There exist stable distributed Kalman filters...?

% The covariance \(R_e\) is dense for a centralized Kalman filter. There exist
% stable distributed Kalman filters for networked systems of the form found in
% large-scale chemical plants.

%% On sparse discrete-time Lyapunov equations: ``...entries of the solution are
%% spatially localized or decaying away from a banded
%% pattern''~\cite{haber:verhaegen:2016}.

\ifthenelse{\boolean{LongVersion}}{%
\subsection{Other parameterizations}\label{ssec:other}
The remainder of this section presents some other formulations of the ML
identification problem. While we do not consider these formulations explicitly
in our algorithm formulation (\Cref{sec:algo}) or case studies
(\Cref{sec:casestudies}), the methods are readily generalized to these
formulations.

\subsubsection{Time-varying Kalman filter formulations}
More generally, we could consider models of the following form:
\begin{subequations}\label{eq:slti}
  \begin{align}
    x_{k+1} &= A(\theta)x_k + B(\theta)u_k + w_k \\
    y_k &= C(\theta)x_k + D(\theta)u_k + v_k \\
    x_0 &\sim \norm(\hat{x}_0(\theta), \hat{P}_0(\theta)) \\
    \begin{bmatrix} w_k \\ v_k \end{bmatrix} &\iid \norm(0, S(\theta))
  \end{align}
\end{subequations}
where \(\mathcal{M} \coloneqq (A, B, C, D, \hat{x}_0, \hat{P}_0, S)\) are the
model functions and \(w\in\real^n\) and \(v\in\real^p\) are the process and
measurement noises. The noise covariance matrix \(S(\theta)\) may be partitioned
as
\begin{equation}\label{eq:noise:cov}
  S(\theta) =
  \begin{bmatrix} Q_w(\theta) & S_{wv}(\theta) \\
    [S_{wv}(\theta)]^\top & R_v(\theta) \end{bmatrix}
\end{equation}
where \(Q_w(\theta)\in\nnegdefm{n}\) is the process noise covariance, \(S_{wv}(\theta)\) is the
cross-covariance, and \(R_v(\theta)\in\nnegdefm{p}\) is the measurement noise covariance.
Throughout, we impose the stronger requirement \(R_v(\theta)\succ 0\) on the
measurement noise covariance.

For the model \cref{eq:slti}, the ML problem is defined as
\begin{equation}\label{eq:ml:tvkf}
  \min_{\theta\in\Theta} L_N(\theta) \coloneqq \frac{1}{2}\sum_{k=0}^{N-1}
  \ln\det\mathcal{R}_k(\theta) + |e_k(\theta)|_{[\mathcal{R}_k(\theta)]^{-1}}^2
\end{equation}
where the \(e_k\) and \(\mathcal{R}_k\) are defined by the Kalman filtering
equations
\begin{subequations}\label{eq:tvkf}
  \begin{align}
    \hat{x}_{k+1} &= A\hat{x}_k + Bu_k + \mathcal{K}_ke_k \\
    y_k &= C\hat{x}_k + Du_k + e_k \\
    e_k &\sim \norm(0, \mathcal{R}_k) \quad \textnormal{(indep.)}
  \end{align}
  where
  \begin{align}
    \hat{P}_{k+1}
    &\coloneqq A\hat{P}_kA^\top + Q_w -
      \mathcal{K}_k\mathcal{R}_k\mathcal{K}_k^\top \\
    \mathcal{K}_k
    &\coloneqq (A\hat{P}_kC^\top + S_{wv})\mathcal{R}_k^{-1} \\
    \mathcal{R}_k
    &\coloneqq C\hat{P}_kC^\top + R_v.
  \end{align}
\end{subequations}
We remark that \(R_v\succ 0\) suffices to guarantee the innovations are
uniformly nondegenerate, i.e., \(\mathcal{R}_k\succ 0\). However, stability of
the filter is more difficult to guarantee as the early iterates
\(A-\mathcal{K}_kC\) may not be stable, even though the overall filter is
stable, or vice versa. Instead, it is necessary to check that a stabilizing
solution to the Riccati equation exists, which we elaborate on in the next
formulation.

\subsubsection{Time-invariant Kalman filter formulations}
In most situations, the state error covariance matrix converges exponentially
fast to a steady-state solution \(\hat{P}_k\rightarrow\hat{P}\), so it suffices
to consider the original steady-state filter model \cref{eq:sskf}. In terms of
the model \cref{eq:slti}, the steady-state filter takes the form \(K\coloneqq
(A\hat{P}C^\top + S_{wv})R_e^{-1}\) and \(R_e\coloneqq C\hat{P}C^\top+R_v\),
where \(\hat{P}\) is the unique, stabilizing solution to the discrete algebraic
Riccati equation (DARE), %
\ifthenelse{\boolean{OneColumn}}{
  \begin{equation}\label{eq:dare}
    \hat{P} = A\hat{P}A^\top + Q_w - (A\hat{P}C^\top + S_{wv})
    (C\hat{P}C^\top + R_v)^{-1} (A\hat{P}C^\top + S_{wv})^\top.
  \end{equation}
}{
  \begin{multline}\label{eq:dare}
    \hat{P} = A\hat{P}A^\top + Q_w - (A\hat{P}C^\top + S_{wv}) \\
    \times (C\hat{P}C^\top + R_v)^{-1} (A\hat{P}C^\top + S_{wv})^\top.
  \end{multline}
}
Recall a solution to the DARE \cref{eq:dare} is stabilizing if the resulting
\(A_K\coloneqq A-KC\) is stable.

Convergence of \(\hat{P}_k\) to \(\hat{P}\) is equivalent to the solution to the
DARE \cref{eq:dare} being unique and stabilizing. We generally assume such a
solution exists, but for completeness, we state the following proposition,
adapted from~\cite[Thm.~18(iii)]{silverman:1976} (see
\ifthenelse{\boolean{LongVersion}}{\Cref{app:dare}}{\cite{kuntz:rawlings:2024a}}
for proof).
\begin{proposition}\label{prop:dare}
  Assume \(R_v\succ 0\) and consider the full rank factorization
  \[
    \begin{bmatrix} Q_w & S_{wv} \\ S_{wv}^\top & R_v \end{bmatrix}
    = \begin{bmatrix} \tilde B \\ \tilde D \end{bmatrix}
    \begin{bmatrix} \tilde B^\top & \tilde D^\top \end{bmatrix}
  \]
  Then the following statements are equivalent:
  \begin{enumerate}
  \item The DARE~\cref{eq:dare} has a unique, stabilizing solution
    \(\hat{P}\succeq 0\).
  \item The error covariance converges exponentially fast
    \(\hat{P}_k\rightarrow\hat{P}\) for any \(\hat{P}_0\succeq 0\).
  \item \((A,C)\) is detectable and \((A-FC,\tilde B-F\tilde D)\) is
    stabilizable for all \(F\in\realm{n}{p}\).
  \end{enumerate}
\end{proposition}

\begin{remark}
  The hypothesis of \Cref{prop:dare} holds if we constrain \(A\) to be stable or
  \((A,C)\) to be observable.
\end{remark}

\begin{remark}
  The cross-covariance \(S_{wv}\) complicates the filter stability analysis.
  With \(S_{wv}=0\), it would suffice to assume \((A,C)\) detectable and
  \((A,Q_w)\) stabilizable. With nonzero \(S_{wv}\), however, a more elaborate
  stabilizability condition is needed. \cite[Thm.~18]{silverman:1976} considers
  the regulation problem with a cross-weighting term and semidefinite input
  weights. \Cref{prop:dare} specializes this result to the filter problem with
  positive definite \(R_v\).
\end{remark}

\begin{remark}
  While \(R_e(\theta)\) and \(K(\theta)\) could be defined via
  \(\hat{P}(\theta)\), taken as the function that returns solutions to the DARE
  \cref{eq:dare} and therefore enforcing filters stability, it is more
  convenient to directly parameterize these matrices as in \cref{eq:sskf}.
\end{remark}

\subsubsection{Minimum determinant formulation}\label{ssec:mindet}
Suppose in the model \cref{eq:sskf}, that \(R_e\) is parameterized fully, and
separately from the other terms, i.e., %
\begin{equation*}
  \mathcal{M}(\tilde\theta,R_e) = \left( A(\tilde\theta), B(\tilde\theta),
    C(\tilde\theta), D(\tilde\theta), \hat{x}_0(\tilde\theta),
    K(\tilde\theta), R_e \right).
\end{equation*}
Moreover, assume \(R_e\) is constrained separately as well, i.e.,
\[
  \Theta = \tilde\Theta \times \posdefm{p}.
\]
Then we can always solve \cref{eq:ml} stagewise, first in \(R_e\), and then in
the remaining variables \(\tilde\theta\), i.e.,
\begin{equation*}
  \min_{\tilde\theta\in\tilde\Theta} \min_{R_e\in\posdefm{p}}  L_N(\tilde\theta,R_e) =
  \frac{N}{2}\ln\det R_e + \frac{1}{2}\sum_{k=0}^{N-1} |e_k(\tilde\theta)|_{R_e^{-1}}^2.
\end{equation*}
Solving the inner problem gives the solution
\[
  \hat R_e(\tilde\theta) \coloneqq \frac{1}{N}\sum_{k=0}^{N-1}
  e_k(\tilde\theta)[e_k(\tilde\theta)]^\top
\]
where we use the fact that \(e_k\) is only dependent on \(\tilde\theta\), and we
assume \(\hat R_e(\tilde\theta)\succ 0\) for all
\(\tilde\theta\in\tilde\Theta\). The outer problem can be written
\begin{equation}\label{eq:ml:mindet}
  \min_{\tilde\theta\in\tilde\Theta} L_N(\tilde\theta,\hat R_e(\tilde\theta)) =
  \frac{N}{2} \det\hat R_e(\tilde\theta) + \frac{p}{2}
\end{equation}
which is the minimum determinant problem.

The problem \cref{eq:ml:mindet} is of relevance because it avoids posing
\cref{eq:ml} as a NSDP.\@ It has been used both in the early ML identification
literature~\cite{%
  astrom:eykhoff:1971,% survey; includes state-space model
  % caines:1978, % confusing
  ljung:1978,% convergence analysis paper
  astrom:1980% survey; includes state-space model
} and in recent works~\cite{%
  mckelvey:helmersson:ribarits:2004,% data-driven local coordinates
  ribarits:deistler:hanzon:2005,% data-driven local coordinates
  li:postlethwaite:turner:2007% helicopter dynamics
}. None of these works consider filter stability constraints. To the best of our
knowledge, only \cite{umenberger:wagberg:manchester:schon:2018} consider the ML
problem \cref{eq:ml:tvkf} with stability constraints, but they consider
open-loop stability (i.e., \(\rho(A)<1\)) and use the EM algorithm.

\begin{remark}
  For real-world data, \(\det\hat R_e(\tilde\theta,\hat R_e(\tilde\theta))=0\)
  is not attainable because that would imply some direction of \(y_k\) were
  perfectly modeled. Therefore, \(\hat R_e(\tilde\theta)\succ 0\) for all
  \(\tilde\theta\in\tilde\Theta\) is a reasonable assumption.
\end{remark}

}{}

\section{Algorithm outline}\label{sec:algo}
\subsection{Constraint set formulation}
More generally, we seek to (i) impose eigenvalue
constraints on any model function \(\tilde{A}(\theta)\) and (ii) impose a
sparsity structure on any semidefinite model function \(\tilde Q(\theta)\).

\subsubsection{Eigenvalue constraints}
First, we define a LMI region.
\begin{definition}\label{defn:lmi:region}
  We call \(\mathcal{D}\subseteq\cplx\) an \emph{LMI region} if
  \[
    \mathcal{D} = \set{ z\in\cplx | f_{\mathcal{D}}(z) \coloneqq M_0 + M_1z +
      M_1^\top\overline{z} \succ 0 }
  \]
  for some \emph{generating matrices}
  \((M_0,M_1)\in\symm{m}\times\realm{m}{m}\). We call
  \(f_{\mathcal{D}}:\cplx\rightarrow\symm{m}\) the \emph{characteristic
    function} of \(\mathcal{D}\).
\end{definition}

\ifthenelse{\boolean{LongVersion}}{%
  The following lemma defines the four basic LMI regions: shifted half-planes,
  circles centered on the real axis, conic sections, and horizontal bands. %
}{%
  %% TODO Bad practice. Maybe find a better solution than \pagebreak
  The following lemma defines a few basic LMI regions: circles, shifted
  half-planes, and shifted conic sections, \pagebreak each symmetric about the
  real axis (see \cite{kuntz:rawlings:2024a} for proof). %
}%
For a general discussion of LMI regions properties,
see~\cite{chilali:gahinet:1996,kushel:2019}.

\begin{lemma}\label{lem:lmi:basic}
  For each \(s,x_0\in\real\), the subsets
  \begin{align*}
    \mathcal{D}_1(x_0) &\coloneqq \set{ z\in\cplx | \textnormal{Re}(z) > x_0 } \\
    \mathcal{D}_2(s,x_0) &\coloneqq \set{ z\in\cplx | |z-x_0| < s } \\
    \mathcal{D}_3(s,x_0) &\coloneqq \set{ z\in\cplx | |\textnormal{Im}(z)| <
                           s(\textnormal{Re}(z)-x_0) }
                           \ifthenelse{\boolean{LongVersion}}{ \\
    \mathcal{D}_4(s) &\coloneqq \set{ z\in\cplx | |\textnormal{Im}(z)| < s }
                       }{}
  \end{align*}
  are LMI regions with characteristic functions
  \begin{align*}
    f_{\mathcal{D}_1(x_0)}(z)
    &\coloneqq -2x_0 + z + \overline{z} \\
    f_{\mathcal{D}_2(s,x_0)}(z)
    &\coloneqq \begin{bmatrix} s & -x_0 \\ -x_0 & s \end{bmatrix}
      + \begin{bmatrix} 0 & 1 \\ 0 & 0 \end{bmatrix}z
      + \begin{bmatrix} 0 & 0 \\ 1 & 0 \end{bmatrix}\overline{z} \\
    f_{\mathcal{D}_3(s,x_0)}(z)
    &\coloneqq -2sx_0I_2 + \begin{bmatrix} s & 1 \\ -1 & s \end{bmatrix}z
      + \begin{bmatrix} s & -1 \\ 1 & s \end{bmatrix}\overline{z}
      \ifthenelse{\boolean{LongVersion}}{ \\
    f_{\mathcal{D}_4(s)}(z)
    &\coloneqq -2sI_2 + \begin{bmatrix} 0 & 1 \\ -1 & 0 \end{bmatrix}z
      + \begin{bmatrix} 0 & -1 \\ 1 & 0 \end{bmatrix}\overline{z}.
      }{.}
  \end{align*}
\end{lemma}

\ifthenelse{\boolean{LongVersion}}{
  \begin{proof}
    %% 1.
    The first identity follows from the formula \(2\textnormal{Re}(z) =
    z+\overline{z}\).
    %% 2.
    For the second identity, we have
    \(f_{\mathcal{D}_2(s,x_s)}(z) = \begin{bsmallmatrix} s & z-x_0 \\
                                      \overline{z}-x_0 & s \end{bsmallmatrix}\succ
    0\) if and only if \(s>0\) and \(s^2>|z-x_0|^2\), or equivalently,
    \(|z-x_0|<s\).
    %% 3.
    For the third identity, we have \(f_{\mathcal{D}_3(s,x_0)}(z) =
    \begin{bsmallmatrix} 2s(\textnormal{Re}(z)-x_0) & 2\iota\textnormal{Im}(z) \\
      -2\iota\textnormal{Im}(z) & 2s(\textnormal{Re}(z)-x_0) \end{bsmallmatrix}
    \succ 0\) if and only if \(2s(\textnormal{Re}(z)-x_0)>0\) and
    \(4s^2(\textnormal{Re}(z)-x_0)^2 > 4|\textnormal{Im}(z)|^2\), or
    equivalently, \(|\textnormal{Im}(z)| < s(\textnormal{Re}(z)-x_0)\). %
    \ifthenelse{\boolean{LongVersion}}{%
      %% 4.
      For the fourth identity, we have \(f_{\mathcal{D}_4(s)}(z) =
      \begin{bsmallmatrix} 2s & 2\iota\textnormal{Im}(z) \\
        -2\iota\textnormal{Im}(z) & 2s \end{bsmallmatrix} \succ 0\) if and only
      if \(2s>0\) and \(4s^2 > 4|\textnormal{Im}(z)|^2\), or equivalently,
      \(|\textnormal{Im}(z)| < s\). %
    }{}
  \end{proof}

  %% NOTE see Chilali and Gahinet (1996), Kushel (2019), and
  %% https://www.mathworks.com/help/control/ref/dynamicsystem.damp.html#mw_774bba0d-fc75-4b75-a377-8ab21de1d8ee
  \begin{remark}\label{rem:lmi:basic}
    For continuous-time systems, \(-\mathcal{D}_1(\alpha)\) corresponds to a
    minimum decay rate of \(\alpha>0\), \(\mathcal{D}_3(-\tan(\omega),0)\)
    corresponds to a minimum damping ratio \(\cos(\omega)\), and
    \(\mathcal{D}_2(r,0)\cap\mathcal{D}_3(-\tan(\omega),0)\) implies to a
    maximum undamped natural frequency \(r\sin(\omega)\), where \(\alpha,r>0\)
    and \(\omega\in[0,\pi/2]\)~\cite{chilali:gahinet:1996}. For discrete-time
    systems, \(\mathcal{D}_2(r,0)\) corresponds to a minimum decay rate of
    \(-\ln r/\Delta\), and \(\mathcal{D}_2(r,0) \cap
    \mathcal{D}_3(\tan(\omega),0)\) implies a minimum damping ratio
    \(\cos(\tan^{-1}(\omega/\ln r))\) and maximum natural frequency
    \(\sqrt{\ln(r)^2 + \omega^2}/\Delta\), where \(r\in(0,1)\),
    \(\omega\in[0,\pi/2)\), and \(\Delta\) is the sample time.
  \end{remark}

  \begin{remark}
    For any LMI region \(\mathcal{D}\) (including those in
    \Cref{lem:lmi:basic}), the set \(\mathcal{D}\) is convex, open, and
    symmetric about the real axis. The intersection of two LMI regions
    \(\mathcal{D}\coloneqq\mathcal{D}_1\cap\mathcal{D}_2\) is an LMI region with
    the characteristic function \(f_{\mathcal{D}}(z) = f_{\mathcal{D}_1}(z)
    \oplus f_{\mathcal{D}_2}(z)\). By this property, we can construct any convex
    polyhedron that is symmetric about the real axis by intersecting left and
    right half-planes, horizontal strips, and conic sections. Moreover, since
    any convex region can be approximated, to any desired accuracy, by a convex
    polyhedron, the set of LMI regions is dense in the space of convex subsets
    of \(\cplx\) that are symmetric about the real axis. An LMI region
    \(\mathcal{D}\) with characteristic function \(f_{\mathcal{D}}\) also has
    characteristic function \(Mf_{\mathcal{D}}(\cdot)M^\top\) for any
    nonsingular \(M\in\realm{m}{m}\). For an in-depth discussion of LMI region
    geometry and other properties, see \cite{kushel:2019}.
  \end{remark}
}{}

In \cite{chilali:gahinet:1996}, it is shown that a matrix \(\tilde A\in\realm{\tilde
  n}{\tilde n}\) has eigenvalues in a LMI region \(\mathcal{D}\) if and only if
the following system of matrix inequalities is feasible:
\begin{align}\label{eq:lmi:dstable}
  M_{\mathcal{D}}(\tilde A,P) &\succ 0, & P&\succ 0
\end{align}
where the \emph{matrix characteristic function} \(M_{\mathcal{D}}:\realm{\tilde
  n}{\tilde n}\times\symm{\tilde n}\rightarrow\symm{\tilde n\tilde m}\) of
\(\mathcal{D}\) is defined by
\begin{equation}\label{eq:lmi:mat}
  M_{\mathcal{D}}(\tilde A, P) \coloneqq M_0\otimes P + M_1\otimes (\tilde AP) +
  M_1^\top\otimes (\tilde AP)^\top.
\end{equation}

%% TODO bsmallmatrix would be better below, but I'm fudging it to match the
%% spacing from the journal version.
From this equivalence, we can build tractable eigenvalue constraints. For the
constraint \cref{cons:stable:filter}, \Cref{lem:lmi:basic} gives the generating
matrices \((M_0,M_1) \coloneqq \left(\begin{bmatrix} 1 & 0 \\ 0 &
    1 \end{bmatrix}, \begin{bmatrix} 0 & 1 \\ 0 & 0 \end{bmatrix}\right)\) and
we have the matrix inequalities
\begin{align*}
  \begin{bmatrix} P & (A-KC)P \\ P(A-KC)^\top & P \end{bmatrix} &\succ 0,
  & P&\succ 0
\end{align*}
which is a well-known LMI for checking
stability~\cite{boyd:ghaoui:feron:balakrishnan:1994}. Similarly, to implement
\cref{cons:pos}, we can use the generating matrices \((M_0,M_1) \coloneqq
(0,1)\), and to implement \cref{cons:conic}, we can use \((M_0,M_1) \coloneqq
\left(\begin{bsmallmatrix} -2\tan(\omega) & 0 \\ 0 &
    -2\tan(\omega) \end{bsmallmatrix}, \begin{bsmallmatrix} \tan(\omega) & 1 \\ -1 &
    \tan(\omega) \end{bsmallmatrix}\right)\).

The system of matrix inequalities \cref{eq:lmi:dstable} contains only strict
inequalities, but we can ``tighten'' them as follows:
\begin{align}\label{eq:lmi:cont}
  M_{\mathcal{D}}(\tilde A,P)&\succeq M, & P&\succeq 0,
  & \tr(VP)&\leq\varepsilon^{-1}
\end{align}
where \(\varepsilon>0\), \(V\in\posdefm{\tilde n}\), and \(M\in\posdefm{\tilde
  n\tilde m}\). The set of \(\tilde{A}\in\realm{\tilde n}{\tilde n}\) for which
\cref{eq:lmi:cont} is feasible defines a \emph{closed} set for which
\(\lambda(\tilde A)\subseteq\mathcal{D}\). In \Cref{sec:stable}, we show this
fact and other properties of the constraint \cref{eq:lmi:cont}.

\subsubsection{Sparsity structure}
To encode sparsity information, we adapt the notation of
\cite{burer:monteiro:zhang:2002}. Define the index sets
\(\mathcal{L}^n\coloneqq\set{(i,j)\in\intinterval{1}{n}^2 | i\geq j}\) and
\(\mathcal{D}^n\coloneqq\set{(i,i)\in\intinterval{1}{n}^2}\) corresponding to
the sparsity patterns of \(n\times n\) lower triangular and diagonal matrices.
With a slight abuse of notation, we define the direct sum of index sets
\(\mathcal{I}\subseteq\mathcal{L}^n\) and \(\mathcal{J}\subseteq\mathcal{L}^m\)
by
\[
  \mathcal{I} \oplus \mathcal{J} \coloneqq \mathcal{I} \cup
  \set{(i+n,j+n)|(i,j)\in\mathcal{J}} \subseteq \mathcal{L}^{n+m}.
\]
For each \(\mathcal{I}\subseteq\mathcal{L}^n\), define the sets
{\allowdisplaybreaks
  \begin{align*}
    \symm{n}[\mathcal{I}]
    &\coloneqq \set{ S\in\symm{n} | S_{ij}=0 \; \forall \;
      (i,j)\in\mathcal{L}^n\setminus\mathcal{I} } \\
    \trilm{n}[\mathcal{I}]
    &\coloneqq \set{ L\in\trilm{n} | L_{ij} = 0\; \forall \;
      (i,j)\not\in\mathcal{I} } \\
    \postrilm{n}[\mathcal{I}]
    &\coloneqq \set{ L\in\postrilm{n} | L_{ij} = 0\; \forall \;
      (i,j)\not\in\mathcal{I} }.
  \end{align*}
}
Finally, let \(\vecs_{\mathcal{I}} : \symm{n}\rightarrow\real^{|\mathcal{I}|}\)
denote the operator that vectorizes the \(|\mathcal{I}|\) entries of the
argument corresponding to the index set \(\mathcal{I}\).

\subsubsection{Constraint definition}
To combine the LMI region and sparsity constraints, we partition the parameter
into vector and sparse symmetric matrix parts, i.e., \(\theta = (\beta,\Sigma)
% \in \Theta \subseteq \real^{n_\beta}\times\symm{n_\Sigma}[\mathcal{I}_\Sigma]
\),
and define the constraint set \(\Theta\) by
\ifthenelse{\boolean{OneColumn}}{%
  \begin{equation}\label{eq:cons}
    \Theta = \set{ (\beta,\Sigma) \in \real^{n_\beta} \times
      \symm{n_\Sigma}[\mathcal{I}_\Sigma] |
      g(\beta,\Sigma) = 0,\;
      h(\beta,\Sigma)\leq 0,\; \Sigma \succeq H(\beta),\;
      \mathcal{A}(\beta,\Sigma)\succeq 0 }
  \end{equation}
}{%
  \begin{multline}\label{eq:cons}
    \Theta = \{ \; (\beta,\Sigma) \in \real^{n_\beta} \times
    \symm{n_\Sigma}[\mathcal{I}_\Sigma] \;|\;
    g(\beta,\Sigma) = 0,
    \ifthenelse{\boolean{OneColumn}}{\;}{\\}
    h(\beta,\Sigma)\leq 0,\; \Sigma \succeq H(\beta),\;
    \mathcal{A}(\beta,\Sigma)\succeq 0 \;\}
  \end{multline}
}
where \(\mathcal{D}^{n_\Sigma} \subseteq \mathcal{I}_\Sigma \subseteq
\mathcal{L}^{n_\Sigma}\), \(\mathcal{D}^{n_{\mathcal{A}}} \subseteq
\mathcal{I}_{\mathcal{A}} \subseteq \mathcal{L}^{n_{\mathcal{A}}}\), \(g :
\real^{n_\beta} \times \symm{n_\Sigma} \rightarrow \real^{n_g}\), \(h :
\real^{n_\beta} \times \symm{n_\Sigma} \rightarrow \real^{n_h}\), \(H :
\real^{n_\beta} \rightarrow \symm{n_\Sigma}\), and \(\mathcal{A} :
\real^{n_\beta} \times \symm{n_\Sigma} \rightarrow
\symm{n_{\mathcal{A}}}[\mathcal{I}_{\mathcal{A}}]\).
% \begin{align*}
%   (\beta, \Sigma) &\in \Theta &&\Leftrightarrow
%   &&\begin{cases} g(\beta,\Sigma) \leq 0, & h(\beta,\Sigma) = 0, \\
%       \Sigma \in \symm{n_\Sigma}[\mathcal{I}_\Sigma], & \Sigma \succeq H(\beta), \\
%       \mathcal{A}(\beta,\Sigma) \succeq 0. \end{cases}
       %   \end{align*}
The purpose of the partition \(\theta = (\beta,\Sigma)\) is to clearly delineate
the sparse semidefinite matrix argument \(\Sigma\) from the remaining parameters
\(\beta\). The index set \(\mathcal{I}_\Sigma\) defines the sparsity pattern of
\(\Sigma\) and \(H\), and the index set \(\mathcal{I}_{\mathcal{A}}\) defines
the sparsity pattern of \(\mathcal{A}\).
% Finally, for compatibility with
% interior point methods, we assume the constraint functions \(\mathcal{C}
% \coloneqq (g,h,H,\mathcal{A})\) are differentiable.

\ifthenelse{\boolean{LongVersion}}{%
  \begin{remark}
    \Cref{assm:cons} rules out direct use strict inequalities, e.g.,
    \(R_e(\theta)\succ 0\) or \(R_v(\theta)\succ 0\). To satisfy nondegeneracy
    requirements, we use the closed constraint \(R_e(\theta)\succeq \delta I_p\)
    with a small back-off \(\delta>0\).
  \end{remark}

  \begin{remark}
    Typically, the index set \(\mathcal{I}_\Sigma\) encodes block diagonal
    structures, e.g., for the model \cref{eq:slti}, \(\Sigma=\hat{P}_0\oplus
    Q_w\oplus R_v \in \symm{2n+p}[\mathcal{I}_\Sigma]\) where
    \(\mathcal{I}_\Sigma \coloneqq
    \mathcal{L}^n\oplus\mathcal{L}^n\oplus\mathcal{L}^p\). However, more general
    structures (e.g., \cref{eq:cov:sparse}) can be stated. For the time-varying
    formulation \cref{eq:ml:tvkf}, we may further restrict \(Q_w\) and \(R_v\)
    to take block tridiagonal and diagonal structures, e.g.,
    \begin{align*}
      Q_w &= \begin{bmatrix} Q_{1,1}
               & Q_{1,2} \\ Q_{1,2}^\top & Q_{2,2} & \ddots \\
               & \ddots & \ddots & Q_{\tilde n-1,\tilde n} \\
               && Q_{\tilde n-1,\tilde n}^\top & Q_{\tilde n,\tilde n}
             \end{bmatrix},
            \ifthenelse{\boolean{OneColumn}}{&}{\\}
      R_v &= R_1 \oplus \ldots \oplus R_{\tilde n}
    \end{align*}
    that arise in sequentially interconnected processes such as chemical plants.
    Adding a \(Q_{1,\tilde n}\) block can account for an overall recycle loop.
    Note that if we parameterize the block tridiagonal \(Q_w\) via a sparse
    shaping matrix (i.e., \(Q_w=G_wG_w^\top\)), then there are more parameters
    than if the sparsity of \(Q_w\) is known.
  \end{remark}

  \begin{remark}
    As alluded to in \Cref{sec:ml}, the Riccati equation solution has a dense
    solution, but the entries far from the core sparsity pattern decay rapidly.
    Thus, we can approximate an eigenvalue constraint, e.g., \(P-APA^\top\succ
    0\), as a function that maps to the same sparsity pattern as
    \(A\)~\cite{motee:jadbabaie:2009,haber:verhaegen:2016,motee:sun:2017,%
      massei:saluzzi:2024}.
  \end{remark}
}{}

\subsection{Cholesky factorization and elimination}
At this juncture, the ML and MAP problems \cref{eq:ml,eq:map} with the
constraints \cref{eq:cons} are in standard NSDP form and can be solved with any
dedicated NSDP solver,
e.g.,~\cite{fiala:kocvara:stingl:2013,kocvara:stingl:2015}. However, such
solvers are neither as widely available nor as well-understood as NLP solvers
such as IPOPT~\cite{wachter:biegler:2006}.

The Burer-Monteiro-Zhang (BMZ) method is a Cholesky factorization-based
substitution and elimination algorithm that can convert certain NSDPs to
NLPs~\cite{burer:monteiro:zhang:2002,%
  burer:monteiro:zhang:2002a}. In \Cref{sec:chol}, we consider a generalization
of this algorithm to (approximately) transform a given NSDP into a NLP while
only introducing \(|\mathcal{I}_{\mathcal{A}}|\) new variables. This
generalization requires the following assumption.

\begin{assumption}\label{assm:cons}
  The model functions \(\mathcal{M}\) are twice differentiable and the
  constraint functions \(\mathcal{C} \coloneqq (g,h,H,\mathcal{A})\) are
  differentiable. Moreover, \(\textnormal{cl}(\Theta_{++})=\Theta\) where
  \ifthenelse{\boolean{OneColumn}}{%
    \begin{equation}\label{eq:cons:strict}
      \Theta_{++} \coloneqq \set{ (\beta,\Sigma) \in \real^{n_\beta} \times
        \symm{n_\Sigma}[\mathcal{I}_\Sigma] \;|\; g(\beta,\Sigma) = 0,\;
        h(\beta,\Sigma)\leq 0,\; \Sigma \succ H(\beta),\;
        \mathcal{A}(\beta,\Sigma)\succ 0 }.
    \end{equation}
  }{%
    \begin{multline}\label{eq:cons:strict}
      \Theta_{++} \coloneqq \{\; (\beta,\Sigma) \in \real^{n_\beta} \times
      \symm{n_\Sigma}[\mathcal{I}_\Sigma] \;|\; g(\beta,\Sigma) = 0,
      \ifthenelse{\boolean{OneColumn}}{\;}{\\}
      h(\beta,\Sigma)\leq 0,\; \Sigma \succ H(\beta),\;
      \mathcal{A}(\beta,\Sigma)\succ 0 \;\}.
    \end{multline}
  }
\end{assumption}

In \Cref{sec:chol}, we construct functions
\begin{align*}
  \mathcal{T}
  &: \real^{n_\beta} \times \postrilm{n_\Sigma}[\mathcal{I}_\Sigma]
    \times \postrilm{n_{\mathcal{A}}}[\mathcal{I}_{\mathcal{A}}]
    \rightarrow \real^{n_\beta} \times \symm{n_\Sigma}[\mathcal{I}_\Sigma] \\
  \mathcal{A}_{\mathcal{T}}
  &: \real^{n_\beta} \times \postrilm{n_\Sigma}[\mathcal{I}_\Sigma] \times
    \postrilm{n_{\mathcal{A}}}[\mathcal{I}_{\mathcal{A}}] \rightarrow
    \posdefm{n_{\mathcal{A}}}[\mathcal{I}_{\mathcal{A}}]
\end{align*}
% \(\mathcal{T} : \real^{n_\beta} \times \postrilm{n_\Sigma}[\mathcal{I}_\Sigma]
% \times \postrilm{n_{\mathcal{A}}}[\mathcal{I}_{\mathcal{A}}] \rightarrow
% \real^{n_\beta} \times \symm{n_\Sigma}[\mathcal{I}_\Sigma]\) and
% \(\mathcal{A}_{\mathcal{T}} : \real^{n_\beta} \times
% \postrilm{n_\Sigma}[\mathcal{I}_\Sigma] \times
% \postrilm{n_{\mathcal{A}}}[\mathcal{I}_{\mathcal{A}}] \rightarrow
  %   \posdefm{n_{\mathcal{A}}}[\mathcal{I}_{\mathcal{A}}]\)
and define transformed constraint functions
\begin{subequations}\label{eq:cons:trans:func}
  \begin{align}
    g_{\mathcal{T}}(\phi)
    &\coloneqq \begin{bmatrix} g(\mathcal{T}(\phi)) \\
                 \vecs_{\mathcal{I}_{\mathcal{A}}}(
                 \mathcal{A}(\mathcal{T}(\phi)) -
                 \mathcal{A}_{\mathcal{T}}(\phi)) \end{bmatrix} \\
    h_{\mathcal{T}}(\phi) &\coloneqq h(\mathcal{T}(\phi))
  \end{align}
\end{subequations}
and a transformed constraint set
\ifthenelse{\boolean{OneColumn}}{%
  \begin{equation}\label{eq:cons:trans}
    \Phi \coloneqq \set{ \phi \in \real^{n_\beta} \times
    \postrilm{n_\Sigma}[\mathcal{I}_\Sigma] \times
    \postrilm{n_{\mathcal{A}}}[\mathcal{I}_{\mathcal{A}}] |
    g_{\mathcal{T}}(\phi) = 0,\; h_{\mathcal{T}}(\phi) \leq 0 }
  \end{equation}
}{%
  \begin{multline}\label{eq:cons:trans}
    \Phi \coloneqq \{\; \phi \in \real^{n_\beta} \times
    \postrilm{n_\Sigma}[\mathcal{I}_\Sigma] \times
    \postrilm{n_{\mathcal{A}}}[\mathcal{I}_{\mathcal{A}}] \;| \\
    g_{\mathcal{T}}(\phi) = 0,\; h_{\mathcal{T}}(\phi) \leq 0 \;\}
  \end{multline}
}
such that, under \Cref{assm:cons}, \(\mathcal{T}\) is an invertible map from
\(\Phi\) to \(\Theta_{++}\).
% transformed constraint
% set
% \begin{multline*}
%   \Phi \coloneqq \{\; (\beta,L_\Sigma,L_{\mathcal{A}}) \in \real^{n_\beta} \times
%   \postrilm{n_\Sigma}[\mathcal{I}_\Sigma] \times
%   \postrilm{n_{\mathcal{A}}}[\mathcal{I}_{\mathcal{A}}] \;|
%   \ifthenelse{\boolean{OneColumn}}{\;}{\\}
%   g_{\mathcal{T}}(\beta,L_\Sigma,L_{\mathcal{A}}) = 0,\;
%   h_{\mathcal{T}}(\beta,L_\Sigma)\leq 0 \;\}.
% \end{multline*}
% such that \(\Phi\) to \(\Theta_{++}\).
Finally, to eliminate the strict inequalities on the diagonal entries of
\((L_\Sigma,L_{\mathcal{A}}) \in \postrilm{n_\Sigma}[\mathcal{I}_\Sigma] \times
\postrilm{n_{\mathcal{A}}}[\mathcal{I}_{\mathcal{A}}] \), we introduce a fixed
lower bound \(\varepsilon>0\) on the diagonal entries,
\ifthenelse{\boolean{OneColumn}}{%
  \begin{equation}\label{eq:cons:approx}
    \Phi_\varepsilon \coloneqq \set{ \phi \in \real^{n_\beta} \times
    \trilm{n_\Sigma}_\varepsilon[\mathcal{I}_\Sigma] \times
    \trilm{n_{\mathcal{A}}}_\varepsilon[\mathcal{I}_{\mathcal{A}}] |
    g_{\mathcal{T}}(\phi) = 0,\; h_{\mathcal{T}}(\phi) \leq 0 }
  \end{equation}
}{%
  \begin{multline}\label{eq:cons:approx}
    \Phi_\varepsilon \coloneqq \{\; \phi \in \real^{n_\beta} \times
    \trilm{n_\Sigma}_\varepsilon[\mathcal{I}_\Sigma] \times
    \trilm{n_{\mathcal{A}}}_\varepsilon[\mathcal{I}_{\mathcal{A}}] \;|
    \ifthenelse{\boolean{OneColumn}}{\;}{\\}
    g_{\mathcal{T}}(\phi) = 0,\; h_{\mathcal{T}}(\phi) \leq 0 \;\}
  \end{multline}
}
where we have defined, for any \(\varepsilon>0\) and
\(\mathcal{I}\subseteq\mathcal{L}^n\),
\[
  \trilm{n}_\varepsilon[\mathcal{I}] \coloneqq \set{ L\in\trilm{n}[\mathcal{I}]
    | L_{ii}\geq\varepsilon\; \forall i\in\intinterval{1}{n} }.
\]
We define the \emph{approximate} transformed problem as
\begin{equation}\label{eq:map:trans}
  \min_{\phi\in\Phi_\varepsilon} L_N(\mathcal{T}(\phi)) +
  R_0(\mathcal{T}(\phi)).
\end{equation}
If \(\hat{\phi}\) solves the problem \cref{eq:map:trans}, then
\(\hat{\theta}\coloneqq\mathcal{T}(\hat{\phi})\) \emph{approximately} solves the
MAP problem \cref{eq:map}. We recover the ML problem \cref{eq:ml} and its
approximate solutions with \(R_0\equiv 0\).

\subsection{Algorithm summary}
\Cref{algo:main} provides an example of our approach towards solving the
identification problem \cref{eq:map} with eigenvalue constraints and the
Cholesky factor-based substitution and elimination scheme.

\begin{algorithm}[t]
  \caption{Identification of an innovation form model \cref{eq:sskf} with
    eigenvalue constraints and the Cholesky factor-based substitution and
    elimination scheme.}
  \label{algo:main}
  \begin{algorithmic}[1]
    \REQUIRE Model functions \(\mathcal{M} = (A, B, C, D, \hat{x}_0, K, R_e)\),
    regularization term \(R_0\), initial parameters
    \(\theta_0=(\beta,\Sigma_0)\) constraint functions \(\mathcal{C}_0 =
    (g,h_0,H_0,\mathcal{A}_0)\) and sparsity patterns
    \((\mathcal{I}_{\Sigma_0},\mathcal{I}_{\mathcal{A}_0})\), a series of LMI
    region constraints \((\mathcal{D}_i,\tilde{A}_i(\cdot))_{i=1}^{n_c}\) with
    small tightening constants \(\varepsilon_i>0, i\in\mathbb{I}_{1:n_c}\), and
    a small constant \(\varepsilon>0\) (for \cref{eq:cons:approx}).
    % 
    % \ENSURE \(\hat{\mathcal{M}} =
    % (\hat{A},\hat{B},\hat{C},\hat{D},\hat{x}_{0|N},\hat{K},\hat{R}_e)
    % \coloneqq \mathcal{M}(\hat{\theta})\) denote the model coefficient
    % estimates.
    % 
    \STATE For each \(i\in\intinterval{1}{n_c}\), let \(M_{\mathcal{D}_i} :
    \realm{n_i}{n_i}\times\symm{n_i} \rightarrow \symm{n_im_i}\) denote the
    matrix characteristic function for \(\mathcal{D}_i\).
    % , and choose \(\varepsilon_i>0\).
    % 
    \STATE Extend the parameters \(\Sigma \coloneqq \Sigma_0 \oplus \left(
      \bigoplus_{i=1}^{n_c} P_i \right)\) and \(\theta \coloneqq
    (\beta,\Sigma)\) with \(P_i\in\symm{n_i}\).
    \STATE Extend the constraint functions %
    \ifthenelse{\boolean{LongVersion}}{%
      \begin{align*}
        h(\theta)
        &\coloneqq
          \begin{bmatrix} h_0(\theta_0) \\ \tr(V_1P_1) - \varepsilon_1^{-1} \\
            \ldots \\ \tr(V_{n_c}P_{n_c}) - \varepsilon_{n_c}^{-1} \end{bmatrix}, \\
        H(\beta) &\coloneqq H_0(\beta) \oplus \textstyle \left(
                   \bigoplus_{i=1}^{n_c} 0_{n_i\times n_i}
                   \right), \\
        \mathcal{A}(\theta)
        &\coloneqq \mathcal{A}_0(\theta_0) \oplus \textstyle \left(
          \bigoplus_{i=1}^{n_c}
          M_{\mathcal{D}_i}(\tilde{A}_i(\theta_0),P_i) - \varepsilon_iI \right).
      \end{align*}
    }{%
      \(h(\theta) \coloneqq \left[
      [h_0(\theta_0)]^\top \; \tr(V_1P_1) - \varepsilon_1^{-1} \; \ldots \;
      \tr(V_{n_c}P_{n_c}) - \varepsilon_{n_c}^{-1} \right]^\top\), \(H(\beta)
    \coloneqq H_0(\beta) \oplus \left( \bigoplus_{i=1}^{n_c} 0_{n_i\times n_i}
    \right)\), and \( \mathcal{A}(\theta) \coloneqq \mathcal{A}_0(\theta_0)
    \oplus \textstyle \left( \bigoplus_{i=1}^{n_c}
      M_{\mathcal{D}_i}(\tilde{A}_i(\theta_0),P_i) - \varepsilon I \right)\). %
    }%
    \STATE Extend the index sets \(\mathcal{I}_\Sigma \coloneqq
    \mathcal{I}_{\Sigma_0} \oplus \textstyle \left( \bigoplus_{i=1}^{n_c}
      \mathcal{L}^{n_i} \right)\) and \(\mathcal{I}_{\mathcal{A}} \coloneqq
    \mathcal{I}_{\mathcal{A}_0} \oplus \textstyle \left( \bigoplus_{i=1}^{n_c}
      \mathcal{L}^{n_im_i} \right)\).
    % \begin{align*}
    %   \mathcal{I}_\Sigma
    %   &\coloneqq \mathcal{I}_{\Sigma_0} \oplus \textstyle \left(
    %     \bigoplus_{i=1}^{n_c} \mathcal{L}^{n_i} \right) \\
    %   \mathcal{I}_{\mathcal{A}}
    %   &\coloneqq \mathcal{I}_{\mathcal{A}_0} \oplus \textstyle \left(
    %     \bigoplus_{i=1}^{n_c} \mathcal{L}^{n_im_i} \right).
    % \end{align*}
    % 
    \STATE Form the functions \(\mathcal{T}\), \(\mathcal{T}^{-1}\), and
    \(\tilde{\mathcal{A}}\) as in \Cref{sec:chol}.
    \STATE Form the transformed constraint functions \cref{eq:cons:trans:func}.
    \STATE Solve \cref{eq:map:trans,eq:cons:approx}, and let \(\hat{\phi}\)
    denote the solution.
    \STATE Let \(\hat{\theta}\coloneqq\mathcal{T}(\hat{\phi})\).
  \end{algorithmic}
\end{algorithm}

\section{Eigenvalue constraints}\label{sec:stable}
In this section, we elaborate on the LMI region constraints previewed in
\Cref{sec:algo}. Throughout, assume the LMI region \(\mathcal{D}\) is nonempty,
not equal to \(\cplx\), and its characteristic function \(f_{\mathcal{D}}\) and
generating matrices \((M_0,M_1)\) are fixed. Our goal in this section is to
define, using only matrix inequalities, a \emph{closed} set of matrices
\(A\in\realm{n}{n}\) such that \(\lambda(A)\subseteq\mathcal{D}\). For this
section, the matrix \(A\in\realm{n}{n}\) need not have any relation to the model
function in \cref{eq:sskf}, and can in fact be any square matrix of any
dimension (e.g., the filter stability matrix \(A-KC\), the plant stability
matrix \(A_s\) from \cref{eq:ladm}, or any submatrix thereof). Throughout this
section, we assume the matrix characteristic function \(M_{\mathcal{D}}\) is
fixed.

% We compare the eigenvalue interpretations and topologies of three LMI region
% constraints based on \(M_{\mathcal{D}}\): the strict and nonstrict LMI region
% constraints proposed by \cite{chilali:gahinet:1996,miller:decallafon:2013},
% and a novel tightened LMI region constraint formulation.

\ifthenelse{\boolean{LongVersion}}{%
\subsection{LMI region constraints}\label{ssec:lmi}
}{%
\subsubsection{LMI region constraints}\label{ssec:lmi}
} %
Originally, Chilali and Gahinet~\cite{chilali:gahinet:1996} proved the following
theorem relating the eigenvalues of \(A\in\realm{n}{n}\) to feasibility of a
system of matrix inequalities.
\begin{theorem}[\protect{\cite[Thm.~2.2]{chilali:gahinet:1996}}]\label{thm:lmi:open}
  For any \(A\in\realm{n}{n}\), we have \(\lambda(A)\subseteq\mathcal{D}\) if
  and only if
  \begin{align}\label{eq:lmi:open}
    M_{\mathcal{D}}(A,P) &\succ 0, & P&\succ 0.
  \end{align}
  holds for some \(P\in\symm{n}\).
\end{theorem}

Ultimately, we seek matrix inequalities that define a \emph{closed} set of
constraints. Due to the strictness of the inequalities \cref{eq:lmi:open}, it is
unlikely that \Cref{thm:lmi:open} achieves this goal.

\ifthenelse{\boolean{LongVersion}}{%
\subsection{Relaxed constraints}\label{ssec:lmi:relax}
}{%
\subsubsection{Relaxed constraints}\label{ssec:lmi:relax}
}
In \cite{miller:decallafon:2013}, the following relaxation of \cref{eq:lmi:open}
was considered,
\begin{align}\label{eq:lmi:relax}
  M_{\mathcal{D}}(A,P)&\succeq 0, & P&\succ 0.
\end{align}
Since \(M_{\mathcal{D}}(A,P)\) is linear in \(P\), feasibility of
\cref{eq:lmi:relax:closed} is equivalent to feasibility of
\begin{align}\label{eq:lmi:relax:closed}
  M_{\mathcal{D}}(A,P)&\succeq 0, & P&\succeq P_0
\end{align}
for some fixed \(P_0\in\posdefm{n}\).\footnote{For any \(P_0\succ 0\) and \(P\)
  satisfying \cref{eq:lmi:relax}, define the scaling factor
  \(\gamma\coloneqq\|P_0\|_2\|P^{-1}\|_2\) and a rescaled solution \(P^*
  \coloneqq \gamma P\). Then \(P^*\succeq P_0\) and \(M_{\mathcal{D}}(A,P^*) =
  \gamma M_{\mathcal{D}}(A,P) \succeq 0\).}
% While the system of inequalities
% \cref{eq:lmi:relax:closed} is suitable for constraint sets of the form
% \cref{eq:cons}, the set of matrices \(A\in\realm{n}{n}\) such that
% \cref{eq:lmi:relax:closed} (and equivalently, \cref{eq:lmi:relax}) is feasible
% is not closed. To see why this is the case, we seek a characterization of the
% feasiblity of \cref{eq:lmi:relax} in terms of the eigenvalues of a given
% \(A\in\realm{n}{n}\).
\ifthenelse{\boolean{LongVersion}}{%

  An attempt was made in \cite[Thm.~1]{miller:decallafon:2013} to characterize
  the eigenvalues of matrices \(A\in\realm{n}{n}\) for which
  \cref{eq:lmi:relax} is feasible, but this theorem does not
  correctly treat eigenvalues on the LMI region's boundary
  \(\partial\mathcal{D}\). We restate \cite[Thm.~1]{miller:decallafon:2013}
  below as a conjecture and disprove it with a simple counterexample.
  \begin{conjecture}[\protect{\cite[Thm.~1]{miller:decallafon:2013}}]\label{conj:lmi:wrong}
    The matrix \(A\in\realm{n}{n}\) satisfies
    \(\lambda(A)\subset\textnormal{cl}(\mathcal{D})\) if and only if
    \cref{eq:lmi:relax} holds for some \(P\in\symm{n}\).
  \end{conjecture}
  \begin{proof}[Counterexample]
    Let \(\mathcal{D}\) be the left half-plane, consider the Jordan block
    \(A=\begin{bsmallmatrix} 0 & 1 \\ 0 & 0 \end{bsmallmatrix}\), and suppose
    \(P=\begin{bsmallmatrix} p_{11} & p_{12} \\ p_{12} &
      p_{22} \end{bsmallmatrix}\in\symm{2}\) such that
    \cref{eq:lmi:relax} holds. Then
    \(\lambda(A)\subset\textnormal{cl}(\mathcal{D})\) and
    \[
      0\preceq M_{\mathcal{D}}(A,P) = -\begin{bmatrix} 2p_{12} & p_{22} \\
        p_{22} & 0 \end{bmatrix}
    \]
    which implies \(p_{12}=p_{22}=0\), a contradiction of
    \cref{eq:lmi:relax}.
    \renewcommand{\qedsymbol}{\textreferencemark}
  \end{proof}

  The correction to \Cref{conj:lmi:wrong} requires a more careful treatment of
  eigenvalues lying on the the LMI region's boundary \(\partial\mathcal{D}\).
  Specifically, we show in the following proposition that feasibility of
  \cref{eq:lmi:relax} for a given \(A\in\realm{n}{n}\) is equivalent to the
  eigenvalues of \(A\) being in \(\textnormal{cl}(\mathcal{D})\), with all
  non-simple eigenvalues lying in \(\mathcal{D}\) %
}{%
  In the following proposition we establish a characterizion of the eigenvalues
  of matrices \(A\in\realm{n}{n}\) for which \cref{eq:lmi:relax} is feasible %
}%
(see
\ifthenelse{\boolean{LongVersion}}{\Cref{app:lmi}}{\cite{kuntz:rawlings:2024a}}
for proof).
\begin{proposition}\label{prop:lmi:closed}
  The matrix \(A\in\realm{n}{n}\) satisfies \(\lambda(A) \subseteq
  \textnormal{cl}(\mathcal{D})\) and \(\lambda\in\mathcal{D}\) for all
  non-simple eigenvalues \(\lambda\in\lambda(A)\) if and
  only if \cref{eq:lmi:relax} holds for some \(P\in\symm{n}\).
\end{proposition}

\ifthenelse{\boolean{LongVersion}}{}{
  \begin{remark}
    A similar, but incorrect, claim is made in
    \cite[Thm.~1]{miller:decallafon:2013}: ``A matrix \(A\in\realm{n}{n}\)
    satisfies \(\lambda(A)\subset\textnormal{cl}(\mathcal{D})\) if and only if
    \cref{eq:lmi:relax} holds for some \(P\in\symm{n}\).'' A simple
    counterexample is the Jordan block \(A=\begin{bsmallmatrix} 0 & 1 \\ 0 &
      0 \end{bsmallmatrix}\) for which no \(P\succ 0\) satisfying
    \(-AP-PA^\top\succeq 0\) exists. \Cref{prop:lmi:closed} is distinguished
    from \cite[Thm.~1]{miller:decallafon:2013} by the fact that nonsimple
    eigenvalues must remain in LMI region \(\mathcal{D}\) rather than lie on its
    boundary \(\partial\mathcal{D}\).
  \end{remark}
}

\ifthenelse{\boolean{LongVersion}}{%
\subsection{Tightened constraints}\label{ssec:lmi:tight}
}{%
\subsubsection{Tightened constraints}\label{ssec:lmi:tight}
}
Instead of the ``relaxed'' constraints \cref{eq:lmi:relax}, we consider
``tightened'' constraints of the form
\begin{align}\label{eq:lmi:closed}
  M_{\mathcal{D}}(A,P)&\succeq M, & P&\succeq 0,
  & \tr(VP)&\leq\varepsilon^{-1}
\end{align}
where \(M\in\nnegdefm{nm}\) and \(V\in\posdefm{n}\) are fixed and chosen in a
way that \cref{eq:lmi:closed} implies \cref{eq:lmi:open}. While we allow \(M\)
to be semidefinite,\footnote{For some LMI regions, \(M\succeq 0\) is
  advantageous. For example, we can always take \(M
  \coloneqq \begin{bsmallmatrix} 1 & 0 \\ 0 & 0 \end{bsmallmatrix} \otimes Q\)
  with \(Q\succ 0\) for circular LMI regions. Then we can reduce the constraint
  dimension by taking the Schur complement.} in the following proposition, we
show \(M\succ 0\) always suffices.
\begin{proposition}\label{prop:lmi:posdef}
  Suppose \(M\in\posdefm{nm}\) and \(V\in\posdefm{n}\). Then
  \cref{eq:lmi:closed} implies \cref{eq:lmi:open} for all \(A\in\realm{n}{n}\)
  and \(\varepsilon>0\).
\end{proposition}
\begin{proof}
  With \(M\succ 0\) and \cref{eq:lmi:closed}, we automatically have
  \(M_{\mathcal{D}}(A,P)\succ 0\). It remains to show \cref{eq:lmi:closed}
  implies \(P\succ 0\). For contradiction suppose \cref{eq:lmi:closed} and
  \(M\succ 0\), but \(P\not\succ 0\). Then there exists a nonzero
  \(v\in\real^n\) such that \(Pv=0\), and
  \ifthenelse{\boolean{OneColumn}}{%
    \begin{equation*}
      (I_m\otimes v)^\top M_{\mathcal{D}}(A,P) (I_m\otimes v)
      = M_0\otimes (v^\top Pv)
      + M_1\otimes (v^\top APv) + M_1^\top\otimes (v^\top PA^\top v) = 0
    \end{equation*}
  }{%
    \begin{multline*}
      (I_m\otimes v)^\top M_{\mathcal{D}}(A,P) (I_m\otimes v)
      = M_0\otimes (v^\top Pv) \\
      + M_1\otimes (v^\top APv) + M_1^\top\otimes (v^\top PA^\top v) = 0
    \end{multline*}
  }
  a contradiction of the assumption \(M_{\mathcal{D}}(A,P)\succeq M\succ 0\).
\end{proof}

\ifthenelse{\boolean{LongVersion}}{%
  \begin{remark}\label{rem:diehl}
    The tightened constraint \cref{eq:lmi:closed} was inspired by a similar set
    of constraints was introduced by Diehl and
    colleagues~\cite{diehl:mombaur:noll:2009} to ``smooth'' the spectral radius.
    Specifically, feasibility of the nonlinear system
    \begin{align}\label{eq:sprad}
      s^2P - APA^\top &= W, & P&\succeq 0, & \tr(VP)&\leq\varepsilon^{-1}
    \end{align}
    implies \(\rho(A)<s\) where \(W,V\in\posdefm{n}\) and \(s,\varepsilon>0\)
    are fixed~\cite[Thms.~5.4,~5.6]{diehl:mombaur:noll:2009}. Similarly, the
    spectral abscissa was ``smoothed''
    in~\cite[Thms.~2.5,~2.6]{vanbiervliet:vandereycken:michiels:vandewalle:diehl:2009},
    and it is straightforward to
    generalize~\cite[Thms.~5.4,~5.6]{diehl:mombaur:noll:2009} to show
    feasibility of
    \begin{align}\label{eq:spabs}
      (A-sI)P + P(A-sI)^\top &= -W, & P&\succeq 0, & \tr(VP)&\leq\varepsilon^{-1}
    \end{align}
    implies \(\alpha(A)<s\) where \(W,V\in\posdefm{n}\), \(s\in\real\), and
    \(\varepsilon>0\) are fixed. The authors do not discuss LMI regions and the
    results are not obviously generalizable to them.
  \end{remark}
}{
  \begin{remark}\label{rem:diehl}
    A similar set of constraints was considered by Diehl and
    colleagues~\cite{diehl:mombaur:noll:2009,%
      vanbiervliet:vandereycken:michiels:vandewalle:diehl:2009} to construct
    ``smooth'' spectral radii and abscissa for optimization. While
    \cref{eq:lmi:closed} generalizes the application of constraints to LMI
    regions, there are other uses (e.g., minimizing the spectral radius) that we
    do not consider.
  \end{remark}
}

\ifthenelse{\boolean{LongVersion}}{%
\subsection{Constraint topology}\label{ssec:lmi:topo}
}{%
\subsubsection{Constraint topology}\label{ssec:lmi:topo}
}
Consider the constraint sets,
\begin{align*}
  \mathbb{A}_{\mathcal{D}}^n
  &\coloneqq \set{ A\in\realm{n}{n} | \exists P\in\symm{n} : \cref{eq:lmi:open}
    \textnormal{ holds} } \\
  \tilde{\mathbb{A}}_{\mathcal{D}}^n
  &\coloneqq \set{ A\in\realm{n}{n} | \exists P\in\symm{n} : \cref{eq:lmi:relax}
    \textnormal{ holds} } \\
  \mathbb{A}_{\mathcal{D}}^n(\varepsilon)
  &\coloneqq \set{ A\in\realm{n}{n} | \exists P\in\symm{n} : \cref{eq:lmi:closed}
    \textnormal{ holds} }.
\end{align*}
The following proposition characterizes the topology of
\(\mathbb{A}_{\mathcal{D}}^n\) and \(\tilde{\mathbb{A}}_{\mathcal{D}}^n\) %
\ifthenelse{\boolean{LongVersion}}{%
  (see \Cref{app:lmi:topo} for proof). %
}{%
  (see \Cref{app:lmi:topo} for proof of parts (a) and (d), and
  \cite{kuntz:rawlings:2024a} for proof of parts (b) and (c)). %
}
\begin{proposition}\label{prop:lmi:topo}
  The following holds.
  \begin{enumerate}[(a)]
  \item \(\mathbb{A}_{\mathcal{D}}^n\) is open.
  \item \(\tilde{\mathbb{A}}_{\mathcal{D}}^n\) is not open if (i) \(n\geq 2\) or
    (ii) \(\partial\mathcal{D}\cap\real\) is nonempty.
  \item \(\tilde{\mathbb{A}}_{\mathcal{D}}^n\) is not closed if (i) \(n\geq 4\)
    or (ii) \(\partial\mathcal{D}\cap\real\) is nonempty and \(n\geq 2\).
  \item \(\textnormal{cl}(\mathbb{A}_{\mathcal{D}}^n) = \set{
      A\in\realm{n}{n} | \lambda(A) \subset \textnormal{cl}(\mathcal{D}) }\).
  \end{enumerate}
\end{proposition}

%% TODO Reviewers may want a numerical example of this.
\Cref{prop:lmi:topo} reveals a weakness of the relaxed constraints
\cref{eq:lmi:relax,eq:lmi:relax:closed}. Since
\(\tilde{\mathbb{A}}_{\mathcal{D}}^n\) is not closed, any feasible path towards
a matrix \(A\in\textnormal{cl}(\mathbb{A}_{\mathcal{D}}^n) \setminus
\tilde{\mathbb{A}}_{\mathcal{D}}^n\) has no feasible limiting \(P\). In fact,
\(P\) will grow unbounded along the path of iterates.

To analyze the topology of \(\mathbb{A}_{\mathcal{D}}^n(\varepsilon)\), we take
a barrier function approach. Consider the parameterized linear SDP,
\begin{equation}\label{eq:lmi:sdp}
  \phi_{\mathcal{D}}(A) \coloneqq \inf_{P\in\nnegdefm{n}} \tr(VP)
  \;\textnormal{subject to}\; M_{\mathcal{D}}(A,P)\succeq M.
\end{equation}
The optimal value function \(\phi_{\mathcal{D}} : \realm{n}{n} \rightarrow
\nnegreal\cup\set{\infty}\) is a barrier function for the constraint
\(A\in\mathbb{A}_{\mathcal{D}}^n\). \Cref{prop:lmi:cont} establishes properties of
\(\phi_{\mathcal{D}}\) and its \(\varepsilon^{-1}\)-sublevel sets (see
\Cref{app:lmi:cont} for proof).

\begin{theorem}\label{prop:lmi:cont}
  Let \(V\in\posdefm{n}\) and \(M\in\nnegdefm{n}\) such that
  \(M_{\mathcal{D}}(A,P)\succeq M\) implies \(M_{\mathcal{D}}(A,P)\succ 0\).
  The following statements hold.
  \begin{enumerate}[(a)]
  \item \(\phi_{\mathcal{D}}\) is continuous on \(\mathbb{A}_{\mathcal{D}}\).
  \item For each \(\varepsilon>0\), \(\mathbb{A}_{\mathcal{D}}^n(\varepsilon)\)
    is equivalent to the \(\varepsilon^{-1}\)-sublevel set of
    \(\phi_{\mathcal{D}}\), i.e.,
    \begin{equation}\label{eq:lmi:sublevel}
      \mathbb{A}_{\mathcal{D}}^n(\varepsilon) = \set{ A\in\realm{n}{n}
        | \phi_{\mathcal{D}}(A)\leq\varepsilon^{-1} }
    \end{equation}
    % \begin{align}
    %   \mathbb{A}_{\mathcal{D}}^n(\varepsilon) \coloneqq
    %   &\set{ A\in\realm{n}{n} | \phi_{\mathcal{D}}(A)\leq\varepsilon^{-1} }
    %     \ifthenelse{\boolean{OneColumn}}{}{\\}
    %   = \ifthenelse{\boolean{OneColumn}}{}{
    %   &}\set{ A\in\realm{n}{n} | \exists P\succeq 0 :
    %       \textnormal{\cref{eq:lmi:cont} holds} } \label{eq:lmi:sublevel}
    % \end{align}
    and both are closed.
  \item \(\mathbb{A}_{\mathcal{D}}^n(\varepsilon) \nearrow
    \mathbb{A}_{\mathcal{D}}^n\) as \(\varepsilon\searrow 0\).
  \end{enumerate}
\end{theorem}

\ifthenelse{\boolean{LongVersion}}{%
  \begin{remark}
    To reconstruct \cref{eq:sprad} via \Cref{prop:lmi:cont}, we set \(M=sW\oplus
    0_{n\times n}\) for any \(W,V\succ 0\) and \(s>0\) and apply the Schur
    complement lemma to \(M_{\mathcal{D}_2}(A,P)/s-M/s\), where \(\mathcal{D}_2\)
    is the circle defined in \Cref{lem:lmi:basic} with \(x_0=0\), and
    \(M_{\mathcal{D}_2}\) is defined by the generating matrices used in
    \Cref{lem:lmi:basic}. Then the \(\varepsilon^{-1}\)-sublevel set of
    \(\phi_{\mathcal{D}_2}\) equals the set of \(A\in\realm{n}{n}\) for which
    \cref{eq:sprad} is feasible.

    Similarly, we can reconstruct the set of \(A\in\realm{n}{n}\) for which
    \cref{eq:spabs} is feasible as \(\varepsilon^{-1}\)-sublevel sets of
    \(\phi_{\mathcal{D}_1}\), where \(\mathcal{D}_1\) is the shifted half-plane
    defined in \Cref{lem:lmi:basic}, and \(M=W\) for any \(W,V\succ 0\).
  \end{remark}
}{}
\section{Cholesky substitution and elimination}\label{sec:chol}
In this section, we seek to approximate certain NSDPs by NLPs. Specifically, we
consider the NSDP
\begin{equation}\label{eq:nsdp}
  \min_{(\beta,\Sigma)\in\Theta} f(\beta,\Sigma)
\end{equation}
where \(\Theta\) is defined as in \cref{eq:cons}. This covers both ML
\cref{eq:ml} and MAP \cref{eq:map} problems with constraints \cref{eq:cons}. We
combine Cholesky factor-based substitution with an elimination scheme to convert
the NSDP to a NLP while adding just \(|\mathcal{I}_{\mathcal{A}}|\) variables to
the optimization problem.

For this section, we define the following notation. For each
\(\mathcal{I}\subseteq\mathcal{L}^n\), let
\(\pi_{\mathcal{I}}^L:\realm{n}{n}\rightarrow \trilm{n}[\mathcal{I}]\) and
\(\pi_{\mathcal{I}} : \realm{n}{n} \rightarrow \symm{n}[\mathcal{I}]\) denote
the orthogonal projections (in the Frobenius norm) from \(\realm{n}{n}\) onto
the subspaces \(\trilm{n}[\mathcal{I}]\) and \(\symm{n}[\mathcal{I}]\),
respectively. Let \(\textnormal{chol} : \posdefm{n} \rightarrow \postrilm{n}\)
denote the invertible function that maps a positive definite matrix to its
Cholesky factor.

\subsection{Burer-Monteiro-Zhang method}\label{ssec:bmz}
We first consider the simplified constraint set
\begin{equation}\label{eq:cons:simple}
  \mathcal{P} \coloneqq \set{ (x,Q)\in\real^m\times\symm{n}[\mathcal{I}] |
    Q \succeq H(x) }
\end{equation}
where \(\mathcal{D}^n\subseteq\mathcal{I}\subseteq\mathcal{L}^n\) and
\(H:\real^m\rightarrow\symm{n}\). As in \cite{burer:monteiro:zhang:2002}, we
parameterize the matrix argument \(Q\) in a way that automatically enforces the
constraint \(Q\succ H(x)\) while introducing just \(n\) scalar inequality
constraints.

Recall \(Q\succ H\) if and only if \(Q=H+LL^\top\) for the \emph{unique} matrix
\(L=\textnormal{chol}(Q-H)\in\postrilm{n}\). With
\(\mathcal{J}\coloneqq\mathcal{L}^n\setminus\mathcal{I}\), we can split \(L\)
into a sum of \(L^{\mathcal{I}}\in\postrilm{n}[\mathcal{I}]\) and
\(L^{\mathcal{J}}\in\trilm{n}[\mathcal{J}]\), giving
\begin{equation}\label{eq:bmz:Q}
  Q = H + (L^{\mathcal{I}} + L^{\mathcal{J}})
  (L^{\mathcal{I}} + L^{\mathcal{J}})^\top.
\end{equation}
But \(Q\in\symm{n}[\mathcal{I}]\), so we can apply the vectorization operator
\(\vecs_{\mathcal{J}}\) on both sides to give
\begin{equation}\label{eq:bmz:proj}
  \vecs_{\mathcal{J}}(H + (L^{\mathcal{I}} + L^{\mathcal{J}})
  (L^{\mathcal{I}} + L^{\mathcal{J}})^\top) = 0.
\end{equation}
\Cref{eq:bmz:proj} defines \(|\mathcal{J}|\) equations to solve for the
\(|\mathcal{J}|\) variables of \(L^{\mathcal{J}}\), where each
\(L_{ij}^{\mathcal{J}}\) is fully specified by \(H_{ij}\) and the \(L_{i'j'}\)
with \((i',j')<(i,j)\).\footnote{The lexicographic order \(<\) on
  \(\allint^2\) is defined by \((i,j)<(i',j')\) if \(i<i'\) or
  \((i=i')\land(j<j')\).} In \Cref{algo:chol}, we compute the
\(L_{ij}^{\mathcal{J}}\) in ascending lexicographic order via Cholesky
factorization.

\begin{algorithm}[t]
  \caption{Cholesky factorization algorithm for solving systems of the form
    \cref{eq:bmz:proj} based on \cite[Lem.~1]{burer:monteiro:zhang:2002}.}
  \label{algo:chol}
  \begin{algorithmic}[1]
    \REQUIRE $\mathcal{D}^n \subseteq \mathcal{I} \subseteq \mathcal{L}^n$,
    $L^{\mathcal{I}}\in\postrilm{n}[\mathcal{I}]$, and $H\in\symm{n}$
    % \REQUIRE \(\mathcal{I} \subseteq \mathcal{L}^n\),
    % \(L^{\mathcal{I}}\in\trilm{n}[\mathcal{I}]\), and \(H\in\symm{n}\) such that
    % \(\mathcal{D}^n \subseteq \mathcal{I}\) and \(L^{\mathcal{I}}_{ii}\neq 0\).
    \STATE $(\mathcal{J}, L^{\mathcal{J}}) \leftarrow (\mathcal{L}^n\setminus
    \mathcal{I}, 0_{n\times n})$
    \FOR {each $(i,j)\in\mathcal{J}$ in ascending lexicographic order}
    % \FOR {$i = 1$ to $n-1$}
    % \FOR {$j = i + 1$ to $n$}
    % \IF {$(i,j)\in\mathcal{J}$}
    \STATE \(L^{\mathcal{J}}_{ij} \leftarrow -\frac{1}{L_{jj}^{\mathcal{I}}}
    ( H_{ij} + \sum_{k=1}^{j-1} (L^{\mathcal{I}}_{ik} +
      L^{\mathcal{J}}_{ik}) (L^{\mathcal{I}}_{jk} +
      L^{\mathcal{J}}_{jk}) )\)
    % \ENDIF
    % \ENDFOR
    \ENDFOR
    \RETURN $L^{\mathcal{J}}$
  \end{algorithmic}
\end{algorithm}

\ifthenelse{\boolean{LongVersion}}{%
  Notice that each \(L^{\mathcal{J}}\) is fully defined by \(H\) and
  \(L^{\mathcal{I}}\) via \cref{algo:chol}, so we have proven the following
  lemma.
  \begin{lemma}[\protect{\cite[Lem.~1]{burer:monteiro:zhang:2002}}]\label{lem:bmz:J}
    For each \((H, L^{\mathcal{I}}) \in \symm{n} \times \trilm{n}[\mathcal{I}]\)
    such that \(L^{\mathcal{I}}_{ii}\neq 0\) for each
    \(i\in\intinterval{1}{n}\), there is a unique \(L^{\mathcal{J}} \in
    \trilm{n}[\mathcal{J}]\) satisfying \cref{eq:bmz:proj}.
  \end{lemma}
}{}

With a slight abuse of notation, we let \(L^{\mathcal{J}} =
L^{\mathcal{J}}(H,L^{\mathcal{I}})\) denote the function defined by
\Cref{algo:chol}, which maps each \((H,L^{\mathcal{I}}) \in \symm{n} \times
\postrilm{n}[\mathcal{I}]\) to the matrix \(L^{\mathcal{J}} \in
\trilm{n}[\mathcal{J}]\) \emph{uniquely} satisfying \cref{eq:bmz:proj}.
Moreover, we let
% \begin{align*}
%   L(H,L^{\mathcal{I}}) &\coloneqq L^{\mathcal{I}} + L^{\mathcal{J}}(H,L^{\mathcal{I}}) \\
%   Q(H,L^{\mathcal{I}}) &\coloneqq H + L(H,L^{\mathcal{I}})[L(H,L^{\mathcal{I}})]^\top
% \end{align*}
\[
  Q(H,L^{\mathcal{I}}) \coloneqq H + (L^{\mathcal{I}} +
  L^{\mathcal{J}}(H,L^{\mathcal{I}})) (L^{\mathcal{I}} +
  L^{\mathcal{J}}(H,L^{\mathcal{I}}))^\top
\]
as in \cref{eq:bmz:Q}. Clearly \(Q(H,L^{\mathcal{I}}) \succ H\) is
satisfied by definition. Finally, we define the transformation
\begin{equation}\label{eq:bmz:trans}
  T(x,L^{\mathcal{I}}) \coloneqq \left( x, Q(H(x),L^{\mathcal{I}}) \right)
\end{equation}
which has the inverse
\begin{equation}\label{eq:bmz:trans:inv}
  T^{-1}(x,Q) \coloneqq \left( x, \pi_{\mathcal{I}}^L[
    \textnormal{chol}(Q-H(x))] \right)
\end{equation}
and we have the following lemma.
\begin{lemma}[\protect{\cite[Lem.~2]{burer:monteiro:zhang:2002}}]\label{lem:bmz}
  The function \(T\) defined by \cref{eq:bmz:trans} is a bijection between
  \(\real^m\times\postrilm{n}[\mathcal{I}]\) and
  \(\textnormal{int}(\mathcal{P})\).
\end{lemma}

Differentiability of \(T\) and \(T^{-1}\) follow from differentiability of \(H\)
and \cref{algo:chol}. In fact, these functions are as smooth as \(H\).
\ifthenelse{\boolean{LongVersion}}{%
  More importantly, the bijection \(T\) allows us to transform the minimum of a
  continuous function over \(\mathcal{P}\) to an infimum over
  \(\real^m\times\postrilm{n}[\mathcal{I}]\), given by the following theorem.
  \begin{theorem}[\protect{\cite[Thm.~1]{burer:monteiro:zhang:2002}}]\label{thm:bmz}
    If \(f:\mathcal{P}\rightarrow\real\) is continuous and attains a minimum in
    \(\mathcal{P}\), then
    \begin{equation}\label{eq:bmz}
      \min_{(x,Q)\in\mathcal{P}} f(x,Q) =
      \inf_{(x,L^{\mathcal{I}})\in\real^m\times\postrilm{n}[\mathcal{I}]}
      f(T(x,L^{\mathcal{I}})).
    \end{equation}
  \end{theorem}

  We reiterate the proof of \Cref{thm:bmz} for illustrative purposes.

  \begin{proof}
    Continuity of \(f\) implies its minimum over \(\mathcal{P}\) equals its
    infimum over \(\textnormal{int}(\mathcal{P})\), i.e.,
    \[
      \min_{(x,Q)\in\mathcal{P}} f(x,Q) =
      \inf_{(x,Q)\in\textnormal{int}(\mathcal{P})} f(x,Q)
    \]
    Since \(T\) is a bijection, we can transform the optimization variables as
    follows:
    \[
      \inf_{(x,Q)\in\textnormal{int}(\mathcal{P})} f(x,Q) =
      \inf_{(x,L^{\mathcal{I}})\in T^{-1}(\textnormal{int}(\mathcal{P}))}
      f(T(x,L^{\mathcal{I}})).
    \]
    Finally, since \(\real^m\times\postrilm{n}[\mathcal{I}] =
    T^{-1}(\textnormal{int}(\mathcal{P}))\), we have \cref{eq:bmz}.
  \end{proof}
}{%
  It is shown in \cite[Thm.~1]{burer:monteiro:zhang:2002} that the minimum of
  any continuous function \(f\) over \(\mathcal{P}\) (if it exists) is equal to
  the infimum of \(f\circ T\) over \(\real^m\times\postrilm{n}[\mathcal{I}]\). %
}

\subsection{Generalized Burer-Monteiro-Zhang method}\label{ssec:gbmz}
We return to constraints of the form \cref{eq:cons}.
% We move on to the general NSDP,
% \begin{equation}\label{eq:nsdp}
%   \min_{(\beta,\Sigma)\in\Theta} f(\beta,\Sigma)
% \end{equation}
% where \(f:\real^{n_\beta}\times\symm{n_\Sigma}\rightarrow\real\) and \(\Theta\)
% is defined as in \cref{eq:cons}.
Recall \Cref{assm:cons} requires the matrix inequalities are strictly feasible
in the constraint set. We use a similar procedure to \Cref{ssec:bmz}, but
\Cref{algo:chol} must be applied to \emph{each} strict inequality \(\Sigma\succ
H\) and \(\mathcal{A}(\beta,\Sigma)\succ 0\).

For the sparse symmetric matrix \(\Sigma\) and matrix inequality \(\Sigma\succ
H(\beta)\), the procedure is the same as in \Cref{ssec:bmz}. Let
\(L^{\mathcal{J}_\Sigma} = L^{\mathcal{J}_\Sigma}(H,L^{\mathcal{I}_\Sigma})\)
denote the function defined by \Cref{algo:chol} with \(L^{\mathcal{I}} =
L^{\mathcal{I}_\Sigma}\), \(\mathcal{I}=\mathcal{I}_\Sigma\), and
\(n=n_\Sigma\). Then
% \begin{align*}
%   L_\Sigma(H,L^{\mathcal{I}_\Sigma})
%   &\coloneqq L^{\mathcal{I}_\Sigma} + L^{\mathcal{J}_\Sigma}(H,L^{\mathcal{I}_\Sigma}) \\
%   \Sigma(H,L^{\mathcal{I}_\Sigma})
%   &\coloneqq H + L_\Sigma(H,L^{\mathcal{I}_\Sigma})[L_\Sigma(H,L^{\mathcal{I}_\Sigma})]^\top
% \end{align*}
\ifthenelse{\boolean{OneColumn}}{%
  \[
    \Sigma(\beta,L^{\mathcal{I}_\Sigma}) \coloneqq H + (L^{\mathcal{I}_\Sigma} +
    L^{\mathcal{J}_\Sigma}(H,L^{\mathcal{I}_\Sigma}))(L^{\mathcal{I}_\Sigma} +
    L^{\mathcal{J}_\Sigma}(H,L^{\mathcal{I}_\Sigma}))^\top
  \]
}{%
  \begin{multline*}
    \Sigma(\beta,L^{\mathcal{I}_\Sigma}) \coloneqq H + (L^{\mathcal{I}_\Sigma} +
    L^{\mathcal{J}_\Sigma}(H,L^{\mathcal{I}_\Sigma})) \\
    \times (L^{\mathcal{I}_\Sigma} + L^{\mathcal{J}_\Sigma}(H,L^{\mathcal{I}_\Sigma}))^\top
  \end{multline*}
}%
guarantees \(\Sigma(H,L^{\mathcal{I}_\Sigma}) \succ H\) and
\(\Sigma(H,L^{\mathcal{I}_\Sigma})\in\symm{n_\Sigma}[\mathcal{I}_\Sigma]\) for
all \((H,L^{\mathcal{I}_\Sigma})\in\symm{n_\Sigma} \times
\postrilm{n_\Sigma}[\mathcal{I}_\Sigma]\). In other words, \(\Sigma\) is fully
defined and the constraint \(\Sigma\succ H\) automatically satisfied by
\((H,L^{\mathcal{I}_\Sigma}) \in \symm{n_\Sigma} \times
\postrilm{n_\Sigma}[\mathcal{I}_\Sigma]\).

For the general matrix inequality \(\mathcal{A}(\beta,\Sigma)\succeq 0\), the
procedure is slightly different. Let \(L^{\mathcal{J}_{\mathcal{A}}} =
L^{\mathcal{J}_{\mathcal{A}}}(L^{\mathcal{I}_{\mathcal{A}}})\) denote function
defined by \Cref{algo:chol} with \(L^{\mathcal{I}} =
L^{\mathcal{I}_{\mathcal{A}}}\), \(\mathcal{I}=\mathcal{I}_{\mathcal{A}}\),
\(n=n_{\mathcal{A}}\), and \(H=0\). Define the functions
% \begin{align*}
%   L_{\mathcal{A}}(L^{\mathcal{I}_{\mathcal{A}}})
%   &\coloneqq L^{\mathcal{I}_{\mathcal{A}}} +
%     L^{\mathcal{J}_{\mathcal{A}}}(L^{\mathcal{I}_{\mathcal{A}}}) \\
%   \mathcal{A}(L^{\mathcal{I}_{\mathcal{A}}})
%   &\coloneqq L_{\mathcal{A}}(L^{\mathcal{I}_{\mathcal{A}}})
%     [L_{\mathcal{A}}(L^{\mathcal{I}_{\mathcal{A}}})]^\top
      %   \end{align*}
\[
  \mathcal{A}(L^{\mathcal{I}_{\mathcal{A}}}) \coloneqq
  (L^{\mathcal{I}_{\mathcal{A}}} +
  L^{\mathcal{J}_{\mathcal{A}}}(L^{\mathcal{I}_{\mathcal{A}}}))(L^{\mathcal{I}_{\mathcal{A}}}
  + L^{\mathcal{J}_{\mathcal{A}}}(L^{\mathcal{I}_{\mathcal{A}}}))^\top
\]
which guarantees \(\mathcal{A}(L^{\mathcal{I}_{\mathcal{A}}}) \in
\posdefm{n_{\mathcal{A}}}[\mathcal{I}_{\mathcal{A}}]\) for all
\(L^{\mathcal{I}_{\mathcal{A}}} \in
\postrilm{n_{\mathcal{A}}}[\mathcal{I}_{\mathcal{A}}]\). However, the constraint
is not fully eliminated; we are left with \(|\mathcal{I}_{\mathcal{A}}|\)
equality constraints in the transform space,
\[
  \vecs_{\mathcal{I}_{\mathcal{A}}}(\mathcal{A}(\beta,
  \Sigma(H(\beta),L^{\mathcal{I}_\Sigma})) -
  \mathcal{A}(L^{\mathcal{I}_{\mathcal{A}}})) = 0
\]
with the other \(|\mathcal{L}^{n_{\mathcal{A}}} \setminus
\mathcal{I}_{\mathcal{A}}|\) constraints automatically guaranteed by
\Cref{algo:chol}.

To define the new constraints, we require the variable transformations
\begin{subequations}\label{eq:gbmz:trans}
  \begin{align}
    \mathcal{T}(\beta,L^{\mathcal{I}_\Sigma},L^{\mathcal{I}_{\mathcal{A}}})
    &\coloneqq \left( \beta, \Sigma(H(\beta),L^{\mathcal{I}_\Sigma}) \right) \\
    \mathcal{A}_{\mathcal{T}}(\beta,L^{\mathcal{I}_\Sigma},L^{\mathcal{I}_{\mathcal{A}}})
    &\coloneqq \mathcal{A}(L^{\mathcal{I}_{\mathcal{A}}})
  \end{align}
\end{subequations}
which are well-defined for all
\((\beta,L^{\mathcal{I}_\Sigma},L^{\mathcal{I}_{\mathcal{A}}}) \in
\real^{n_\beta} \times \trilm{n_\Sigma}[\mathcal{I}_\Sigma] \times
\trilm{n_{\mathcal{A}}}[\mathcal{I}_{\mathcal{A}}]\). With the functions
\cref{eq:gbmz:trans}, we define the transformed constraint functions
\((g_{\mathcal{T}},h_{\mathcal{T}})\) and the transformed constraint set
\(\Phi\subseteq\real^{n_\beta} \times \trilm{n_\Sigma}[\mathcal{I}_\Sigma]
\times \trilm{n_{\mathcal{A}}}[\mathcal{I}_{\mathcal{A}}]\) according to
\cref{eq:cons:trans:func,eq:cons:trans}. The inverse transform is
\ifthenelse{\boolean{OneColumn}}{%
  \begin{equation}\label{eq:gbmz:trans:inv}
    \mathcal{T}^{-1}(\beta,\Sigma) \coloneqq \left( \beta,
      \pi_{\mathcal{I}_\Sigma}^L[\textnormal{chol}(\Sigma - H(\beta))],
      \pi_{\mathcal{I}_{\mathcal{A}}}^L
      [\textnormal{chol}(\mathcal{A}(\beta,\Sigma))] \right)
  \end{equation}
}{%
  \begin{multline}\label{eq:gbmz:trans:inv}
    \mathcal{T}^{-1}(\beta,\Sigma) \coloneqq \left( \beta,
      \pi_{\mathcal{I}_\Sigma}^L[\textnormal{chol}(\Sigma - H(\beta))], \right. \\
    \left. \pi_{\mathcal{I}_{\mathcal{A}}}^L
      [\textnormal{chol}(\mathcal{A}(\beta,\Sigma))] \right)
  \end{multline}
}
for all \((\beta,\Sigma)\in\Theta_{++}\), and we have the following lemma.
\begin{lemma}\label{lem:gbmz}
  The function \(\mathcal{T}\) defined by \cref{eq:gbmz:trans} is a bijection
  between \(\Phi\) and \(\Theta_{++}\).
\end{lemma}
\ifthenelse{\boolean{LongVersion}}{%
  \begin{proof}
    First, we have \(\mathcal{T}(\Phi)\subseteq\Theta_{++}\) since the
    transformed constraints guarantee the constraints \(g(\beta,\Sigma)=0\),
    \(h(\beta,\Sigma)\leq 0\), \(\Sigma\succ H(\beta)\), and
    \(\mathcal{A}(\beta,\Sigma)\succ 0\) for any \((\beta,\Sigma) \coloneqq
    \mathcal{T}(\phi) \) and \(\phi\in\Phi\). Next, it is clear by construction
    that \(\mathcal{T}^{-1}\circ\mathcal{T}\) is the identity map on \(\Phi\).
    Therefore \(\mathcal{T}\) is injective. Similarly, we have
    \(\mathcal{T}^{-1}(\Theta_{++})\subseteq\Phi\) by construction, and
    \(\mathcal{T}\circ\mathcal{T}^{-1}\) is the identity map on \(\Theta_{++}\),
    so \(\mathcal{T}:\Phi\) is surjective.
  \end{proof}
}{}

Under \Cref{assm:cons}, the functions \(\mathcal{T}\), \(\mathcal{T}^{-1}\), and
\(\mathcal{A}_{\mathcal{T}}\) are as smooth as \(H\), and moreover, %
\ifthenelse{\boolean{LongVersion}}{%
  the bijection \(\mathcal{T}\) transforms a minimum over \(\Theta\) into an
  infimum over \(\Phi\).

  \begin{proposition}\label{prop:gbmz}
    If \Cref{assm:cons} holds and \(f:\Theta\rightarrow\real\) is continuous and
    attains a minimum in \(\Theta\), then
    \[
      \min_{\theta\in\Theta} f(\theta) = \inf_{\phi \in \Phi}
      f(\mathcal{T}(\phi)).
    \]
  \end{proposition}
  \begin{proof}
    The proof follows that of \Cref{thm:bmz}, noting that \Cref{assm:cons} gives
    \(\textnormal{cl}(\Theta_{++})=\Theta\) and therefore the minimum of \(f\)
    over \(\Theta\) equals the infimum of \(f\) over \(\Theta_{++}\).
  \end{proof}
}{%
  for any continuous \(f\), we have \(\min_{\theta\in\Theta} f(\theta) =
  \inf_{\phi\in\Phi} f(\mathcal{T}(\phi))\) if the former
  exists~\cite{kuntz:rawlings:2024a}. %
}

\subsection{Approximate solutions}\label{ssec:gbmz:approx}
As mentioned in \Cref{sec:algo}, we consider a lower bound \(\varepsilon>0\) on
the diagonal elements of
\((L^{\mathcal{I}_\Sigma},L^{\mathcal{I}_{\mathcal{A}}})\). We define the
tightened constraint set \(\Phi_\varepsilon\) by \cref{eq:cons:approx}. %
\ifthenelse{\boolean{LongVersion}}{%
  In the following theorem we show, under \Cref{assm:cons} and continuity of
  \(f\), the infimum of \(f\circ\mathcal{T}\) over \(\Phi_\varepsilon\)
  converges to the minimum of \(f\) over \(\Theta\) (see \Cref{app:bmz:approx}
  for proof).
  \begin{theorem}\label{prop:gbmz:value}
    Suppose \(f\) is continuous and attains a minimum in \(\Theta\). Define
    \[
      \mu_0\coloneqq\min_{\theta\in\Theta} f(\theta)
    \]
    and
    \begin{equation}\label{eq:gbmz:eps}
      \mu_\varepsilon \coloneqq \inf_{\phi\in\Phi_\varepsilon}
      f(\mathcal{T}(\phi))
    \end{equation}
    for each \(\varepsilon>0\). If \Cref{assm:cons} holds, then
    \(\mu_\varepsilon\searrow\mu\) as \(\varepsilon\searrow 0\).
  \end{theorem}

}{%
  Under \Cref{assm:cons} and continuity of \(f\), it can be shown the infimum of
  \(f\circ\mathcal{T}\) over \(\Phi_\varepsilon\) converges to the minimum of
  \(f\) over \(\Theta\) as \(\varepsilon\searrow
  0\)~\cite{kuntz:rawlings:2024a}. %
}%
In fact, with a few additional requirements on the objective \(f\), convergence
of approximate problem solutions to the solution of the original problem is
guaranteed by the following theorem (see \Cref{app:bmz:approx} for proof).
%% TODO move to notation section?
% As in \cite{rockafellar:wets:1998}, we define the
% \(\limsup\) of a set-valued function \(S:\mathcal{X}\rightarrow\mathcal{Y}\) by
% \[
%   \limsup_{x\rightarrow\overline x} S(x) \coloneqq \set{ y\in\mathcal{Y} |
%     \exists x_k\rightarrow\overline x, y_k\rightarrow \overline y : y_k\in
%     S(x_k) }
% \]
% and we say \(S\) is \emph{outer semicontinuous at \(\overline
% x\in\mathcal{X}\)} if \(\limsup_{x\rightarrow\overline x} S(x)\subseteq
% S(\overline x)\).
\begin{theorem}\label{prop:gbmz:cont}
  Suppose \(f\) is continuous and \Cref{assm:cons} holds. Consider the
  set-valued function \(\hat\theta : \nnegreal \rightarrow
  \mathcal{P}(\Theta)\), defined as \(\hat\theta_\varepsilon \coloneqq
  \argmin_{\theta\in\mathcal{T}(\Phi_\varepsilon)} f(\theta)\) for all
  \(\varepsilon>0\), and \(\hat\theta_0 \coloneqq \argmin_{\theta\in\Theta}
  f(\theta)\). If there exists \(\alpha\in\real\) and compact
  \(C\subseteq\Theta\) such that
  \[
    \Theta_{f\leq\alpha} \coloneqq \set{ \theta \in \Theta | f(\theta)\leq\alpha
    }
  \]
  is contained in \(C\) and \(\Theta_{f\leq\alpha}\cap\Theta_{++}\) is nonempty,
  then there exists \(\overline\varepsilon > 0\) such that, for all
  \(\varepsilon_0 \in [0,\overline\varepsilon)\),
  \begin{enumerate}[(a)]
  \item \(f\) achieves a minimum in \(\Theta\) and \(\hat\theta_0\) is nonempty;
  \item if \(\varepsilon_0>0\), then \(f\) achieves a minimum in
    \(\mathcal{T}(\Phi_{\varepsilon_0})\) and \(\hat\theta_{\varepsilon_0}\)
    is nonempty;
  \item \(\mu_\varepsilon\) is continuous and \(\hat\theta_\varepsilon\) is
    outer semicontinuous at \(\varepsilon=\varepsilon_0\); and
  \item if \(\hat\theta_0\) is a singleton, then \(\limsup_{\varepsilon\searrow
      0} \hat\theta_{\varepsilon} = \hat\theta_0\).
  \end{enumerate}
\end{theorem}

\begin{remark}
  A log-barrier approach was used by Burer et
  al.~\cite{burer:monteiro:zhang:2002} to handle strict inequalities in
  \(L\in\postrilm{n}[\mathcal{I}]\) and achieve global convergence for linear
  SDPs. For \ifthenelse{\boolean{LongVersion}}{the problem
    \cref{eq:bmz}}{problems with constraints of the form~\cref{eq:cons:simple}},
  the log-barrier term eliminates all remaining constraints. However, for the
  problem~\cref{eq:nsdp}, many constraints remain in addition to the strict
  inequalities on the diagonal elements of
  \((L^{\mathcal{I}_\Sigma},L^{\mathcal{I}_{\mathcal{A}}}) \in
  \postrilm{n_\Sigma}[\mathcal{I}_\Sigma] \times
  \postrilm{n_{\mathcal{A}}}[\mathcal{I}_{\mathcal{A}}]\).
\end{remark}

\section{Case Studies}\label{sec:casestudies}
\begin{figure}[t]
  \begin{center}
    \ifthenelse{\boolean{OneColumn}}{%
      \includegraphics[width=0.75\linewidth]{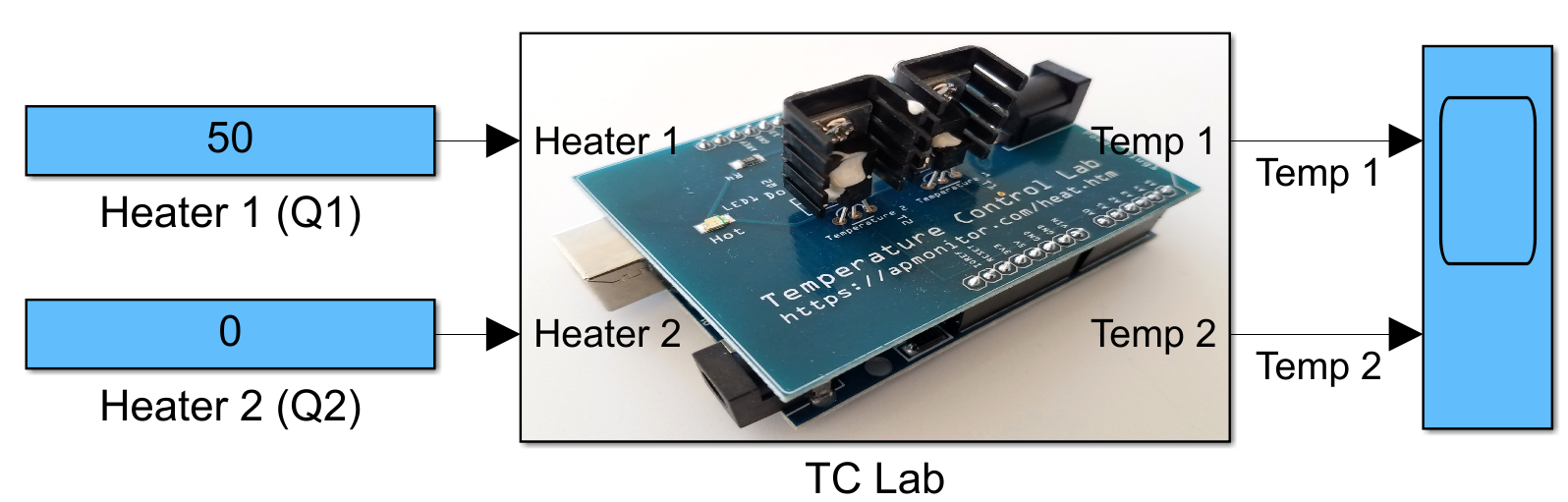}
    }{
      \includegraphics[width=\linewidth]{tclab_crop.png}
    }
    \caption{Benchmark temperature Control Laboratory (TCLab)
      \cite{park:martin:kelly:hedengren:2020}.}
    \label{fig:tclab}
  \end{center}
\end{figure}

In this section, we present two real-world case studies in which
\Cref{algo:main} is used to identify the LADM~\cref{eq:ladm} and implement
offset-free MPC.\@ In the first case study, we consider the temperature control
laboratory (TCLab) (see \Cref{fig:tclab}), an Arduino-based temperature control
laboratory that serves as a low-cost\footnote{The TCLab is available for under
  \$40 from
  \href{https://apmonitor.com/heat.htm}{\url{https://apmonitor.com/heat.htm}}
  and
  \href{https://www.amazon.com/gp/product/B07GMFWMRY}{\url{https://www.amazon.com/gp/product/B07GMFWMRY}}.}
benchmark for linear MIMO control~\cite{park:martin:kelly:hedengren:2020}. We
identify the TCLab from open-loop data and use the resulting model to design an
offset-free MPC.\@ We compare closed-loop control and estimation performance of
these models to that of offset-free MPCs designed with the identification
methods from \cite{kuntz:rawlings:2022,kuntz:downs:miller:rawlings:2023a}. In
the second case study, data from an industrial-scale chemical reactor is used to
design Kalman filters for the linear augmented disturbance model, and the
closed-loop estimation performance is compared to that of the designs proposed
in \cite{kuntz:downs:miller:rawlings:2023a}.

Throughout these experiments, we use an \(\ell_2\) regularization term in the
transformed space,\footnote{With \(L_\Sigma(\overline\beta,\overline
  L^{\mathcal{I}_\Sigma})=0\), the last term of \cref{eq:ml:reg:chol} becomes
  proportional to \(\tr(L_\Sigma L_\Sigma^\top) = \tr(\Sigma)\) where \(L_\Sigma
  = L_\Sigma(\beta,L^{\mathcal{I}_\Sigma})\) and \(\Sigma =
  \Sigma(\beta,L^{\mathcal{I}_\Sigma})\).}$^,$\footnote{With
  \(L^{\mathcal{J}_\Sigma}(\beta, L^{\mathcal{I}_\Sigma}) \equiv 0\) (e.g.,
  \(\Sigma\) is block diagonal and \(H(\beta)\equiv 0\)) the last term of
  \cref{eq:ml:reg:chol} is proportional to \(\|L^{\mathcal{I}_\Sigma} -
  \overline L^{\mathcal{I}_\Sigma}\|_{\mathrm{F}}^2 =
  |\vect_{\mathcal{I}_\Sigma}(L^{\mathcal{I}_\Sigma} - \overline
  L^{\mathcal{I}_\Sigma})|^2\).}%
\ifthenelse{\boolean{OneColumn}}{%
  \begin{equation}\label{eq:ml:reg:chol}
    -\ln p_0(\beta,L^{\mathcal{I}_\Sigma}) \propto R_0(\beta,L^{\mathcal{I}_\Sigma})
    \coloneqq \frac{\rho}{2}\left(
      |\beta-\overline\beta|^2 + \|L_\Sigma(\beta,L^{\mathcal{I}_\Sigma}) -
      L_\Sigma(\overline\beta,\overline L^{\mathcal{I}_\Sigma})\|_{\mathrm{F}}^2 \right).
  \end{equation}
}{%
  \begin{multline}\label{eq:ml:reg:chol}
    -\ln p_0(\beta,L^{\mathcal{I}_\Sigma}) \propto R_0(\beta,L^{\mathcal{I}_\Sigma})
    \coloneqq \frac{\rho}{2}|\beta-\overline\beta|^2 \\
    + \frac{\rho}{2}\|L_\Sigma(\beta,L^{\mathcal{I}_\Sigma}) -
    L_\Sigma(\overline\beta,\overline L^{\mathcal{I}_\Sigma})\|_{\mathrm{F}}^2.
  \end{multline}
} %
where \(\rho\geq 0\) is the regularization weight and
\((\overline\beta,\overline L^{\mathcal{I}_\Sigma},\overline
L^{\mathcal{I}_{\mathcal{A}}})\) denote the initial guess for the optimizer. The
variable \(L_{\mathcal{A}}\) is not regularized. With \(\rho=0\), the MAP
problem \cref{eq:map} with the regularizer \cref{eq:ml:reg:chol} simplifies to
the standard ML identification problem \cref{eq:ml}.

The initial guess for the ML and MAP problems is based on a nested ML
estimation approach described
in~\cite{kuntz:rawlings:2022,kuntz:downs:miller:rawlings:2023a}. The initial
guess methods effectively augment standard identification methods (e.g.,
principal component analysis (PCA), Ho-Kalman (HK), canonical correlation
analysis (CCA) algorithms), so we refer to the initial guess models as
``augmented'' versions of the standard method being used (e.g., augmented PCA,
augmented HK, augmented CCA). Each optimization problem is formulated in CasADi
via \Cref{algo:main} and solved with IPOPT.\@ Information about each model fit
and configuration is presented in \Cref{table:tclab}. Wall times for a
single-thread of an Intel Core i9-10850K processor are reported.

\subsection{Benchmark temperature controller}\label{ssec:tclab}
\begin{table*}
  \renewcommand{\arraystretch}{1.3}
  \ifthenelse{\boolean{OneColumn}}{\vspace{0.5em}}{}
  \caption{TCLab model fitting results. $^*$~The augmented PCA/ARX
    identification methods are not iterative. $^{**}$~The maximum number of
    iterations was set at 500.}
  \label{table:tclab}
  \centering
  \ifthenelse{\boolean{OneColumn}\AND\boolean{LongVersion}}{\tiny}{}
  
\begin{tabular}{c||c|c|c|c|c|c|c|c}
\hline
\multicolumn{1}{c||}{\multirow{2}{*}{\bfseries Model}}
& \multicolumn{3}{c|}{\bfseries Results}
& \multicolumn{5}{c}{\bfseries Configuration} \\
\cline{2-9}
& \bfseries Time (s) & \bfseries Iterations & $L_N(\hat{\theta})$
& \bfseries{Method} & $\rho$ & $\mathcal{D}$
& $\varepsilon$ & $\varepsilon_i$ \\
\hline\hline
Augmented PCA & 0.02 & N/A$^*$ & 3823.4 & \cite{kuntz:rawlings:2022} & N/A & N/A & N/A & N/A \\
Augmented ARX & 0.03 & N/A$^*$ & 3807.3 & see text & N/A & N/A & N/A & N/A \\
Unregularized ML & 125.48 & 500$^{{**}}$ & -9430.9 & Algo.~\ref{algo:main} & 0 & $\cplx$ & 10$^{-6}$ & N/A \\
Regularized ML 1 & 126.05 & 500$^{{**}}$ & -9431.7 & Algo.~\ref{algo:main} & 0.002 & $\cplx$ & 10$^{-6}$ & N/A \\
Regularized ML 2 & 9.23 & 21 & -9416.6 & Algo.~\ref{algo:main} & 0.005 & $\cplx$ & 10$^{-6}$ & N/A \\
Constrained ML 1 & 74.88 & 97 & -9347.2 & Algo.~\ref{algo:main} & 0 & $\mathcal{D}_1(0.3)\cap\mathcal{D}_2(0.998,0)$ & 10$^{-6}$ & 0.03 \\
Constrained ML 2 & 51.17 & 62 & -9358.2 & Algo.~\ref{algo:main} & 0 & $\mathcal{D}_1(0.3)\cap\mathcal{D}_2(0.999,0)$ & 10$^{-6}$ & 0.03 \\
Reg. \& Cons. ML & 37.07 & 40 & -9338.4 & Algo.~\ref{algo:main} & 0.001 & $\mathcal{D}_1(0.3)\cap\mathcal{D}_2(0.998,0)$ & 10$^{-6}$ & 0.03 \\
\hline
\end{tabular}

\end{table*}

\begin{figure*}
  \centering
  \includegraphics[page=1,width=\linewidth]{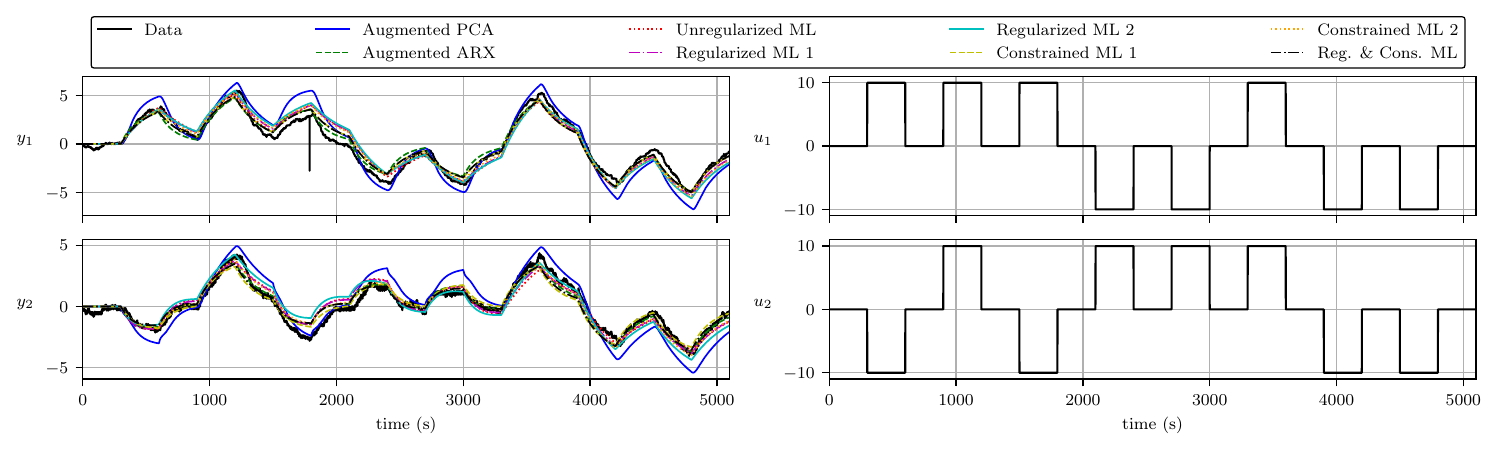}
  \caption{TCLab identification data and noise-free responses \(\hat y_k =
    \sum_{j=1}^k \hat C\hat A^{j-1}\hat Bu_{k-j}\) of a few selected models.}
  \label{fig:tclab:yfit}
\end{figure*}

Unless otherwise specified, the TCLab is modeled as a two-state, two-disturbance
system of the form \cref{eq:ladm}, with internal temperatures as plant states
\(s=\begin{bmatrix} T_1 & T_2 \end{bmatrix}^\top\), heater voltages as inputs
\(u=\begin{bmatrix} V_1 & V_2 \end{bmatrix}^\top\), and measured temperatures
\(y=\begin{bmatrix} T_{m,1} & T_{m_2}\end{bmatrix}^\top\) as outputs.
Throughout, we choose \(n_d=p\) to satisfy the offset-free necessary conditions
in~\cite{muske:badgwell:2002,pannocchia:rawlings:2003}, and we consider output
disturbance models \((B_d,C_d)=(0_{2\times 2},I_2)\). We use \((A_s,B_s)\) fully
parameterized and \(C=I_2\) to guarantee model identifiability and make the
states interpretable as internal temperatures. For the remaining model terms, we
have \((K_x,K_d,R_e)\) fully parameterized and
\((D,\hat{s}_0,\hat{d}_0)=(0,0,0)\).

Eight TCLab models are presented.
\begin{enumerate}[leftmargin=2.5em]
\item \textbf{Augmented PCA}: the 6-state TCLab model used in
  \cite{kuntz:rawlings:2022}, where principle component analysis on a
  \(400\times 5100\) data Hankel matrix is used to determine the states in the
  disturbance-free model.
\item \textbf{Augmented ARX}: a VARX\((1,1)\) model, equivalent to a stochastic
  LTI model with process noise but zero measurement noise.
\item[3--5)] \textbf{Unregularized ML}, \textbf{Regularized ML 1} and
  \textbf{2}: classic ML and MAP models.
\item[6--8)] \textbf{Constrained ML 1} and \textbf{2},
  \textbf{Reg.~\&~Cons.~ML}: eigenvalue-constrained ML and MAP models. LMI
  region constraints enforce filter stability and impose a lower bound on the
  real part of the filter eigenvalues.
\end{enumerate}
Each ML model uses Augmented ARX as the initial guess. See \Cref{table:tclab}
for specifics of the formulation of models 3--8.

\begin{figure*}
  \centering
  \includegraphics[page=4,width=\linewidth]{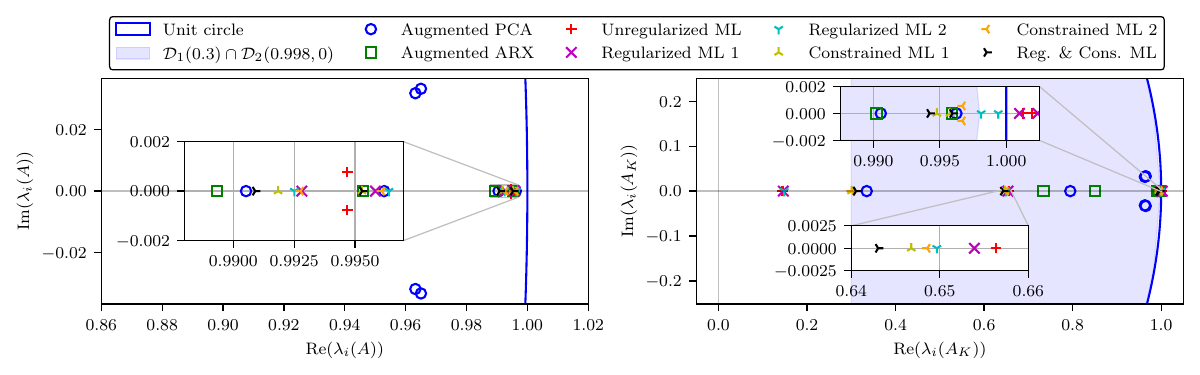}
  \caption{TCLab models open-loop and closed-loop (filter) eigenvalues.}
  \label{fig:tclab:eigs}
\end{figure*}

In \Cref{fig:tclab:yfit}, the identification data is presented along with the
noise-free responses \(\hat y_k = \sum_{j=1}^k \hat C\hat A^{j-1}\hat Bu_{k-j}\)
of a few selected models. Note that the noise-free responses are a fit to
\emph{training} data and do not reflect \emph{testing} performance. Computation
times, numbers of IPOPT iterations, and \emph{unregularized} log-likelihood
\(L_N(\hat{\theta})\) values are reported in \Cref{table:tclab}. The open-loop
\(A\) and closed-loop \(A_K \coloneqq A-KC\) eigenvalues of each model are
plotted in \Cref{fig:tclab:eigs}.

Except for the augmented PCA model, all of the open-loop eigenvalues cluster
around the same region of the complex plane (\cref{fig:tclab:eigs}). The
closed-loop filter eigenvalues are also placed similarly, although the classic
ML models (Unregularized ML, Regularized ML 1 and 2) suffer from slow or even
unstable filter eigenvalues. The models with unstable eigenvalues fail to
converge (see \Cref{table:tclab}) as the unstable filter modes make the problem
extremely sensitive to changes in the parameter values. While a high \(\rho\) is
sufficient to achieve filter stability, there is no clear minimum value of
\(\rho\) to achieve this. On the other hand, the constrained ML models have
stable filter eigenvalues \emph{without regularization}, and have well-defined
estimator performance guarantees based on the applied constraints. We remark the
relative performance is unchanged by removal of the bad sensor value (around
\(t=1800\) s in \Cref{fig:tclab:yfit}).

\begin{figure*}
  \centering
  \includegraphics[page=1,width=\linewidth]{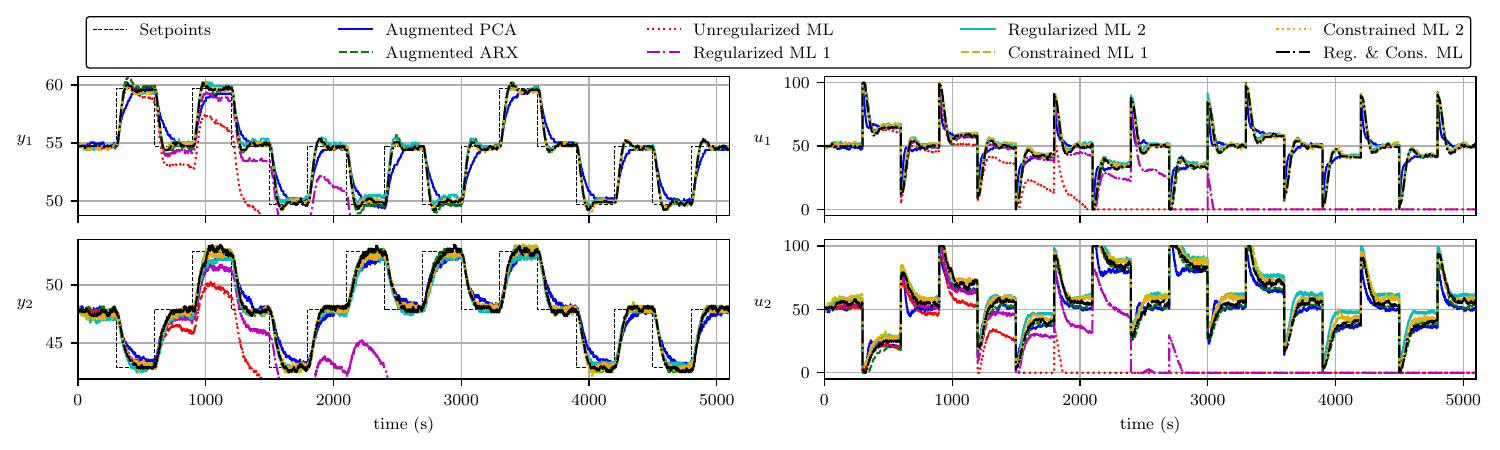}
  \caption{TCLab setpoint tracking tests.}
  \label{fig:tclab:setpoints}
\end{figure*}

\begin{figure*}
  \centering
  \includegraphics[page=1,width=\linewidth]{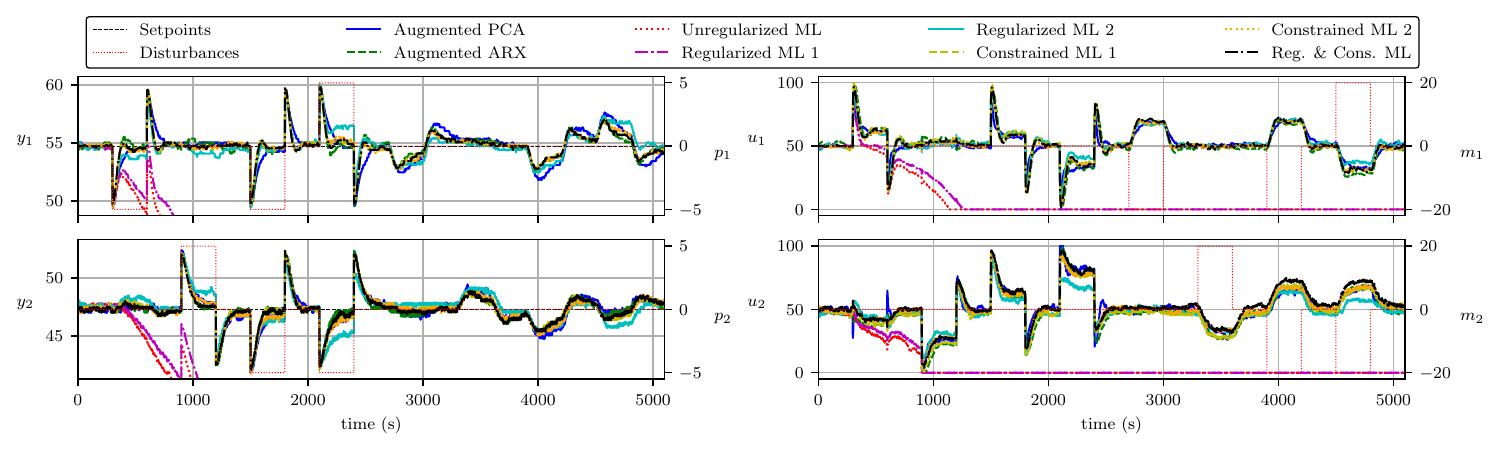}
  \caption{TCLab disturbance rejection tests.}
  \label{fig:tclab:disturbances}
\end{figure*}

\begin{figure}
  \centering
  \includegraphics[page=10,width=\linewidth]{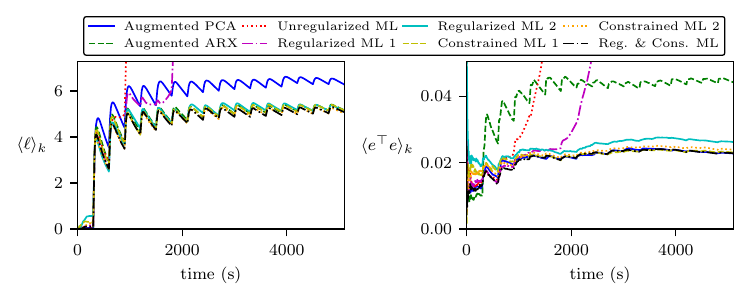}
  \caption{TCLab setpoint tracking test performance.}
  \label{fig:tclab:setpoints:performance}
\end{figure}

\begin{figure}
  \centering
  \includegraphics[page=10,width=\linewidth]{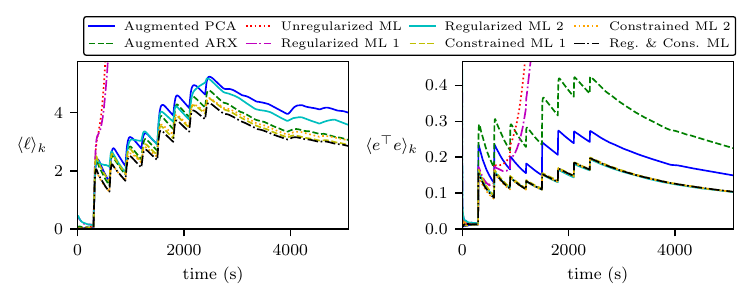}
  \caption{TCLab disturbance rejection test performance.}
  \label{fig:tclab:disturbances:performance}
\end{figure}

To test offset-free control performance, we performed two sets of closed-loop
experiments on offset-free MPCs designed with the models. In
\Cref{fig:tclab:setpoints}, identical setpoint changes were applied to a TCLab
running at a steady-state power output of 50\%. The setpoint changes were
tracked with the offset-free MPC design described in~\cite{kuntz:rawlings:2022}.
In \Cref{fig:tclab:disturbances}, step disturbances in the output \(p_i\) and
the input \(m_i\) are injected into a plant trying to maintain a given
steady-state temperature. The setpoints are tracked with the offset-free MPC
design described in~\cite{kuntz:rawlings:2022}.

Control performance is quantified by the squared distance from the setpoint
\(\ell_k \coloneqq |y_k-y_{\textnormal{sp},k}|^2\). Estimation performance is
quantified by the squared filter errors \(e_k^\top e_k\). For any signal
\(a_k\), we define a \(T\)-sample moving average by \(\langle a_k \rangle_T
\coloneqq T^{-1}\sum_{j=0}^{T-1} a_{k-j}\). Setpoint tracking performance is
reported in \Cref{fig:tclab:setpoints:performance}, and disturbance rejection
performance is reported in \Cref{fig:tclab:disturbances:performance}. The worst
performing models are those with unstable filters (Unregularized ML and
Regularized ML 1). These models shut off over the course of the experiment as
the integrating disturbance estimates grow unbounded. The remaining classic ML
model (Regularized ML 2) has slow filter eigenvalues that contribute to poor
control performance on the disturbance rejection test
(see \Cref{fig:tclab:disturbances:performance}, left). The augmented models
(Augmented PCA/ARX) perform poorly in either control or estimation aspect on
both tests. The best performance is achieved by the remaining ML models, which
all perform approximately the same across the tests.

\begin{figure}
  \centering
  \includegraphics[page=11,width=0.49\linewidth]{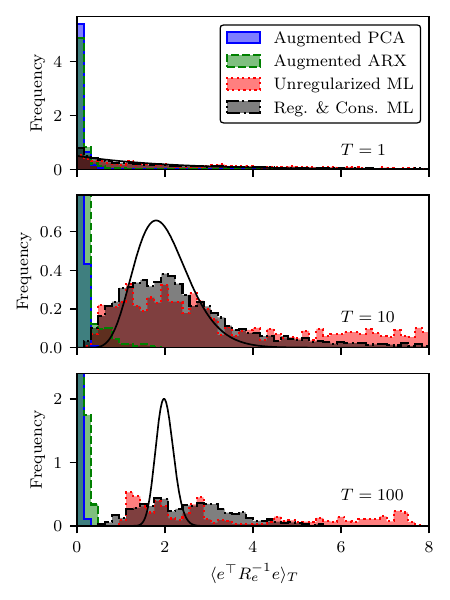}
  \includegraphics[page=11,width=0.49\linewidth]{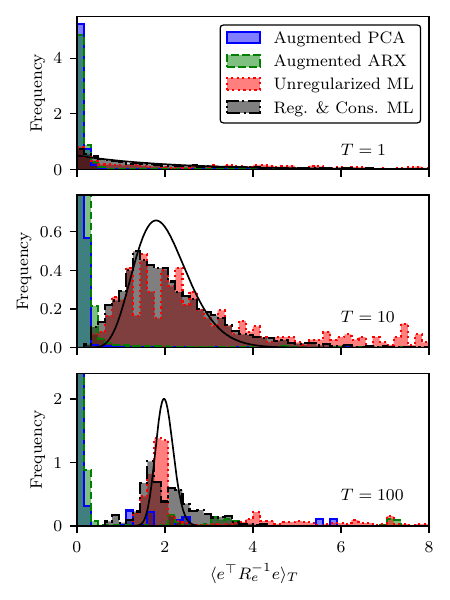}
  \caption{TCLab identification index data for (left) setpoint tracking and
    (right) disturbance rejection tests.}
  \label{fig:tclab:filterindex}
\end{figure}

To investigate the \emph{distributional} accuracy of the models, we consider the
identification index \(q\coloneqq e^\top R_e^{-1}e\). Recall the signal
\(e_k\in\real^2\) is an i.i.d., zero-mean Gaussian process, i.e.,
\(e_k\iid\norm(0,R_e)\), and therefore the index \(q_k\) is i.i.d.~with a
\(\chi^2_2\) distribution. Moreover, the moving average \(\langle q_k
\rangle_T\) is distributed as \(\chi^2_{2T}/T\), although it is no longer
independent in time. In \Cref{fig:tclab:filterindex}, histograms of \(\langle
q\rangle_T,T\in\set{1,10,100}\) are plotted against their expected distribution
for a few selected models (Augmented PCA/ARX, Unregularized ML, and
Reg.~\&~Cons.~ML). The extreme discrepancies between the augmented models'
performance index \(\langle q \rangle_T\) and the reference distribution
\(\chi^2_{2T}/T\) are primarily due to the augmented models significantly
overestimating \(R_e\) compared to the ML models,
{\allowdisplaybreaks
  \begin{align*}
    \hat R_e^{\textnormal{Aug.~PCA}}
    &= \begin{bmatrix} 0.5871 & 0.3365 \\ 0.3365 & 0.2878 \end{bmatrix}, \\
    \hat R_e^{\textnormal{Aug.~ARX}}
    &= \begin{bmatrix} 0.5084 & 0.2198 \\ 0.2198 & 0.2980 \end{bmatrix}, \\
    \hat R_e^{\textnormal{Unreg.~ML}}
    &= \begin{bmatrix} 0.0106 & 0.0007 \\ 0.0007 & 0.008 \end{bmatrix}, \\
    \hat R_e^{\textnormal{Reg.~Cons.~ML}}
    &= \begin{bmatrix} 0.0107 & 0.0007 \\ 0.0007 &  0.008  \end{bmatrix}.
  \end{align*}
}
The reference distribution and the ML models' \(\langle q \rangle_T\)
distribution diverge at large \(T\) since, due to plant-model mismatch, the
filter's innovation errors are autocorrelated. Filter instability of the
unregularized ML model causes frequent right-tail errors.

\subsection{Eastman reactor}\label{ssec:reactor}
\ifthenelse{\boolean{LongVersion}}{%
  %% the schematic is a large figure, only for report version
  \begin{figure*}
    \centering
    \ifthenelse{\boolean{OneColumn}}{%
      \footnotesize
      \def\svgwidth{\linewidth}
    }{
      \def\svgwidth{0.9\linewidth}
    }
    \input{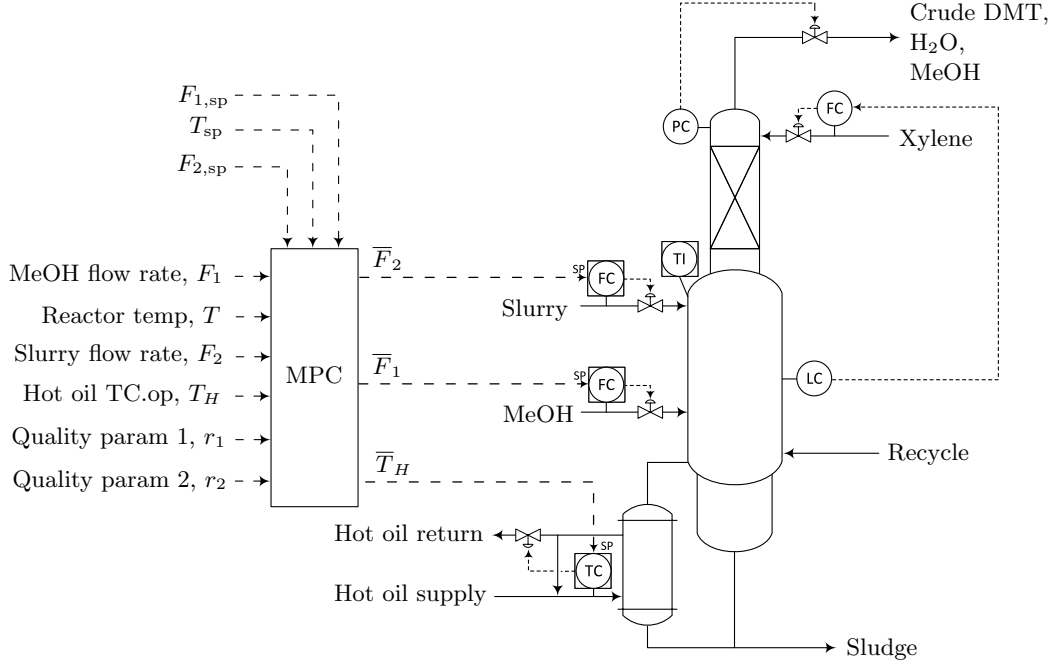}
    \caption{Schematic of the DMT reactor and MPC control
      strategy.}\label{fig:reactor}
  \end{figure*}
}{}

\begin{table*}
  \renewcommand{\arraystretch}{1.3}
  \ifthenelse{\boolean{OneColumn}}{\vspace{0.5em}}{}
  \caption{Eastman reactor model fitting results. The data is a reduced version
    of the closed-loop data from~\cite{kuntz:downs:miller:rawlings:2023a}.
    $^*$~The augmented identification methods are not iterative. $^{**}$~The
    maximum number of iterations was set at 500.}
  \label{table:reactor}
  \centering
  \ifthenelse{\boolean{OneColumn}\AND\boolean{LongVersion}}{\tiny}{}
  
\begin{tabular}{c||c|c|c|c|c|c|c|c}
\hline
\multicolumn{1}{c||}{\multirow{2}{*}{\bfseries Model}}
& \multicolumn{3}{c|}{\bfseries Results}
& \multicolumn{5}{c}{\bfseries Configuration} \\
\cline{2-9}
& \bfseries Time (s) & \bfseries Iterations & $L_N(\hat{\theta})$
& \bfseries{Method} & $\rho$ & $\mathcal{D}$
& $\varepsilon$ & $\varepsilon_i$ \\
\hline\hline
Augmented CCA & 0.06 & N/A$^*$ & -11399.3 & \cite{kuntz:downs:miller:rawlings:2023a} & N/A & N/A & N/A & N/A \\
Unregularized ML & 5.47 & 19 & -14383.1 & Algo.~\ref{algo:main} & 0 & $\cplx$ & 10$^{-6}$ & N/A \\
Regularized ML 1 & 5.28 & 17 & -14362.5 & Algo.~\ref{algo:main} & 0.01 & $\cplx$ & 10$^{-6}$ & N/A \\
Regularized ML 2 & 5.61 & 20 & -14346.7 & Algo.~\ref{algo:main} & 0.10 & $\cplx$ & 10$^{-6}$ & N/A \\
Regularized ML 3 & 4.80 & 13 & -14108.0 & Algo.~\ref{algo:main} & 1.00 & $\cplx$ & 10$^{-6}$ & N/A \\
Constrained ML 1 & 20.03 & 92 & -13944.9 & Algo.~\ref{algo:main} & 0 & $\mathcal{D}_1(0.3)$ & 10$^{-6}$ & 0.01 \\
Constrained ML 2 & 16.60 & 73 & -13941.1 & Algo.~\ref{algo:main} & 0 & $\mathcal{D}_1(0.3)$ & 10$^{-6}$ & 0.02 \\
Constrained ML 3 & 13.99 & 58 & -13928.5 & Algo.~\ref{algo:main} & 0 & $\mathcal{D}_1(0.3)$ & 10$^{-6}$ & 0.04 \\
\hline
\end{tabular}

\end{table*}

\begin{figure*}
  \includegraphics[page=1,width=\linewidth]{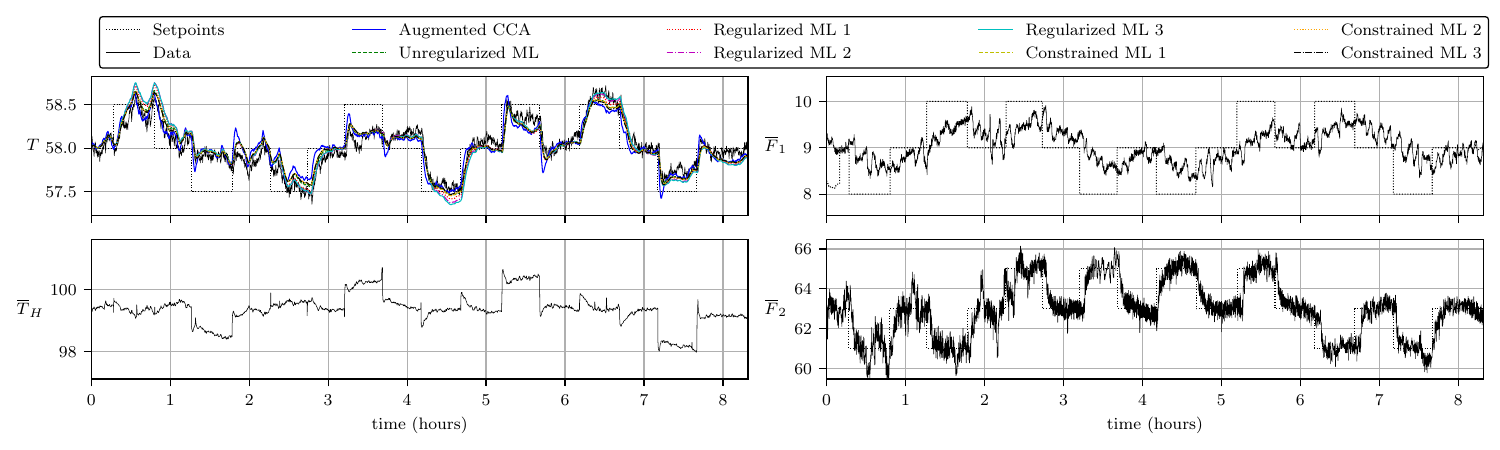}
  \caption{Training data and noise-free responses for the Eastman reactor models
    (Augmented HK and ML models using Augmented HK as the initial
    guess).}\label{fig:reactor:train}
\end{figure*}

\begin{figure*}
  \centering
  \includegraphics[page=4,width=\linewidth]{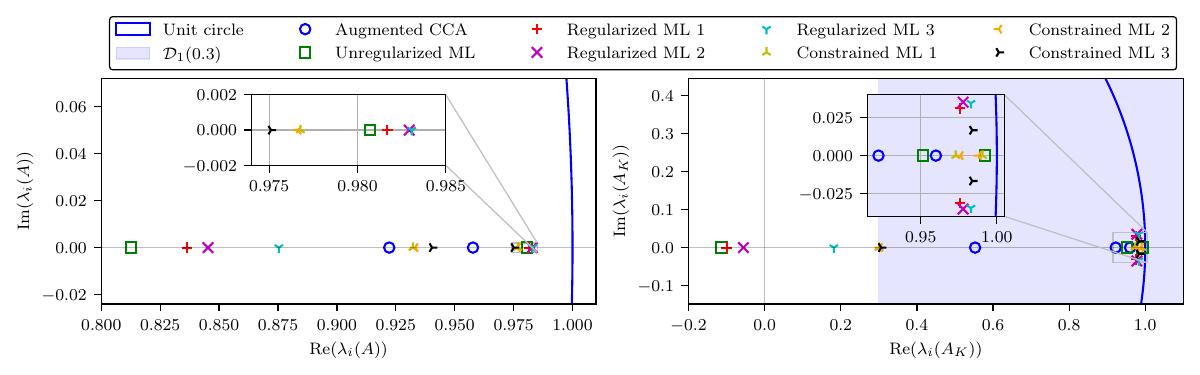}
  \caption{Eastman reactor models open-loop and closed-loop (filter) eigenvalues.}
  \label{fig:reactor:eigs}
\end{figure*}

\begin{figure}
  \centering
  \includegraphics[page=9,width=\linewidth]{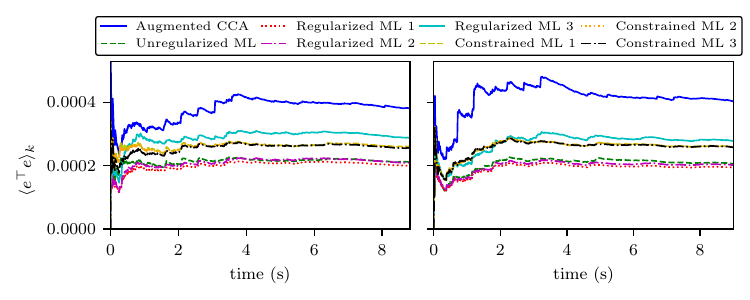}
  \caption{Test performance for the Eastman reactor models on the test data
    sets from~\cite{kuntz:downs:miller:rawlings:2023a}.}\label{fig:reactor:test}
\end{figure}

\begin{figure}
  \centering
  \includegraphics[page=6,width=\linewidth]{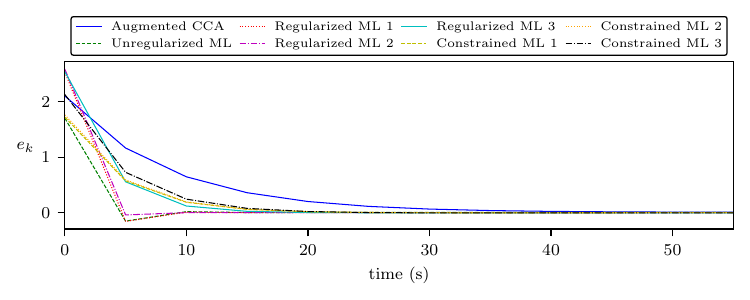}
  \caption{Eastman reactor models' closed-loop (filter) response to the
    eigenvector corresponding to the fastest eigenvalue.}
  \label{fig:reactor:errs}
\end{figure}

\ifthenelse{\boolean{LongVersion}}{%
%% the schematic is a large figure, only for report version
  A schematic of the chemical reactor considered in the next case study is
  presented in \Cref{fig:reactor}.%
}{}%
The control objective of the chemical reactor is to track three setpoints (the
output, a specified reactor temperature \(y=T\), and the flowrates
\(\begin{bmatrix} u_1 & u_3 \end{bmatrix}^\top = \begin{bmatrix} F_1 &
  F_2 \end{bmatrix}^\top\)), without offset, by controlling the three inputs
(the reactant flow rates and utility temperatures \(u=\begin{bmatrix} F_1 & T_H
  & F_2 \end{bmatrix}^\top\)).\footnote{The flowrates are both manipulated
  variables and controlled variables. At steady-state, we should reach the
  setpoints in \(y=T\) and \(\begin{bmatrix} u_1 & u_3 \end{bmatrix}^\top
  = \begin{bmatrix} F_1 & F_2 \end{bmatrix}^\top\), but \(u_2=T_H\) will not
  reach a predefined setpoint.} See \cite{kuntz:downs:miller:rawlings:2023a} for
more details about the reactor operation. As in \Cref{ssec:tclab}, we choose
\(n_d=p\) and consider output disturbance models \((B_d,C_d)=(0_{2\times
  1},1)\). This time, we use an observability canonical form~\cite{denham:1974}
with \(A_s=\begin{bsmallmatrix} 0 & 1 \\ a_1 & a_2 \end{bsmallmatrix}\) and
\(C_s=\begin{bmatrix} 1 & 0 \end{bmatrix}\). For the remaining model terms, we
have \((B_s,K_x,K_d,R_e)\) fully parameterized and
\((D,\hat{s}_0,\hat{d}_0)=(0,0,0)\).

Eight reactor models were fit to closed-loop data:
\begin{enumerate}[leftmargin=2.5em]
\item \textbf{Augmented CCA}: a CCA model~\cite{larimore:1990} augmented with a
  disturbance model, as detailed in
  \cite{kuntz:downs:miller:rawlings:2023a}.\footnote{This is not the same model
    used in~\cite{kuntz:downs:miller:rawlings:2023a}, as a different
    input-output model is considered, although the same data is used.
    Specifically, Kuntz et al.~\cite{kuntz:downs:miller:rawlings:2023a}
    considered a model with both regulatory-layer setpoints and measured
    values.}
\item[2--5)] \textbf{Unregularized ML}, \textbf{Regularized ML 1} to \textbf{3}:
  classic ML and MAP models.
\item[6--8)] \textbf{Constrained ML 1} to \textbf{3}:
  eigenvalue-constrained ML and MAP models. LMI region constraints impose a
  lower bound on the real part of the filter eigenvalues.
\end{enumerate}
Each ML model uses the augmented CCA model as the initial guess. See
\Cref{table:reactor} for specifics of the formulation of models 2--8. In
\Cref{fig:reactor:train}, the closed-loop identification data and noise-free
responses are presented. Again, these responses do not reflect \emph{testing}
performance. Computational details, the \emph{unregularized} log-likelihood
value, and model configurations are reported in \Cref{table:reactor}. The
open-loop \(A_s\) and closed-loop \(A_K\) eigenvalues are plotted in
\Cref{fig:reactor:eigs}.

The main difference between eigenvalues of the unconstrained ML models
(Unregularized ML and Regularized ML 1--3) and the constrained ML models
(Constrained ML 1--3) are faster open-loop eigenvalues and closed-loop
eigenvalues with possibly negative real part (see \Cref{fig:reactor:eigs}). For
the constrained ML models, the real part of this fast filter eigenvalue is
bounded from below using the LMI region constraint \(\mathcal{D}_1(0.3)\). As in
the TCLab case study, high \(\rho\) is sufficient to avoid negative eigenvalues,
but there is no clear cutoff.

In \Cref{fig:reactor:test}, the estimation performance for these filters is
compared on two test data sets (from~\cite{kuntz:downs:miller:rawlings:2023a}).
While the unconstrained models appear to have the best test performance, it is
at a cost of undesirable estimate dynamics. In~\Cref{fig:reactor:errs}, we plot
the filter response to an initial guess equal to the eigenvector corresponding
to the smallest eigenvalue of \(A_K\). Those filters with eigenvalues having
negative real parts exhibit overshoot in the estimate. The best performing
filters \emph{without} this behavior are the constrained ML models.

\ifthenelse{\boolean{LongVersion}}{%
  \begin{figure*}
    \centering
    \includegraphics[page=5,width=\linewidth]{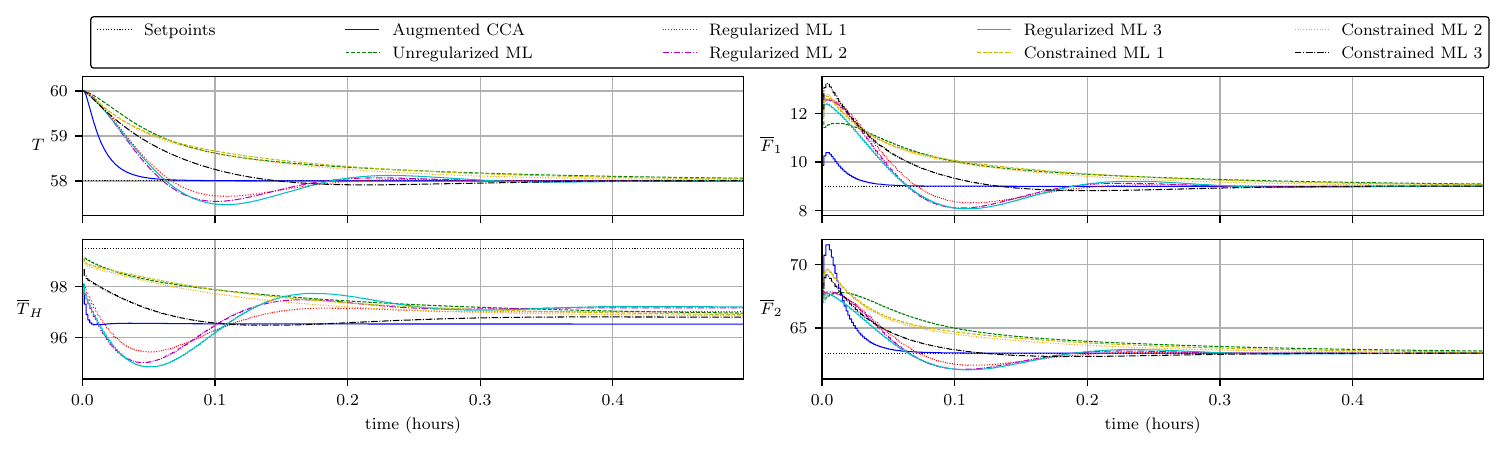}
    \caption{Eastman reactor models simulated closed-loop test performance.}
    \label{fig:reactor:sim}
  \end{figure*}

  Control performance could not be compared on the real plant due to cost and
  safety considerations. However, the closed-loop responses can be compared in
  simulation. In \Cref{fig:reactor:sim},~we plot simulated responses to a setpoint
  change. Each simulation considers the nominal closed-loop response (i.e., plant
  as the model, no noise) using the offset-free MPC design in
  \cite{kuntz:downs:miller:rawlings:2023a} with \(Q_s=1\) and
  \(R_s=\diag(0.01,1,0.01)\). The regularized ML models exhibit significant
  overshoot in the response, whereas the unregularized ML model and constrained ML
  models do not. %
}{}

\subsection{Discussion}
The main limitation of eigenvalued-constrained ML is computational cost. While
constrained ML retains linear scaling in sample size \(N\),\footnote{It is
  well-known that computation of the log-likelihood \(L_N(\theta)\) and its
  derivatives scale linearly in \(N\)~\cite{astrom:1979}. Moreover, the
  constraint function computation scales independently of \(N\).} each LMI
region constraint on an arbitrary system matrix \(\tilde A\in\realm{\tilde
  n}{\tilde n}\) requires an additional \(O(\tilde n^2(m^2+1))\) variables and
\(O(\tilde n^2m^2)\) equality constraints. These requirements can be
significantly reduced for spectral abscissa bounds \(\mathcal{D}_1(s)\) and
stability constraints \(\mathcal{D}_2(s,0)\). As mentioned in~\cref{rem:diehl},
these constraints are quite similar to the ``smooth'' spectral radii and
abscissa constraints of \cite{diehl:mombaur:noll:2009,%
  vanbiervliet:vandereycken:michiels:vandewalle:diehl:2009}, which only add
\(O(\tilde n^2)\) variables and \(O(\tilde n^2)\) equality constraints. For
eigenvalues constrained to the LMI regions \(\mathcal{D}_1(s)\) or
\(\mathcal{D}_2(s,x_0)\), implementing these constraints as a special case can
reduce the computational cost significantly.

For a standard, black-box LADM~\cref{eq:ladm} with \(n_d=p\), a canonical form
for \((A_s,B_s,C_s)\), and \((D,\hat{s}_0,\hat{d}_0)=(0,0,0)\), there are
\(O(n_s(p+m)+p^2)\) variables before constraints are added, and \(O(n_s^2)\)
variables after. Thus, fitting black-box models of large-scale systems is
computationally prohibitive. However, as discussed in \Cref{sec:ml}, networked
systems may be represented by significantly fewer variables:
\(O(N_un_u(p_u+m_u)+N_up_u^2)\) without constraints, or \(O(N_un_u^2)\) with
constraints, where \(N_u\) is the number of units or nodes, and \(n_u,m_u,p_u\)
are the number of states, inputs, and outputs per unit or node.

\section{Conclusion}\label{sec:conclusion}
%% NEW CONCLUSION
We propose an algorithm for identifying offset-free MPC-relevant models with ML
identification. The algorithm is validated on real-world data in two case
studies: a low-cost benchmark temperature controller, and an industrial-scale
reactor. Our method maintains LADM filter stability, avoids fitting artificial
high-frequency dynamics, and outperforms standard ML identification in both
control and estimation performance.

%% availability of code
The code, including sample TCLab datasets and scripts for model fitting, is made
available
at~\href{https://github.com/rawlings-group/mlid_2024}{\url{github.com/rawlings-group/mlid_2024}}.
% This repository also includes sample TCLab datasets and scripts for model
% fitting.
The Eastman reactor data is proprietary and is not made available. We will
develop this code further for reliable applications on large-scale systems.

%% Future directions
We conclude with some suggestions of future research. Since ML identified models
are more distributionally accurate, they are more suitable to the performance
monitoring technique of~\cite{zagrobelny:ji:rawlings:2013}. Integrated
identification and offset-free controller validation may be possible by
combining this method with ours. Recall that there are limitations to the
performance of the Kalman filter for LADMs, as shown by Bageshwar and
Borrelli~\cite{bageshwar:borrelli:2009} and in the case studies. Filter designs
with eigenvalue constraints may overcome these limitations and deliver superior
offset-free MPC performance. Finally, we have found ML identification sometimes
produces poor open-loop gain estimates. Using prediction error minimization with
a least squares objective (as a regularizer or in a multiobjective approach) may
alleviate these issues.

\appendix
\ifthenelse{\boolean{LongVersion}}{%
\section{Proof of \Cref{prop:dare}}\label[appendix]{app:dare}
Silverman~\cite{silverman:1976} contains a more complete characterization of the
DARE solutions for regulation problems with cross terms. However, this admits
additional nullspace terms into the gain matrix which the Kalman filtering
problem does not allow. We avoid nullspace terms through the assumption
\(R_v\succ 0\) and therefore streamline the proof of \Cref{prop:dare}.

For the following definitions and lemmas, we consider the matrices
\(\mathcal{W}\coloneqq (A,B,C,D)\) corresponding to a noise-free system.
\begin{definition}
  The system \(\mathcal{W}\) is \emph{left invertible on
    \(\intinterval{0}{k-1}\)} if
  \[
    0 =
    \begin{bmatrix} D \\ CB & D \\
      \vdots & \ddots & \ddots \\ CA^{k-2}B & \hdots
             & CB & D
    \end{bmatrix}
    \begin{bmatrix} u_0 \\ \vdots \\ u_{k-1} \end{bmatrix}
  \]
  % \(x_0=0\) and \(y_i=0\) for each \(i\in\intinterval{0}{k-1}\)
  implies \(u_0=0\). The system \(\mathcal{W}\) is \emph{left invertible} if
  there is some \(j\in\posint\) such that \(\mathcal{W}\) is left invertible on
  \(\intinterval{0}{k-1}\) for all \(k\geq j\).
\end{definition}
\begin{definition}
  The system \(\mathcal{W}\) is \emph{strongly detectable} if \(y_k\rightarrow
  0\) implies \(x_k\rightarrow 0\).
\end{definition}
% Next we state the dual versions of these definitions:
% \begin{definition}
%   A system \((A,B,C,D)\) is \emph{right invertible} (on
%   \(\intinterval{0}{k-1}\)) if \((A^\top,C^\top,B^\top,D^\top)\) is left
%   invertible (on \(\intinterval{0}{k-1}\)).
% \end{definition}
% \begin{definition}
%   A system \((A,B,C,D)\) is \emph{strongly stabilizable} if
%   \((A^\top,C^\top,B^\top,D^\top)\) is strongly detectable.
% \end{definition}
The following lemmas are taken directly from
\cite[Thms.~8,~18(iii)]{silverman:1976}, but the proofs are omitted for the sake
of brevity.
\begin{lemma}[\protect{\cite[Thm.~8]{silverman:1976}}]\label{lem:strong:det}
  If \(\mathcal{W}\) is left invertible, then \(\mathcal{W}\) is strongly
  detectable if and only if \((A-BF,C-DF)\) is detectable for all \(F\) of
  appropriate dimension.
\end{lemma}
\begin{lemma}[\protect{\cite[Thm.~18(iii)]{silverman:1976}}]\label{lem:dare:reg}
  If \(\mathcal{W}\) is left invertible, then the DARE
  \ifthenelse{\boolean{OneColumn}}{%
    \begin{equation*}
      P = A^\top PA - (A^\top PB+C^\top D)(B^\top PB+D^\top D)^{-1}(B^\top PA+D^\top C)
    \end{equation*}
  }{%
    \begin{multline*}
      P = A^\top PA - (A^\top PB+C^\top D)(B^\top PB+D^\top D)^{-1}\\
      \times (B^\top PA+D^\top C)
    \end{multline*}
  }%
  has a unique, stabilizing solution\footnote{Contrary to in \Cref{sec:ml}, here
    we mean the solution \(P\) is stabilizing when \(A-BK(P)\) is stable, where
    \(K(P)\coloneqq (B^\top PB+D^\top D)^{-1}B^\top P\).} if and only if
  \(\mathcal{W}\) is stabilizable and semistrongly detectable.
\end{lemma}

For the remainder of this section, we consider the full rank factorization
\[
  \begin{bmatrix} Q_w & S_{wv} \\ S_{wv}^\top & R_v \end{bmatrix}
  = \begin{bmatrix} \tilde B \\ \tilde D \end{bmatrix}
  \begin{bmatrix} \tilde B^\top & \tilde D^\top \end{bmatrix}
\]
and the dual system \(\tilde{\mathcal{W}} \coloneqq (A^\top,C^\top,\tilde
B^\top,\tilde D^\top)\) to analyze the properties of the original system
\cref{eq:slti}. The following lemma relates the properties \(R_v\succ 0\) and
left invertability of \(\tilde{\mathcal{W}}\).
\begin{lemma}\label{lem:left:inv}
  If \(R_v\succ 0\) then \(\tilde{\mathcal{W}}\) is left invertible.
\end{lemma}
\begin{proof}
  Left invertability on \(\intinterval{0}{k-1}\) is equivalent to
  \begin{equation}\label{eq:left:inv}
    0 =
    \begin{bmatrix} \tilde D^\top \\ \tilde B^\top C^\top & \tilde D^\top \\
      \vdots & \ddots & \ddots \\ \tilde B^\top (A^\top)^{k-2} C^\top &
                                                                        \hdots
             & \tilde B^\top C^\top & \tilde D^\top
    \end{bmatrix}
    \begin{bmatrix} u_0 \\ \vdots \\ u_{k-1} \end{bmatrix}
  \end{equation}
  implying \(u_0=0\). But \(R_v=\tilde D\tilde D^\top\succ 0\), so \(\tilde
  D^\top\) has a zero nullspace. For each \(k\in\posint\), the coefficient
  matrix of \cref{eq:left:inv} has a zero nullspace. Thus, \(u_0=0\) and
  \(\tilde{\mathcal{W}}\) is left invertible.
\end{proof}
% \begin{lemma}\label{lem:strong:det}
%   If \(R_v\succ 0\) then \(\tilde{\mathcal{W}}\) is strongly detectable if and
%   only if \((A-FC,\tilde B-F\tilde D)\) is stabilizable for all \(F\).
% \end{lemma}
% \begin{proof}
%   This follows straightforwardly from \Cref{lem:left:inv} and
%   \cite[Thm.~8]{silverman:1976}.
% \end{proof}
Finally, we can prove \Cref{prop:dare}.
\begin{proof}[Proof of \Cref{prop:dare}]
  By \Cref{lem:left:inv}, we have that \(\tilde{\mathcal{W}}\) is left
  invertible. Therefore, by \Cref{lem:dare:reg}, the DARE \cref{eq:dare} has a
  unique, stabilizing solution if and only if \(\tilde{\mathcal{W}}\) is
  stabilizable and strongly detectable. But by \Cref{lem:strong:det} and
  duality, the latter statement is true if and only if \((A,C)\) is detectable
  and \((A-FC,\tilde B-F\tilde D)\) is stabilizable for all
  \(F\in\realm{n}{p}\).
\end{proof}
}{}
\ifthenelse{\boolean{LongVersion}}{%
\section{Proof of \Cref{prop:lmi:closed}}\label[appendix]{app:lmi}
Throughout this appendix, we define the set of \(n\times n\) Hermitian,
Hermitian positive definite, and Hermitian positive semidefinite matrices as
\(\herm{n}\), \(\posdefhm{n}\), and \(\nnegdefhm{n}\). Notice that
\(f_{\mathcal{D}}\) maps to Hermitian matrices so we can write it as
\(f:\cplx\rightarrow\herm{m}\). We define the extension of \(M_{\mathcal{D}}\)
to complex arguments \(M_{\mathcal{D}} : \cplxm{n}{n}\times\nnegdefhm{n}
\rightarrow \herm{nm}\) as
\[
  M_{\mathcal{D}}(A,P) \coloneqq M_0\otimes P + M_1\otimes(AP) +
  M_1^\top\otimes(AP)^{\mathrm{H}}.
\]
To show \Cref{prop:lmi:closed}, we need a preliminary result about Hermitian
positive semidefinite matrices, generalized from Lemma A.1 in
\cite{chilali:gahinet:1996}.
\begin{lemma}\label{lem:posdefm}
  For any \(M\in\herm{n}\), if \(M\succeq 0\) (\(M\succ 0\)) then
  \(\textnormal{Re}(M)\succeq 0\) (\(\textnormal{Re}(M)\succ 0\)).
\end{lemma}
\begin{proof}
  With \(M=\textnormal{Re}(M) + \iota\textnormal{Im}(M)\), it is clear \(M\)
  Hermitian implies \(\textnormal{Re}(M)\) is symmetric and
  \(\textnormal{Im}(M)\) is skew-symmetric. Thus \(v^\top
  Mv=v^\top\textnormal{Re}(M)v\) for all \(v\in\real^n\), and positive
  (semi)definiteness of \(M\) implies positive (semi)definiteness of
  \(\textnormal{Re}(M)\).
\end{proof}
In proving \Cref{prop:lmi:closed}, we take the approach of
\cite{chilali:gahinet:1996} but are careful to distinguish eigenvalues on the
interior \(\mathcal{D}\) from those on the boundary \(\partial\mathcal{D}\).
\begin{proof}[Proof of \Cref{prop:lmi:closed}]
  (\(\Leftarrow\)) Suppose that \(M_{\mathcal{D}}(A,P)\succeq 0\) for some
  \(P\succ 0\) and let \(\lambda\in\lambda(A)\). Then there exists
  a nonzero \(v\in\cplx^n\) for which \(v^{\mathrm{H}}A=\lambda
  v^{\mathrm{H}}\). Consider the identity
  \begin{align*}
    \ifthenelse{\boolean{OneColumn}}{%
    (I_m\otimes v)^{\mathrm{H}} M_{\mathcal{D}}(A,P)(I_m\otimes v)
    }{%
    (I_m&\otimes v)^{\mathrm{H}} M_{\mathcal{D}}(A,P)(I_m\otimes v) \\
    }%
    % &= (I_m\otimes v^{\mathrm{H}})
    % (M_0\otimes P+M_1\otimes(AP)+M_1^\top\otimes(PA^\top))
    % (I_m\otimes v) \\
    &= M_0\otimes v^{\mathrm{H}}Pv + M_1\otimes(v^{\mathrm{H}}APv) +
      M_1^\top\otimes(v^{\mathrm{H}}PA^\top v) \\
          &= M_0\otimes v^{\mathrm{H}}Pv + M_1\otimes(\overline\lambda v^{\mathrm{H}}Pv) +
            M_1^\top\otimes(\lambda v^{\mathrm{H}}Pv) \\
          &= v^{\mathrm{H}}Pv(M_0 + M_1\lambda + M_1^\top\overline\lambda) \\
          &= v^{\mathrm{H}}Pv f_{\mathcal{D}}(\lambda).
    \end{align*}
    The assumption \(P\succ 0\) implies \(v^{\mathrm{H}}Pv>0\), and
    \(M_{\mathcal{D}}(A,P)\succeq 0\) further implies
    \(f_{\mathcal{D}}(\lambda)\succeq 0\). Therefore
    \(\lambda\in\textnormal{cl}(\mathcal{D})\).

    Next suppose \(\lambda\in\lambda(A)\) is non-simple and
    \(\lambda\in\partial\mathcal{D}\). Then there exists nonzero
    \(v_1,v_2\in\cplx^n\) (linearly independent) such that
    \(v^{\mathrm{H}}f_{\mathcal{D}}(\lambda)v=0\), \(v_1^{\mathrm{H}} A=\lambda
    v_1^{\mathrm{H}}\), and \(v_2^{\mathrm{H}}A=\lambda v_2^{\mathrm{H}}+v_1\).
    Because \(\mathcal{D}\) is open, \(\lambda \in \partial\mathcal{D} =
    \textnormal{cl}(\mathcal{D}) \setminus \mathcal{D}\) must satisfy both
    \(f_{\mathcal{D}}(\lambda)\succeq 0\) and \(f_{\mathcal{D}}(\lambda)\not\succ
    0\). Therefore \(f_{\mathcal{D}}(\lambda)\) is singular, and there exists a
    nonzero vector \(v\in\cplx^m\) such that
    \(v^{\mathrm{H}}f_{\mathcal{D}}(\lambda)v=0\). With the \(2\times 2\) matrices
    \begin{align*}
      \tilde P &= \begin{bmatrix} p_{11} & p_{12} \\ p_{12} & p_{22} \end{bmatrix}
      \coloneqq \begin{bmatrix} v_1^{\mathrm{H}} \\ v_2^{\mathrm{H}} \end{bmatrix}
                 P \begin{bmatrix} v_1 & v_2 \end{bmatrix} \succ 0 \\
      \tilde J
      &\coloneqq \lambda I_2 + \begin{bmatrix} 0 & 1 \\ 0 & 0 \end{bmatrix}
    \end{align*}
    we have \(\begin{bmatrix} v_1 & v_2 \end{bmatrix}^{\mathrm{H}}A = \tilde J
    \begin{bmatrix} v_1 & v_2 \end{bmatrix}^{\mathrm{H}}\) and therefore
    \begin{align*}
      (I_m\ifthenelse{\boolean{OneColumn}}{}{
      &} \otimes \begin{bmatrix} v_1 & v_2 \end{bmatrix})^{\mathrm{H}}M_{\mathcal{D}}(A,P)
        (I_m\otimes \begin{bmatrix} v_1 & v_2 \end{bmatrix})
        \ifthenelse{\boolean{OneColumn}}{}{\\}
      &= M_0\otimes \tilde P + M_1\otimes \tilde J\tilde P +
        M_1^\top\otimes(\overline{\tilde J}\tilde P)^\top \\
      &= M_{\mathcal{D}}(\tilde J,\tilde P) \succeq 0.
    \end{align*}
    Next, we have
    \begin{align*}
      \tilde M
      &\coloneqq K_{2,m}M_{\mathcal{D}}(\tilde J,\tilde P)K_{2,m}^\top \\
      &= \tilde P\otimes M_0 + \tilde J\tilde P \otimes M_1 +
        (\overline{\tilde J}\tilde P)^\top\otimes M_1^\top \\
      &= \tilde P\otimes f_{\mathcal{D}}(\lambda) +
        \begin{bmatrix} p_{12}(M_1+M_1^\top) & p_{22}M_1 \\ p_{22}M_1^\top & 0
        \end{bmatrix} \succeq 0.
    \end{align*}
    Finally,
    \ifthenelse{\boolean{OneColumn}}{
      \begin{equation*}
        (I_2\otimes v)^{\mathrm{H}} \tilde M (I_2\otimes v) =
        \begin{bmatrix} p_{12}v^{\mathrm{H}}(M_1+M_1^\top)v
          & p_{22}v^{\mathrm{H}}M_1v \\ p_{22}v^{\mathrm{H}}M_1^\top v & 0
        \end{bmatrix} \succeq 0.
      \end{equation*}
    }{
      \begin{multline*}
        (I_2\otimes v)^{\mathrm{H}} \tilde M (I_2\otimes v) \\ =
        \begin{bmatrix} p_{12}v^{\mathrm{H}}(M_1+M_1^\top)v
          & p_{22}v^{\mathrm{H}}M_1v \\ p_{22}v^{\mathrm{H}}M_1^\top v & 0
        \end{bmatrix} \succeq 0.
      \end{multline*}
    }
    But \(\tilde P\succ 0\) implies \(p_{22}>0\), so the above matrix inequality
    implies \(v^{\mathrm{H}}M_1v=0\). Moreover, with
    \(v^{\mathrm{H}}f_{\mathcal{D}}(\lambda)v=0\), we also have
    \(v^{\mathrm{H}}M_0v=0\) and therefore \(f(z)\equiv 0\) and \(\mathcal{D}\) is
    empty, a contradiction. Therefore each \(\lambda\in\lambda(A)\) non-simple
    implies \(\lambda\in\mathcal{D}\).

    (\(\Rightarrow\)) Suppose \(\lambda(A)\subset\textnormal{cl}(\mathcal{D})\)
    and \(\lambda\in\lambda(A)\) non-simple implies \(\lambda\in\mathcal{D}\).

    If \(A=\lambda\) is a (possibly complex) scalar, then it lies in
    \(\textnormal{cl}(\mathcal{D})\) by assumption, with
    \(M_{\mathcal{D}}(\lambda,p)=pf_{\mathcal{D}}(\lambda)\succeq 0\) for all
    \(p>0\).

    If \(A=\lambda I_n+N\) is a (possibly complex) Jordan block, where
    \(N\in\realm{n}{n}\) is a shift matrix and \(n>1\), then
    \(\lambda\in\mathcal{D}\) and \(f_{\mathcal{D}}(\lambda)\succ 0\). Let
    \(T_k\coloneqq\diag(k^{n-1},\ldots,k,1)\) for each \(k\in\posint\). Then
    \(T_k^{-1}AT_k=\lambda I_n + k^{-1}N\rightarrow\lambda I_n\) as
    \(k\rightarrow\infty\). Moreover, because \(M_{\mathcal{D}}\) is continuous, we have
    \[
      M_{\mathcal{D}}(T_k^{-1}AT_k, I_n) \rightarrow M_{\mathcal{D}}(\lambda
      I_n, I_n) = f_{\mathcal{D}}(\lambda) \otimes I_n \succ 0.
    \]
    Therefore there exists some \(k_0\in\posint\) such that
    \(M_{\mathcal{D}}(T_k^{-1}AT_k, I_n)\succ 0\) for all \(k\geq k_0\). With
    \(P\coloneqq T_kT_k^\top\), we have
    \begin{align*}
      M_{\mathcal{D}}(A,P)
      &= M_0\otimes T_kT_k^\top + M_1\otimes(AT_kT_k^\top)
        \ifthenelse{\boolean{OneColumn}}{}{\\ &\qquad}
        + M_1^\top\otimes(AT_kT_k^\top)^\top \\
      &= (I_m\otimes T_k)(M_0\otimes I_n + M_1\otimes T_k^{-1}AT_k
        \ifthenelse{\boolean{OneColumn}}{}{\\ &\qquad}
        + M_1^\top\otimes (T_k^{-1}AT_k)^\top)(I_m\otimes T_k)^\top \\
      &= (I_m\otimes T_k)M_{\mathcal{D}}(T_k^{-1}AT_k,I_n)(I_m\otimes T_k)^\top
        \succ 0.
    \end{align*}

    Finally, for any \(A\in\realm{n}{n}\), let \(A=V(\bigoplus_{i=1}^p J_i)V^{-1}\)
    denote the Jordan decomposition of \(A\), where \(J_i=\lambda_iI_{n_i}+N_i\),
    \(\lambda_i\in\lambda(A)\), \(N_i\) are shift matrices, and
    \(n=\sum_{i=1}^p n_i\). We have already shown that for each
    \(i\in\intinterval{1}{p}\), there exists \(P_i\succ 0\) such that
    \(M_{\mathcal{D}}(J_i,P_i)\succeq 0\). Then with \(\tilde P\coloneqq V(\bigoplus_{i=1}^p
    P_i)V^{-1}\), we have
    \begin{align*}
      (I_m&\otimes V^{-1})M_{\mathcal{D}}(A,\tilde P)(I_m\otimes V^{-1})^{\mathrm{H}} \\
          &= M_0\otimes\left(\bigoplus_{i=1}^p P_i\right) +
            M_1\otimes\left(\bigoplus_{i=1}^p J_iP_i\right)
            \ifthenelse{\boolean{OneColumn}}{}{\\ &\qquad}
            + M_1\otimes\left(\bigoplus_{i=1}^p J_iP_i\right)^\top \\
          % &= K_{n,m}\left( \bigoplus_{i=1}^p M_0\otimes P_i + M_1\otimes J_iP_i \right. \\
          % &\qquad \left. + M_1^\top\otimes (J_iP_i)^\top \right)K_{n,m}^\top \\
        &= K_{n,m}\left( \bigoplus_{i=1}^p K_{m,n_i}M_{\mathcal{D}}(J_i,P_i)K_{m,n_i}^\top
          \right)K_{n,m}^\top \succeq 0
    \end{align*}
    and therefore \(M_{\mathcal{D}}(A,\tilde P)\succeq 0\). Last,
    \Cref{lem:posdefm} gives \(M_{\mathcal{D}}(A,P)\succeq 0\) with
    \(P\coloneqq\textrm{Re}(\tilde P)\) since
    \[
      M_{\mathcal{D}}(A,P) = M_{\mathcal{D}}(A,\textrm{Re}(\tilde P)) =
      \textrm{Re}(M_{\mathcal{D}}(A,\tilde P)). \qedhere
    \]
  \end{proof}
}{}

\ifthenelse{\boolean{LongVersion}}{
\section{Proof of \Cref{prop:lmi:topo}}
}{
\section{Proof of \Cref{prop:lmi:topo}(\textnormal{a,d})}
}\label[appendix]{app:lmi:topo}
To show \Cref{prop:lmi:topo}(a), we first require the following eigenvalue
sensitivity result due to~\cite[Thm.~7.2.3]{golub:vanloan:2013}.
\begin{theorem}[\protect{\cite[Thm.~7.2.3]{golub:vanloan:2013}}]\label{thm:eig:cont}
  For any \(A\in\cplxm{n}{n}\), denote its Schur decomposition by
  \(A=Q(D+N)Q^{\mathrm{H}}\), where \(Q\in\cplxm{n}{n}\) is unitary,
  \(D\in\cplxm{n}{n}\) is diagonal, and \(N\in\cplxm{n}{n}\) is strictly upper
  triangular.\footnote{A matrix \(U\) is strictly upper triangular if
    \(U_{ij}=0\) for all \(i\geq j\).} Let \(p\) be the smallest positive
  integer for which \(M^p=0\) where \(M_{ij}\coloneqq|N_{ij}|\). Then, for any
  \(E\in\realm{n}{n}\) and \(\mu\in\lambda(A+E)\),
  \[
    \min_{\lambda\in\lambda(A)} |\mu-\lambda| \leq \max\set{ c\|E\|,
      (c\|E\|)^{1/p} }
  \]
  where \(c\coloneqq\sum_{k=0}^{p-1} \|N\|^k\).
\end{theorem}

\begin{proof}[Proof of \Cref{prop:lmi:topo}\ifthenelse{\boolean{LongVersion}}{}{(a,d)}]
  \ifthenelse{\boolean{LongVersion}}{%
    Throughout this proof, we show a set \(S\) is not open (or not closed) by
    demonstrating that \(S^c\) (or \(S\)) does not contain all its limit points.

  }{}%
  (a)---For any \(A\in\mathbb{A}_{\mathcal{D}}^n\), continuity of
  \(f_{\mathcal{D}}\) gives the existence of a function \(\delta(\lambda)>0\)
  such that \(f_{\mathcal{D}}(z)\succ 0\) for all
  \(|z-\lambda|<\delta(\lambda)\) and \(\lambda\in\lambda(A)\). Let
  \(\delta\coloneqq\min_{\lambda\in\lambda(A)} \delta(\lambda)\). By
  \Cref{thm:eig:cont} and norm equivalence, there exist \(c>0\) and
  \(p\in\intinterval{1}{n}\) such that
  \[
    \max_{\mu\in\lambda(A+E)} \min_{\lambda\in\lambda(A)} |\lambda-\mu| \leq
    \max\set{ c\|E\|_{\textnormal{F}}, (c\|E\|_{\textnormal{F}})^{1/p}}
  \] for all \(E\in\realm{n}{n}\). Therefore there exists a \(\varepsilon>0\)
  such that
  \[ \max_{\mu\in\lambda(A+E)} \min_{\lambda\in\lambda(A)} |\lambda-\mu| <
    \delta
  \]
  for all \(E\in\mathcal{B} \coloneqq \set{ E'\in\realm{n}{n} |
    \|E'\|_{\textnormal{F}} < \varepsilon }\). Finally, \(A+\mathcal{B}\) is a
  neighborhood of \(A\) contained in \(\mathbb{A}_{\mathcal{D}}^n\), and, since
  \(A\in\mathbb{A}_{\mathcal{D}}^n\) was chosen arbitrarily,
  \(\mathbb{A}_{\mathcal{D}}^n\) is open.

  \ifthenelse{\boolean{LongVersion}}{%
    (b)(i)---Because \(\mathcal{D}\) is open, nonempty, and not equal to
    \(\mathcal{D}\), \(\partial\mathcal{D}\) is nonempty. Let
    \(\lambda\in\partial\mathcal{D}\) and \(\lambda_k\in\mathcal{D}^c\) be a
    sequence for which \(\lambda_k\rightarrow\lambda\). By symmetry, we also have
    \(\overline\lambda\in\mathcal{D}\) and \(\overline\lambda_k\in\mathcal{D}^c\).

    For \(n=2\), we have \(A\coloneqq
    \begin{bsmallmatrix}
      \textnormal{Re}(\lambda) & -\textnormal{Im}(\lambda) \\
      \textnormal{Im}(\lambda) & \textnormal{Re}(\lambda) \end{bsmallmatrix} \in
    \realm{2}{2}\) has eigenvalues \(\lambda,\overline\lambda\in\mathcal{D}\), and
    \(A_k\coloneqq
    \begin{bsmallmatrix}
      \textnormal{Re}(\lambda_k) & -\textnormal{Im}(\lambda_k) \\
      \textnormal{Im}(\lambda_k) & \textnormal{Re}(\lambda_k)
    \end{bsmallmatrix} \in \realm{2}{2}\) has eigenvalues
    \(\lambda_k,\overline{\lambda_k}\in\mathcal{D}^c\) for each
    \(k\in\posint\). The corresponding eigenvectors are
    \(\begin{bsmallmatrix} \pm\iota \\ 1 \end{bsmallmatrix}\in\cplx^2\). Therefore
    \(A\in\tilde{\mathbb{A}}_{\mathcal{D}}^2\) but
    \(A_k\in(\tilde{\mathbb{A}}_{\mathcal{D}}^2)^c\) for each \(k\in\posint\),
    and the limit \(A_k\rightarrow A\) gives us that
    \((\tilde{\mathbb{A}}_{\mathcal{D}}^2)^c\) does not contain all its limit
    points.

    For \(n>2\), let \(A_0\in\tilde{\mathbb{A}}_{\mathcal{D}}^{n-2}\), and we can
    extend the prior argument with the sequence \(B_k\coloneqq A_k\oplus A_0\in
    (\tilde{\mathbb{A}}_{\mathcal{D}}^n)^c,k\in\posint\) that converges to \(B
    \coloneqq A \oplus A_0 \in \tilde{\mathbb{A}}_{\mathcal{D}}^n\).

    (b)(ii)---By part (b)(i), it suffices to consider the case \(n=1\). By closure
    and convexity of \(\mathcal{D}\), \(\mathcal{D}\cap\real\) is either a closed
    line segment, a closed ray, or \(\real\) itself. In other words,
    \(\mathcal{D}\cap\real\) is open if and only if it has no endpoints. Moreover,
    since \(\partial\mathcal{D}\cap\real\) is the set of the endpoints of
    \(\mathcal{D}\cap\real\), \(\mathcal{D}\cap\real\) is open if and only if
    \(\partial\mathcal{D}\cap\real\) is empty. Finally, since
    \(\tilde{\mathbb{A}}_{\mathcal{D}}^1 = \mathcal{D}\cap\real\),
    \(\tilde{\mathbb{A}}_{\mathcal{D}}^1\) is open if and only if
    \(\partial\mathcal{D}\cap\real\) is empty.

    (c)(i)---Let \(\lambda\in\partial\mathcal{D}\). Suppose \(n=4\). Then
    \(\overline{\lambda}\in\partial\mathcal{D}\) by symmetry. Because
    \(\mathcal{D}\) is open, there exists a sequence \(\lambda_k\in\mathcal{D}\)
    such that \(\lambda_k\rightarrow\lambda\), and by symmetry, we also have
    \(\overline\lambda_k\in\mathcal{D}\) and
    \(\overline\lambda_k\rightarrow\overline\lambda\). Consider again the
    \(2\times 2\) matrices \(A\) and \(A_k\) from part (b)(i), which have
    eigenvalues \(\lambda,\overline\lambda\in\mathcal{D}\) and
    \(\lambda_k,\overline{\lambda_k}\in\mathcal{D}^c\), respectively. Then the
    block matrices \(B\coloneqq
    \begin{bsmallmatrix} A & I_2 \\ 0 & A \end{bsmallmatrix}\in\realm{4}{4}\)
    and \(B_k\coloneqq\begin{bsmallmatrix} A_k & I_2 \\ 0 &
      A_k \end{bsmallmatrix}\in\realm{4}{4}\) have the same eigenvalues, but
    this time the eigenvectors are \(\begin{bsmallmatrix} \pm\iota \\ 1 \\ 0 \\
      0 \end{bsmallmatrix},\begin{bsmallmatrix} 0 \\ 0 \\ \pm\iota \\
      1 \end{bsmallmatrix}\in\cplx^4\) and the eigenvalues are non-simple. Since
    \(\lambda\) is a non-simple eigenvalue on the boundary of \(\mathcal{D}\),
    we have \(B\not\in\tilde{\mathbb{A}}_{\mathcal{D}}^4\). However,
    \(\lambda_k\) are all in the interior of \(\mathcal{D}\), so
    \(B_k\in\tilde{\mathbb{A}}_{\mathcal{D}}^4\). Since \(B_k\rightarrow B\),
    the set \(\tilde{\mathbb{A}}_{\mathcal{D}}^4\) does not contain all its
    limit points.

    On the other hand, let \(\lambda\in\partial\mathcal{D}\) and suppose
    \(n>4\). Similarly to part (b)(i), with any \(\tilde
    A_0\in\tilde{\mathbb{A}}_{\mathcal{D}}^{n-4}\), we can extend the argument
    for the \(n=4\) case with the sequence \(\tilde A_k\coloneqq B_k\oplus
    \tilde A_0\in \tilde{\mathbb{A}}_{\mathcal{D}}^n,k\in\posint\) that
    converges to \(\tilde A \coloneqq B \oplus \tilde A_0 \in
    (\tilde{\mathbb{A}}_{\mathcal{D}}^n)^c\).

    (c)(ii)---Let \(\lambda\in\partial\mathcal{D}\cap\real\) and \(n\geq 2\).
    Because \(\mathcal{D}\) is convex, open, and nonempty, there exists
    \(\varepsilon>0\) such that exactly one of the real intervals
    \((\lambda,\lambda+\varepsilon)\) or \((\lambda-\varepsilon,\lambda)\) is
    contained in \(\mathcal{D}\), whereas the other is contained in
    \(\textnormal{int}(\mathcal{D}^c)\). Without loss of generality, assume
    \((\lambda-\varepsilon,\lambda) \subseteq \mathcal{D}\).\footnote{Otherwise,
      take the reflection about the imaginary axis \(-\mathcal{D}\) and
      \(-\tilde{\mathbb{A}}_{\mathcal{D}}^n\).} Then \(A_k \coloneqq
    (\lambda-\varepsilon/k)I_n + N_n \in \tilde{\mathbb{A}}_{\mathcal{D}}^n\)
    for each \(k\in\posint\), but \(A_k\rightarrow \lambda I_n+N_n \in
    (\tilde{\mathbb{A}}_{\mathcal{D}}^n)^c\) and therefore
    \(\tilde{\mathbb{A}}_{\mathcal{D}}^n\) does not contain all its limit
    points. %
  }{}

  (d)---Since \(\overline{\mathbb{A}}_{\mathcal{D}}^n \coloneqq \set{
    A\in\realm{n}{n} | \lambda(A) \subset \textnormal{cl}(\mathcal{D}) }\)
  contains \(\mathbb{A}_{\mathcal{D}}^n\), it suffices to show any
  \(A\in\overline{\mathbb{A}}_{\mathcal{D}}^n\) is a limit point of
  \(\mathbb{A}_{\mathcal{D}}^n\). Denote the Jordan form by \(A=V\left(
    \bigoplus_{i=1}^p \mu_iI_{n_i} + N_{n_i} \right)V^{-1}\), where
  \(V\in\realm{n}{n}\) is invertible, \(\mu_i\in\lambda(A)\), \(n=\sum_{i=1}^p
  n_i\), and \(N_i\in\realm{n_i}{n_i}\) is a shift matrix. Because
  \(\mu_i\in\textnormal{cl}(\mathcal{D})\), there exists a sequence
  \(\mu_{i,k}\in\mathcal{D}\) such that \(\mu_{i,k}\rightarrow\mu_i\). Then
  \(A_k\coloneqq V\left( \bigoplus_{i=1}^p \mu_{i,k}I_{n_i} + N_i
  \right)V^{-1}\in\mathbb{A}_{\mathcal{D}}^n\) and \(A_k\rightarrow A\).
\end{proof}

\section{Proof of \Cref{prop:lmi:cont}}\label[appendix]{app:lmi:cont}
To prove \Cref{prop:lmi:cont}(a,b), we use sensitivity results on the value
functions of parameterized nonlinear SDPs,
\begin{equation}\label{eq:sdp:par}
  V(y) \coloneqq \inf_{x\in\mathbb{X}(y)} F(x,y)
\end{equation}
where the set-valued function
\(\mathbb{X}:\real^m\rightarrow\mathcal{P}(\real^n)\) is defined by
\[
  \mathbb{X}(y) \coloneqq \set{ x\in\real^n | G(x,y)\succeq 0 }.
\]
Consider also the graph of the set-valued function \(\mathbb{X}\),
\[
  \mathbb{Z} \coloneqq \set{ (x,y)\in\real^{n+m} | G(x,y)\succeq 0 }.
\]
Notice that \(\mathbb{Z}\) is closed if \(G\) is continuous. We say Slater's
condition holds at \(y\in\real^m\) if there exists \(x\in\real^n\) such that
\(x\in\textnormal{int}(\mathbb{X}(y))\), or equivalently, \(G(x,y)\succ 0\).
% In \cite[Prop.~4.4]{bonnans:shapiro:2000}, continuity of a general class of
% optimization problems is considered.
In the following proposition, we specialize
\cite[Prop.~4.4]{bonnans:shapiro:2000} to nonlinear SDPs.
\begin{proposition}[\protect{\cite[Prop.~4.4]{bonnans:shapiro:2000}}]\label{prop:sdp:cont}
  \label{prop:sdp:cont}
  Let \(y_0\in\real^m\) and suppose
  \begin{enumerate}[(i)]
  \item \(F\) and \(G\) are continuous on \(\real^{n+m}\);
  % \item[(b)] there exists a neighborhood \(N_y\) of \(y_0\) such that, for all
  %   \(y\in N_y\) and all sequences \(x_k\in\mathbb{X}(y)\) such that
  %   \(\|x_k\|\rightarrow\infty\), we have \(F(x_k,y)\rightarrow\infty\); and
  \item there exist \(\alpha\in\real\) and compact \(C\subset\real^n\)
    such that, for each \(y\) in a neighborhood of \(y_0\), the level set
    \[
      \textnormal{lev}_{\leq\alpha} F(\cdot,y) \coloneqq \set{ x\in\mathbb{X}(y)
        | F(x,y)\leq\alpha }
    \]
    is nonempty and contained in \(C\); and
  \item Slater's condition holds at \(y_0\).
  \end{enumerate}
  Then \(F(\cdot,y)\) attains a minimum on \(\mathbb{X}(y)\) for all \(y\in
  N_y\), and \(V(y)\) is continuous at \(y=y_0\).
\end{proposition}
\begin{proof}
  See \cite[Prop.~4.4]{bonnans:shapiro:2000} and the discussions in
  \cite[pp.~264,~483--484,~491--492]{bonnans:shapiro:2000}.
  % By \cite[Prop.~4.4]{bonnans:shapiro:2000}, it suffices to show
  % \begin{enumerate}
  % % \item[(i)] \(F\) is continuous on \(\real^{n+m}\);
  % \item[(i)] \(\mathbb{Z}\) is closed,
  % \item[(ii)] there exist \(\alpha\in\real\) and compact \(C\subset\real^n\)
  %   such that, for each \(y\) in a neighborhood of \(y_0\), the level set
  %   \[
  %     \textnormal{lev}_{\leq\alpha} F(\cdot,y) \coloneqq \set{ x\in\mathbb{X}(y)
  %       | F(x,y)\leq\alpha }
  %   \]
  %   is nonempty and contained in \(C\); and
  % \item[(iii)] for any neighborhood \(N_x\) of the set of solutions to
  %   \cref{eq:sdp:par} with \(y=y_0\), there exists a neighborhood \(N_y\) of
  %   \(y_0\) such that \(N_x\cap\mathbb{X}(y)\) is nonempty for all \(y\in N_y\);
  % \end{enumerate}
  % Continuity of \(G\) implies (i). Condition (b) implies that \(F(\cdot,y)\) has
  % bounded level sets for each \(y\in N_y\). Taking any \(\alpha\) for which , for each \(\delta>0\), there is some
  % \(\varepsilon>0\) such that \(F(x_k,y)>\delta\)
  %%
  % (ii), and the
  % discussions in \cite[pp.~264,~483--484,~491--492]{bonnans:shapiro:2000} show
  % the Slater condition (c) and continuity of \(G\) imply condition (iii).
\end{proof}

% Finally, we prove \Cref{prop:lmi:cont}.% with \Cref{prop:sdp:cont}.

\begin{proof}[Proof of \Cref{prop:lmi:cont}]
  Let \(\vect:\realm{n}{n}\rightarrow\real^{n^2}\) and
  \(\vecs:\realm{n}{n}\rightarrow\real^{(1/2)(n+1)n}\) denote the vectorization
  and symmetric vectorization operators, respectively.
  % In this proof, we let \(\vect:\realm{n}{n}\rightarrow\real^{n^2}\) and
  % \(\vecs:\realm{n}{n}\rightarrow\real^{(1/2)(n+1)n}\) denote operators that
  % returns a vectorization of the arugment, and of the lower triangular elements
  % of the argument, respectively.

  (a)---With \(x\coloneqq\vecs(P)\), \(y\coloneqq\vect(A)\),
  \(F(x,y)\coloneqq\tr(VP)\), and \(G(x,y)\coloneqq
  P\oplus(M_{\mathcal{D}}(A,P)-M)\), we can use \Cref{prop:sdp:cont} to show the
  continuity of \(\phi_{\mathcal{D}}\) on \(\mathbb{A}_{\mathcal{D}}^n\). Let
  \(A_0\in\mathbb{A}_{\mathcal{D}}^n\). Condition (i) of \Cref{prop:sdp:cont}
  holds by assumption. Slater's condition (iii) holds because for any \(P\succ
  0\) such that \(M_{\mathcal{D}}(A_0,P)\succ 0\), we can define
  \(P_0\coloneqq\gamma P\succ 0\) for some
  \(\gamma>\gamma_0\coloneqq\|M\| \times \|[M_{\mathcal{D}}(A_0,P)]^{-1}\|\) to give
  \[
    M_{\mathcal{D}}(A_0,P_0) = \gamma M_{\mathcal{D}}(A_0,P) \succ \gamma_0
    M_{\mathcal{D}}(A_0,P) \succeq M.
  \]
  Moreover, by continuity of \(M_{\mathcal{D}}\), there exists a neighborhood
  \(N_A\) of \(A_0\) such that \(M_{\mathcal{D}}(A,P_0)\succ M\) for all \(A\in
  N_A\). Letting \(\alpha\coloneqq\tr(VP_0)>0\), we have that the set %
  \ifthenelse{\boolean{LongVersion}}{%
    \[
      \set{ P\in\nnegdefm{n} | \tr(VP) \leq \alpha }
    \]
  }{%
    \(\set{ P\in\nnegdefm{n} | \tr(VP) \leq \alpha }\)
  }%
  is compact and contains the nonempty level set %
  \ifthenelse{\boolean{LongVersion}}{%
    \[
      \set{ P\in\mathbb{P}(A) | \tr(VP) \leq \alpha }
    \]
  }{%
    \(\set{ P\in\mathbb{P}(A) | \tr(VP) \leq \alpha }\)
  }%
  for all \(A\in N_A\). Taking the image of each of the above sets under the
  \(\vecs\) operation gives condition (ii) of \Cref{prop:sdp:cont}. All the
  conditions of \Cref{prop:sdp:cont} are thus satisfied for each
  \(A_0\in\mathbb{A}_{\mathcal{D}}^n\), and we have \(\phi_{\mathcal{D}}\) is
  continuous on \(\mathbb{A}_{\mathcal{D}}^n\).

  (b)---Continuity of \(\phi_{\mathcal{D}}\) on \(\mathbb{A}_{\mathcal{D}}^n\)
  implies closure of the sublevel sets of \(\phi_{\mathcal{D}}\), and
  \cref{eq:lmi:sublevel} follows by definition of
  \(\mathbb{A}_{\mathcal{D}}^n(\varepsilon)\).

  (c)---First, \(M_{\mathcal{D}}(A,P)\succ 0\) implies \(P\succ 0\) by
  \Cref{prop:lmi:posdef}.
  % since, if
  % \(M_{\mathcal{D}}(A,P)\succ 0\) and \(P\succeq 0\) but \(P\not\succ 0\), there
  % exists a nonzero \(v\in\real^n\) such that \(Pv=0\) and
  % \begin{multline*}
  %   (I_m\otimes v)^\top (M_{\mathcal{D}}(A,P))(I_m\otimes v)
  %   = M\otimes (v^\top Pv) \\
  %   + M_1\otimes (v^\top APv) + M_1^\top\otimes (v^\top PA^\top v) = 0
  % \end{multline*}
  % a contradiction of the assumption \(M_{\mathcal{D}}(A,P)\succ 0\).
  Moreover, for any \(P\succ 0\) such that \(M_{\mathcal{D}}(A,P)\succ 0\), we
  have \(M_{\mathcal{D}}(A,P)\succeq\gamma M_{\mathcal{D}}(A,P)\succeq M\) with
  \(P\coloneqq\gamma P\) and \(\gamma\coloneqq\|M\| \times
  \|[M_{\mathcal{D}}(A,P)]^{-1}\|\), so feasibility of \cref{eq:lmi:dstable} is
  equivalent to feasibility of
  \begin{align*}
    M_{\mathcal{D}}(A,P) &\succ M, & P&\succeq 0
  \end{align*}
  and therefore \(\bigcup_{\varepsilon>0}\mathbb{A}_{\mathcal{D}}^n(\varepsilon)
  = \mathbb{A}_{\mathcal{D}}^n\). But
  \(\mathbb{A}_{\mathcal{D}}^n(\varepsilon)\) is monotonically
  decreasing,\footnote{By ``monotonically decreasing'' we mean
    \(\varepsilon\leq\varepsilon' \Rightarrow
    \mathbb{A}_{\mathcal{D}}^n(\varepsilon) \supseteq
    \mathbb{A}_{\mathcal{D}}^n(\varepsilon')\).} so
  \(\mathbb{A}_{\mathcal{D}}^n(\varepsilon) \nearrow \bigcup_{\varepsilon>0}
  \mathbb{A}_{\mathcal{D}}^n(\varepsilon) = \mathbb{A}_{\mathcal{D}}^n\) as
  \(\varepsilon\searrow 0\).
\end{proof}
\ifthenelse{\boolean{LongVersion}}{%
\section{Proof of \Cref{prop:gbmz:value,prop:gbmz:cont}}\label[appendix]{app:bmz:approx}
}{%
\section{Proof of \Cref{prop:gbmz:cont}}\label[appendix]{app:bmz:approx}
}
\ifthenelse{\boolean{LongVersion}}{%
  Starting with \Cref{prop:gbmz:value}:
  \begin{proof}[Proof of \Cref{prop:gbmz:value}]
    Since \(\mu_\varepsilon\) is nondecreasing and bounded from below by
    \(\mu\), it suffices to show that for each \(\delta>0\), there exists a
    \(\overline\varepsilon>0\) such that
    \(\mu_{\overline\varepsilon}-\mu<\delta\).

    Let \(\theta^*\in\Theta\) denote a point for which \(\mu=f(\theta^*)\). If
    \(\theta^*\in\Theta_{++}\), we could simply choose
    \(\overline{\varepsilon}>0\) large enough to put \(\theta^*\) in
    \(\mathcal{T}(\Phi_{\overline\varepsilon})\) and achieve
    \(\mu_{\overline\varepsilon} - \mu = 0 < \delta\).

    Instead, we assume \(\theta^*\not\in\Theta_{++}\). By \Cref{assm:cons},
    there exists a sequence \(\theta_k \in \Theta_{++}, k\in\posint\) such that
    \(\theta_k\rightarrow\theta\) as \(k\rightarrow\infty\). Defining
    \(\nu_k\coloneqq f(\theta_k)\), we have \(\nu_k\rightarrow\mu\) by
    continuity of \(f\). Therefore, there exists some \(k_0\in\posint\) such
    that \(\nu_k-\mu<\delta\) for all \(k\geq k_0\). For each
    \(\theta_k\in\Theta_{++}\), there exists a unique \(\phi_k = (\beta_k,
    L^{\mathcal{I}_\Sigma}_k, L^{\mathcal{I}_{\mathcal{A}}}_k) \in \Phi\) such
    that \(\theta_k = \mathcal{T}(\phi_k)\) (by \Cref{lem:gbmz}). Let
    \(\overline\varepsilon\) be the minimum over all the diagonal elements of
    \(L^{\mathcal{I}_\Sigma}_{k_0}\) and
    \(L^{\mathcal{I}_{\mathcal{A}}}_{k_0}\). Then
    \((\beta_{k_0},L^{\mathcal{I}_\Sigma}_{k_0},L^{\mathcal{I}_{\mathcal{A}}}_{k_0})
    \in \Phi_{\overline\varepsilon}\) by construction,
    \(\nu_{k_0}\geq\mu_{\overline\varepsilon}\) by optimality, and
    \(\mu_{\overline\varepsilon} - \mu \leq \nu_{k_0} - \mu < \delta\).
  \end{proof}
}{}

As in \Cref{app:lmi:cont}, we use sensitivity results of
\cite{bonnans:shapiro:2000} on optimization problems to prove
\Cref{prop:gbmz:cont}. This time, however, we consider the continuity of the
value function for parameterized NLPs on Banach spaces. Let \(\mathcal{X}\),
\(\mathcal{Y}\), and \(\mathcal{K}\) be Banach spaces and consider the
parameterized NLP,
\begin{equation}\label{eq:nlp:par}
  V(y) \coloneqq \inf_{x\in\mathbb{X}(y)} F(x,y)
\end{equation}
where the set-valued function
\(\mathbb{X}:\mathcal{Y}\rightarrow\mathcal{P}(\mathcal{X})\) is defined by
\[
  \mathbb{X}(y) \coloneqq \set{ x\in\mathcal{X} | G(x,y)\in K }
\]
for some \(G:\mathcal{X}\times\mathcal{Y}\rightarrow\mathcal{K}\) and
\(K\subseteq\mathcal{K}\) is closed. Let \(X^0(y)\) denote the (possibly empty)
set of solutions to \cref{eq:nlp:par}. Define the graph of the set-valued
function \(\mathbb{X}(\cdot)\) by
\[
  \mathbb{Z} \coloneqq \set{ (x,y)\in\mathcal{X}\times\mathcal{Y} |
    G(x,y)\in K }.
\]
Notice that \(\mathbb{Z}\) is closed if \(G\) is continuous and \(K\) is closed.

\begin{proposition}[\protect{\cite[Prop.~4.4]{bonnans:shapiro:2000}}]\label{prop:nlp:cont}
  Let \(y_0\in\mathcal{Y}\) and assume:
  \begin{enumerate}[(i)]
  \item \(F\) and \(G\) are continuous on \(\mathcal{X}\times\mathcal{Y}\) and
    \(K\) is closed;
  \item there exist \(\alpha\in\real\) and a compact set
    \(C\subseteq\mathcal{X}\) such that, for every \(y\) in a neighborhood of
    \(y_0\), the level set
    \[
      \set{ x\in\mathbb{X}(y) | f(x,y)\leq\alpha }
    \]
    is nonempty and contained in \(C\); and
  \item for any neighborhood \(N_x\) of the solution set \(X^0(y_0)\), there
    exists a neighborhood \(N_y\) of \(y_0\) such that \(N_x\cap\mathbb{X}(y)\)
    is nonempty for all \(y\in N_y\);
  \end{enumerate}
  then \(V(y)\) is continuous and \(X^0(y)\) is outer semicontinuous at
  \(y=y_0\).
\end{proposition}

\begin{proof}[Proof of \Cref{prop:gbmz:cont}]
  First, we must specify \(\overline\varepsilon\). For each
  \(\theta\in\Theta_{++}\), let %
  \ifthenelse{\boolean{LongVersion}}{%
    \[
      \varepsilon(\theta) \coloneqq \max \set{ \varepsilon > 0 |
        \theta\in\mathcal{T}(\Phi_\varepsilon) }
    \]
  }{%
    \(\varepsilon(\theta) \coloneqq \max \set{ \varepsilon > 0 |
      \theta\in\mathcal{T}(\Phi_\varepsilon) }\) %
  }%
  where the maximum is achieved since there is a finite number of diagonal
  elements of the Cholesky factors that must be lower bounded. Now we %
  \ifthenelse{\boolean{LongVersion}}{%
    specify \(\overline\varepsilon\) as the supremum of \(\varepsilon(\theta)\)
    over all \(\theta\in\Theta_{f\leq\alpha}\cap\Theta_{++}\),
    \[
      \overline\varepsilon \coloneqq \sup \set{ \varepsilon(\theta) |
        \theta\in\Theta_{f\leq\alpha}\cap\Theta_{++} }
    \]
  }{%
    let \(\overline\varepsilon \coloneqq \sup \set{ \varepsilon(\theta) |
      \theta\in\Theta_{f\leq\alpha}\cap\Theta_{++} }\) %
  }%
  so that, for any \(\varepsilon\in(0,\overline\varepsilon)\),
  \(\Theta_{f\leq\alpha} \cap \mathcal{T}(\Phi_\varepsilon)\) is nonempty and is
  contained in the compact set \(C\).

  (a)---Following the proof of \cite[Prop.~4.4]{bonnans:shapiro:2000}, we have
  (i) \(F\) is continuous and (ii) the level set \(\Theta_{f\leq\alpha}\) is nonempty and
  contained in the compact set \(C\), which implies \(\Theta_{f\leq\alpha}\) is a
  compact level set and therefore the minimum of \(f\) over
  \(\Theta_{f\leq\alpha}\) is achieved and equals the minimum over \(\Theta\).
  Moreover, \(\hat\theta_0\) must be nonempty.

  (b)---Similarly to part (a), we have, for each \(\varepsilon \in
  (0,\overline\varepsilon)\), that the level set
  \(\Theta_{f\leq\alpha}\cap\mathcal{T}(\Phi_\varepsilon)\) is nonempty and contained
  in the compact set \(C\), so \(f\) achieves its minimum over
  \(\mathcal{T}(\Phi_\varepsilon)\) and \(\hat\theta_\varepsilon\) is nonempty.

  (c)---Consider the graph of the constraint function,
  \[
    \mathbb{Z} \coloneqq \set{(\theta,\varepsilon)\in\Theta\times\nnegreal |
      \theta\in\mathcal{T}(\Phi_\varepsilon) \textnormal{ if } \varepsilon>0 }.
  \]
  Consider a sequence \((\theta_k,\varepsilon_k)\in\mathbb{Z},k\in\posint\) that
  is convergent \((\theta_k,\varepsilon_k)\rightarrow(\theta,\varepsilon)\).
  Then \(\varepsilon\geq 0\), otherwise the sequence would not converge.
  Moreover, \(\theta\in\Theta\) since \(\theta_k \in
  \mathcal{T}(\Phi_{\varepsilon_k}) \subseteq \Theta\) for all \(k\in\posint\)
  and \(\Theta\) contains all its limit points. If \(\varepsilon=0\), then
  \((\theta,\varepsilon)\in\mathbb{Z}\) trivially. On the other hand, if
  \(\varepsilon>0\), then \(\varepsilon(\theta_k)\) converges to
  \(\varepsilon(\theta)\) because \(\mathcal{T}\) is continuous and the max can
  be taken over a finite number of elements of \(\mathcal{T}^{-1}(\theta_k)\).
  Moreover, \(\varepsilon(\theta_k)\) upper bounds \(\varepsilon_k\) because
  \(\theta_k\in\mathcal{T}(\Phi_{\varepsilon_k})\), so
  \(\varepsilon(\theta)\geq\varepsilon\). Finally, we have
  \(\theta\in\mathcal{T}(\Phi_\varepsilon)\),
  \((\theta,\varepsilon)\in\mathbb{Z}\), and \(\mathbb{Z}\) is closed.

  Let \(\varepsilon_0\geq 0\) and \(N_\theta\) be a neighborhood
  of \(\hat\theta_{\varepsilon_0}\). With %
  \ifthenelse{\boolean{LongVersion}}{%
    \[
      \delta \coloneqq \sup \set{ \varepsilon(\theta) | \theta\in N_\theta} > 0
    \]
  }{%
    \(\delta \coloneqq \sup \set{ \varepsilon(\theta) | \theta\in N_\theta} >
    0\) %
  }%
  we have \(N_\theta\cap\Theta\) and
  \(N_\theta\cap\mathcal{T}(\Phi_\varepsilon)\) are nonempty for all
  \(\varepsilon\in(0,\varepsilon_0+\delta)\).

  Finally, the requirements of \Cref{prop:nlp:cont} are satisfied for all
  \(\varepsilon_0\in[0,\overline\varepsilon)\), so \(\mu_\varepsilon\) is
  continuous and \(\hat\theta_\varepsilon\) is outer semicontinuous at
  \(\varepsilon=\varepsilon_0\).

  (d)---The last statement follows by the definition of outer semicontinuity and
  the fact that the \(\limsup\) is nonempty.
\end{proof}

\bibliographystyle{IEEEtran}
\bibliography{paper_twcccreport}

\end{document}